\documentclass[epj]{svjour}
\usepackage{graphicx}
\usepackage{amsmath}
\usepackage{amssymb}
\usepackage{latexsym}
\usepackage{enumerate}
\usepackage{amssymb}
\usepackage{slashed} 
\usepackage{color}
\begin{document}
\title{Analysis of Deeply Virtual Compton Scattering Data at Jefferson Lab and Proton Tomography}
\titlerunning{Extraction of CFFs from DVCS data}
\author{R. Dupr\'e$^{1}$ \and M. Guidal$^1$ \and S. Niccolai$^1$ \and M. Vanderhaeghen$^2$\\
}                     % Do not remove
\authorrunning{R. Dupr\'e, M. Guidal, S. Niccolai and M. Vanderhaeghen}
\institute{Institut de Physique Nucl\'eaire d'Orsay, CNRS-IN2P3, 
Universit\'e Paris-Sud, Universit\'e Paris-Saclay, 91406 Orsay, France.
\and Institut f\"ur Kernphysik and PRISMA Cluster of Excellence,
Johannes Gutenberg-Universit\"at, 55099 Mainz, Germany.
}
\date{Received: date / Revised version: date}
\abstract{
The CLAS and Hall A collaborations at Jefferson Laboratory have recently released new results for the $e p\to e p \gamma$ reaction. We analyze these new data within the Generalized Parton Distribution formalism. Employing a fitter algorithm introduced and used in earlier works, we are able to extract from these data new constraints on the kinematical dependence of three Compton Form Factors. Based on experimental data, we subsequently extract the dependence of the proton charge radius on the quarks' longitudinal momentum fraction. 
} %end of abstract
\maketitle
\section{Introduction}
\label{intro}

The past two decades have seen an important progress in the research field of
nucleon structure with the emergence of the Generalized Parton Distribution
(GPDs) formalism and its associated experimental program. The
GPDs are the structure functions of the nucleon (and of hadrons, more generally)
which are accessed in the deeply exclusive leptoproduction of a photon or a meson. They parametrize the complex non-perturbative QCD (Quantum Chromodynamics)
partonic dynamics and structure of the nucleon. In particular, in the light-front frame, where the nucleon 
is moving with large momentum, 
GPDs give access concurrently to the spatial distribution of charges in the plane perpendicular to the average nucleon 
momentum direction, and to  
the longitudinal momentum distribution of the partons in the nucleon. 
The correlation between these two distributions 
is presently still largely unknown. As a result of these position-momentum
interrelations, GPDs also provide a way to measure the unknown orbital momentum 
contribution of quarks to the total spin of the nucleon through Ji's sum 
rule~\cite{Ji97a}. We refer the reader to Refs.~\cite{Ji97a,Mueller:1998fv,Rady96a,Ji97b}
for the original articles on GPDs and to
Refs.~\cite{Goeke:2001tz,Diehl:2003ny,Belitsky:2005qn,Boffi:2007yc,Guidal:2013rya,Berthou:2015oaw,Kumericki:2016ehc} 
for reviews of the field.

\begin{figure*}[htbp] 
\vskip -4.8cm
\includegraphics[width=0.88\textwidth]{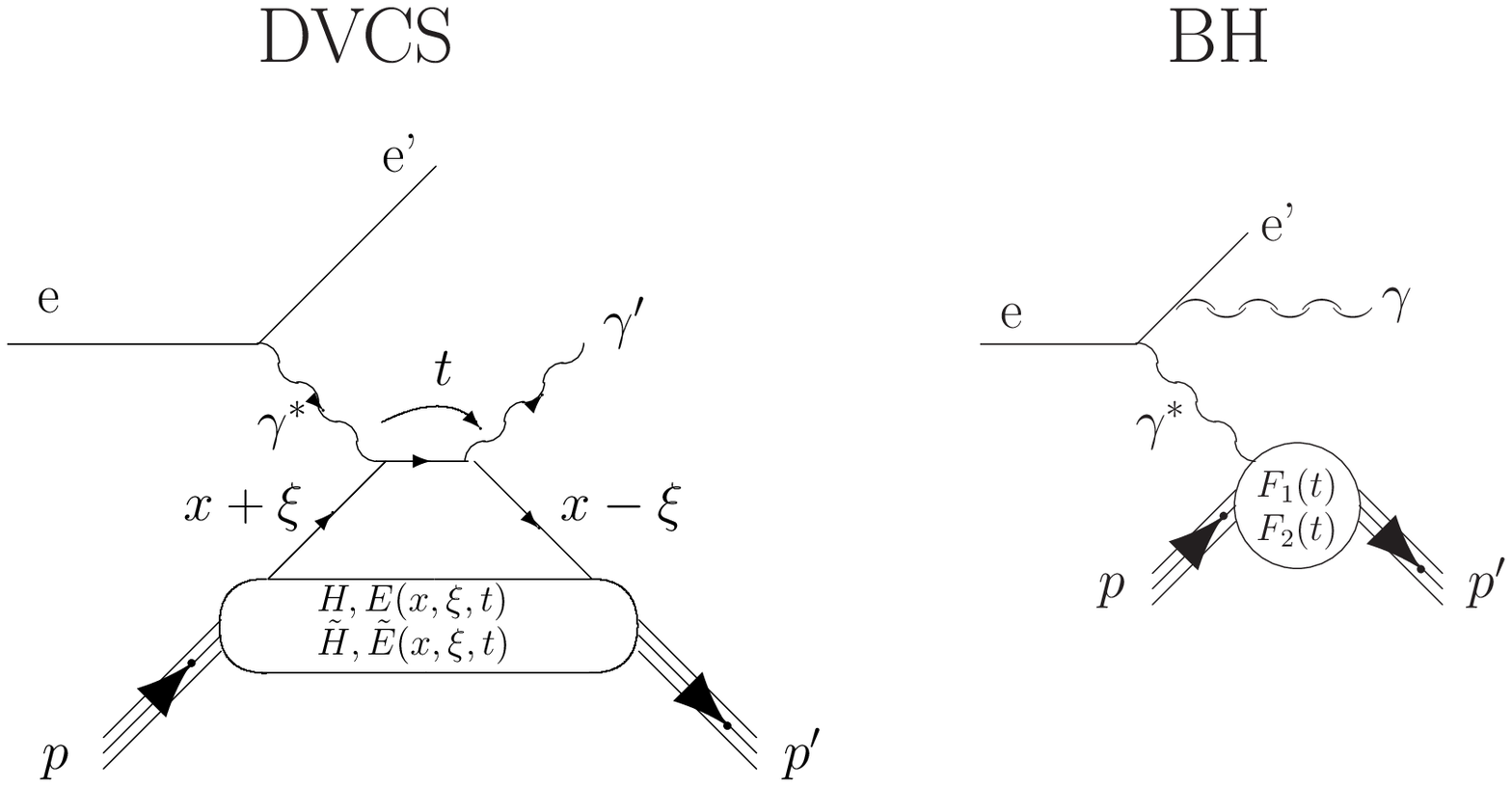}
\vskip -9cm
\caption{Left: the DVCS process (there is also a crossed diagram where the
final state photon is emitted from the initial quark). Right: the BH process
(there is also the process where the final state photon is emitted from the
initial electron). The various variables and quantities are defined in the text.}
\label{fig:dvcsbh}
\end{figure*}

GPDs are most directly accessible in Deeply Virtual Compton Scattering (DVCS).
In this process, an incoming virtual photon, emitted by a high-energy lepton beam, hits a quark of the nucleon which radiates a final real photon (Fig.~\ref{fig:dvcsbh}-left). We consider here and in the following an electron beam
and a proton target and we denote by $e$, $p$ and $\gamma^*$ ($e'$, $p'$,
and $\gamma$) the four-vectors of the initial state (final state) electron, 
proton and photon respectively. QCD states that in this process there is a factorization between the elementary photon-quark Compton scattering, which is precisely calculable in perturbative QCD,  and the GPDs, which encode the complex unknown non-perturbative dynamics of the quarks in the nucleon. This factorization has been shown to hold for sufficiently large $Q^2=(e-e')^2$, the squared momentum transfer between the final and initial leptons, and sufficiently small $t=(p-p')^2=(\gamma^*-\gamma ')^2$, the squared momentum transfer between the final and initial protons (or photons). 

In the QCD leading-twist framework, in which this work is placed, there are
four quark helicity-conserving GPDs, $H$, $E$, $\tilde H$ and $\tilde E$, parametrizing the DVCS process. 
This reflects the four independent helicity-spin
transitions between the initial and final quark-nucleon systems.
The way to disentangle the contributions of the four GPDs is to measure 
unpolarized cross sections and different spin observables for the 
$e p\to e p \gamma$ reaction. This can be done by the use of polarized 
beam, polarized target, or a combination of both.

Over the past few years, the CLAS and Hall A collaborations at Jefferson Lab (JLab), using a $\approx$~5.75 GeV electron beam, have released new results for four observables of the $e p\to e p \gamma$ reaction: unpolarized cross sections and difference of beam-polarized cross sections by the Hall A~\cite{Defurne:2015kxq} and CLAS~\cite{Jo:2015ema} experiments, as well as single and double target-spin asymmetries with longitudinally polarized target and polarized beam by the CLAS experiment~\cite{Seder:2014cdc,Pisano:2015iqa}. 

In this article, we analyze these data and extract new constraints on GPDs. Furthermore, based on DVCS data, 
we will extract the longitudinal momentum dependence ($x$-dependence) of the radius of the transverse charge 
distribution in a proton. 
The present article details and extends our earlier work published in Ref.~\cite{Dupre:2016mai}, where the specifics of the techniques used to extract the GPD information from the experimental data were not presented. 
Furthermore, we extend the analysis of Ref.~\cite{Dupre:2016mai}, where results were presented for one GPD observable, to three GPD observables in the present work. In particular, we demonstrate the constraints between real and imaginary parts of the observables involving the GPD $H$ within a dispersive framework. 

The outline of this paper is as follows. 
Section 2 of this article is devoted to a very concise review of earlier works 
on the GPD formalism and on the fitting technique that we use to extract 
the GPD information from DVCS data. Section 3 details some of
the numerous Monte-Carlo studies that were carried out to demonstrate the reliability of the fitting procedure. In Section 4, we apply the method to the Hall A and CLAS data and extract three (out of eight) Compton form factors, which parametrize the DVCS process at leading twist.  
In Section 5, we provide a physical interpretation of the extracted observables. In particular, we discuss the longitudinal momentum 
dependence of the transverse charge densities in a proton, and show the constraints imposed within a dispersive framework. Finally, we present our conclusions in Section 6.

\section{GPD formalism and fitting technique in brief}
\label{sec:form}

The GPDs are functions of three variables: $x$, $\xi$ and $t$ (Fig.~\ref{fig:dvcsbh}-left), where 
$x+\xi$ ($x-\xi$) represents the longitudinal momentum fraction of the initial (final) quark w.r.t. the average nucleon momentum~\cite{Ji97a}, and $t$ is the conjugate variable of the localization of the quark in the transverse position space (impact parameter)~\cite{Burkardt:2000za,Ralston:2001xs,Diehl:2002he}. Thus, an intuitive interpretation of GPDs is that they describe the amplitude of hitting a quark in the nucleon with momentum fraction $x+\xi$
and putting it back with a different moment fraction $x-\xi$ at a given 
transverse distance, relative to the transverse center of mass, in the nucleon.

As we are considering the DVCS process on a proton target in this work, all GPDs in the following 
stand for the quark flavor combination: 
$H(x,\xi,t) = 4/9 H^u(x,\xi,t) +  1/9 H^d(x,\xi,t) + 1/9 H^s(x,\xi,t)$, and 
similarly for the other GPDs. 

One major difficulty in the study of GPDs is that they appear in the DVCS amplitude as integrals over $x$. This is due to the loop in the DVCS diagram of Fig.~\ref{fig:dvcsbh}-left, which generates convolution terms of the form:
\begin{equation} 
\int_{-1}^{+1}d x {{GPD(x,\xi,t)} \over {x - \xi + i \epsilon}}+..., 
\end{equation}
where the denominator arises from the quark propagator. Using the residue theorem, the following 8 real quantities, hereafter referred to as Compton Form Factors (CFFs)~\footnote{We point out that the original definition of CFFs is slightly different. For instance in Ref.~\cite{Belitsky:2001ns} they are complex quantities, while, for convenience, we use real quantities in this work.}, are directly accessible via DVCS measurements:

\begin{align}
H_{Re}(\xi , t) &\equiv {\cal P} \int_0^1 d x  H_+(x, \xi, t)  
C^+(x, \xi), \label{eq:eighta} \\
E_{Re}(\xi , t) &\equiv {\cal P} \int_0^1 d x  E_+(x, \xi, t)  
C^+(x, \xi), \label{eq:eightb} \\
\tilde H_{Re}(\xi , t) &\equiv  {\cal P}  \int_0^1 d x  \tilde H_+(x, \xi, t) 
C^-(x, \xi), \label{eq:eightc} \\
\tilde E_{Re}(\xi , t) &\equiv  {\cal P}  \int_0^1 d x  
\tilde E_+(x, \xi, t) C^-(x, \xi), \label{eq:eightd} \\
H_{Im}(\xi , t) &\equiv H_+(\xi , \xi, t), \label{eq:eighte} \\
E_{Im}(\xi , t) &\equiv E_+(\xi , \xi, t),  \label{eq:eightf} \\
\tilde H_{Im}(\xi , t) &\equiv \tilde H_+(\xi , \xi, t), 
 \label{eq:eightg} \\
\tilde E_{Im} (\xi , t) &\equiv \tilde E_+(\xi , \xi, t), 
 \label{eq:eighth} 
\end{align}
where the coefficient functions $C^\pm$ are defined as:
\begin{equation}
C^\pm(x, \xi) = \frac{1}{x - \xi} \pm \frac{1}{x + \xi},
\end{equation}
and $\cal P$ denotes the principal value integral.  
The subscript "+" on the GPDs denotes their singlet (quark plus anti-quark) combinations:
\begin{eqnarray}
H_+(x,\xi,t) &\equiv& H(x, \xi, t) - H(-x, \xi, t), \\
E_+(x,\xi,t) &\equiv& E(x, \xi, t) - E(-x, \xi, t), \\
\tilde H_+(x,\xi,t) &\equiv& \tilde H(x, \xi, t) + \tilde H(-x, \xi, t), \\
\tilde E_+(x,\xi,t) &\equiv& \tilde E(x, \xi, t) + \tilde E(-x, \xi, t), 
\end{eqnarray}

Thus, the maximum model-independent information which can be extracted from the $e p\to e p \gamma$ reaction at leading twist are 8 CFFs, which depend on two variables, $\xi$ and $t$, at QCD leading order. There is an additional $Q^2$-dependence in the CFFs (and in the GPDs) if QCD evolution is taken into account. Given the small $Q^2$
ranges dealt with in this work and that the $Q^2$-evolution is in principle calculable (see Ref.~\cite{Mueller:2011xd} for a recent review), we will not consider it in the following.

Kinematically, the $e p\to e' p' \gamma$ reaction depends, for a given electron beam energy, on four independent variables. The most appropriate ones for a GPD analysis are: $\xi$, $t$, $Q^2$ and $\phi$. We already defined $Q^2$ and $t$. The variable $\xi$ is related to the standard $x_B$ variable from inclusive Deep Inelastic Scattering:
\begin{equation}\label{xi_def}
\xi=\frac{x_B}{2-x_B},
\end{equation} 
with $x_B=\frac{Q^2}{2m_p(E_e-E_{e'})}$, where $m_p$ is the proton mass, $E_e$ the incident beam energy, and $E_{e'}$ the scattered electron energy. The angle $\phi$ is the azimuthal angle between the electron scattering plane and the hadronic production plane.  

A further complexity in studying GPDs via DVCS is that there is an additional significant mechanism contributing to the $e p \gamma$ final state, the Bethe-Heitler (BH) process. In this process (Fig.~\ref{fig:dvcsbh}-right) the final state photon is radiated by the incoming or scattered electron, and not by a quark of the nucleon. The BH and DVCS mechanisms interfere at the amplitude level. However, the BH amplitude is precisely calculable theoretically. The only non-QED inputs in the calculation are the nucleon elastic form factors $F_1(t)$ and $F_2(t)$ and these are well known at the small momentum transfers $t$ considered in this work. Consequently, the only unknown theoretical quantities entering the computation of the $e p\to e p \gamma$ observables are therefore the eight CFFs. 

In Refs.~\cite{Guidal:2008ie,Guidal:2009aa,Guidal:2010ig,Guidal:2010de,Boer:2014kya}, we proposed and applied a method to extract CFFs in a quasi model-independent way. It consists in taking the 8 CFFs as free parameters and, knowing the well-established BH and DVCS leading-twist amplitudes, to fit, at a fixed ($x_B$, $t$) kinematics, simultaneously the $\phi$-distributions of several $e p\to e p\gamma$ experimental observables. 
If the range of variation of the CFFs is limited, the dominant CFFs contributing to the observables which are fitted are obtained from the fit procedure with finite error bars. These error bars are mainly due to the correlations between the CFFs. Rather than the error on the experimental data, they reflect the influence of the other subdominant CFFs, as we shall see in the following. The approach of fitting CFFs at fixed ($x_B$, $t$) kinematics is called ``local fitting''. Aside from the limits imposed on the variation of the CFFs, which will be discussed in the following sections, it has the merit of being mostly model-independent as there is no need to assume and hypothesize any functional shape for the CFFs. The method has also its drawbacks, in particular it only makes use of the data available at a particular ($x_B$, $t$) kinematics, without exploiting potentially useful neighbouring data. 
Nevertheless, with this local fitting method, in our earlier works, we managed to derive limits and constraints for the $H_{Im}$, $\tilde H_{Im}$ and $H_{Re}$ CFFs, with an average 40\% relative uncertainty for $H_{Im}$, at JLab~\cite{Guidal:2008ie,Guidal:2010ig} and HERMES~\cite{Guidal:2009aa,Guidal:2010de} kinematics. 

In the following, we analyze with this fitting technique the new CLAS and Hall-A DVCS data. We will denote the unpolarized cross sections, difference of beam-polarized cross sections, longitudinally polarized target single spin asymmetries and beam-longitudinally polarized target double spin asymmetries, respectively, as $\sigma$, $\Delta\sigma_{LU}$, $A_{UL}$ and $A_{LL}$. The two indices refer respectively to the polarization of the beam and of the target ($U$ for unpolarized and $L$ for longitudinally polarized). The Hall-A collaboration has measured the $\phi$ distribution of $\sigma$ and $\Delta\sigma_{LU}$ 
for 20 ($x_B$, $Q^2$, $t$) bins in the phase space 0.34$ \lesssim x_B\lesssim$ 0.40, 1.98 $\lesssim Q^2\lesssim$ 2.36 GeV$^2$, 0.15 $\lesssim -t\lesssim$ 0.40 GeV$^2$. The CLAS collaboration has measured the $\phi$ distribution of $\sigma$ and $\Delta\sigma_{LU}$ for more than 100 ($x_B$, $Q^2$, $t$) bins in the phase space 0.12 $\lesssim x_B\lesssim$ 0.50, 1.11$\lesssim Q^2\lesssim$ 3.90 GeV$^2$, 0.12 $\lesssim -t\lesssim$ 0.45 GeV$^2$, and the $\phi$ distribution of $A_{UL}$ and $A_{LL}$ for 20 ($x_B$, $Q^2$, $t$) bins in approximately the same phase space.

\section{Monte-Carlo studies}
\label{sec:1}

We present in this section some examples of the simulations that we have carried out in order to test and demonstrate the reliability and robustness of our fitting method. 
We consider the least constrained and most challenging case, having at our disposal only two observables: the unpolarized cross section $\sigma$ and the difference of beam-polarized cross sections $\Delta\sigma_{LU}$. Additional observables can of course only improve the situation, as will be shown with real data in the next section. 

Each DVCS observable receives contributions from several CFFs, which are strongly correlated. Thus, the extraction of 8 CFFs from only two observables, with finite experimental uncertainties, is an underconstrained problem. 
However, some observables are dominated by and mostly sensitive to one or two CFFs compared to the others. For instance, it is well known~\cite{Belitsky:2001ns} that $\Delta\sigma_{LU}$ is dominated by the $H_{Im}$ CFF and that $A_{UL}$ is strongly sensitive to $\tilde H_{Im}$. Other CFFs contribute to these two observables, but they are kinematically suppressed, all the more in comparison to the experimental uncertainties. Therefore, in order to progress from the unconstrained problem it was decided to limit, in a conservative and educated way, the range of variation of the CFFs, especially the sub-dominant ones. While keeping the 8 CFFs in the fit, this effectively and essentially reduces the problem to fitting the one or two dominant CFFs to the one or two experimental observables. The influence of the sub-dominant CFFs, over the domain in which they are allowed to vary, is then reflected in the resulting uncertainty on the dominant CFFs extracted. The only model-dependent input in this approach is the definition of the range of variation of the CFFs. We illustrate and clarify the approach in the following sub-sections. 

\subsection{Pseudo-data Generation}

In a first stage, we generate, for a given ($x_B$, $Q^2$, $t$) kinematic bin
and a given beam energy, the unpolarized cross sections and the difference of beam-polarized cross sections of the $e p\to e p \gamma$ process as a function of $\phi$, based on the leading-twist and leading-order DVCS+BH amplitude. 

For our first example, we take the particular kinematics ($x_B$, $Q^2$, $t$)=($0.126, 1.1114, -0.1078$) with a 5.75-GeV beam energy. This corresponds to a kinematic bin measured by the CLAS experiment. We generate 24 $\phi$ points like for the experimental data. Then, the only inputs needed to generate the cross sections are the 8 CFFs entering the DVCS amplitude. We shall generate them randomly. 
In order to keep the problem realistic, we pick them in a bounded 8-fold hypervolume, whose limits are defined as $\pm 5$ times the CFFs predicted by the VGG model. VGG~\cite{Goeke:2001tz,Vanderhaeghen:1998uc,Vanderhaeghen:1999xj,Guidal:2004nd} is a well-known and widely used GPD model which obeys most of the model-independent GPD normalization constraints and which reproduces the general trends of the existing DVCS data (see Refs.~\cite{Jo:2015ema,Seder:2014cdc,Pisano:2015iqa} for instance).
Centering the 8-CFF hypervolume around the VGG model and limiting it to a $\pm 5$ factor prevents the fitter from exploring too unlikely cases.

For obvious symmetry reasons due to this definition of the 8-CFF hypervolume 
, it was chosen not to generate the CFF values themselves but rather their ``multipliers", i.e. their deviations from the VGG CFFs. In other words, we generate 8 random numbers between -5 and +5. The CFFs entering the DVCS amplitude are then the product of these multipliers by the VGG reference CFFs. As an illustration, for this first example, we list here the 8 randomly generated CFFs multipliers that have been generated, which are denoted as $a(CFF)$:

\begin{equation}
\begin{aligned}
a(H_{Re})&=3.191610 &  a(E_{Re})&=2.378950  \\
a(\tilde H_{Re})&=3.167072 &  a(\tilde E_{Re})&=3.091025  \\
a(H_{Im})&=3.124754 &  a(E_{Im})&=-2.095427 \\
a(\tilde H_{Im})&=1.641959 &  a(\tilde E_{Im})&=-3.279582 .
\end{aligned}
\label{eq:cffset}
\end{equation}

The CFFs used for the cross section calculations are then the result of the product of these multipliers by the VGG reference CFFs which are, at the ($x_B$, $Q^2$, $t$)=(0.126, 1.1114 GeV$^2$, -0.1078 GeV$^2$) kinematics:

\begin{equation}
\begin{aligned}
H_{Re}&=3.30098 & E_{Re}&=2.69182 \\
\tilde H_{Re}&=0.116259 & \tilde E_{Re}&=-263.284537 \\
H_{Im}&=5.09888 & E_{Im}&=1.01539 \\
\tilde H_{Im}&=0.590312 & \tilde E_{Im}&=-263.28453. 
\end{aligned}
\label{eq:vggcffs}
\end{equation}

Some of the multipliers in Eq.~\ref{eq:cffset} are very far from 1. They correspond probably to quite
unrealistic CFFs.
For instance, $a(H_{Im})=3.124754$ means that the generated $H_{Im}$ CFF is more than 3 times 
the VGG value. Given that GPDs have to fulfill a certain number of normalization 
constraints~\cite{Goeke:2001tz,Diehl:2003ny,Belitsky:2005qn,Boffi:2007yc,Guidal:2013rya,Berthou:2015oaw,Kumericki:2016ehc}, such a strong deviation from the VGG reference value is quite unlikely. 
We consider however that exploring and scanning such a large range of values should make our case all the more robust and convincing.

\begin{figure}[htbp] 
\vskip -1.8cm
\includegraphics[width=0.58\textwidth]{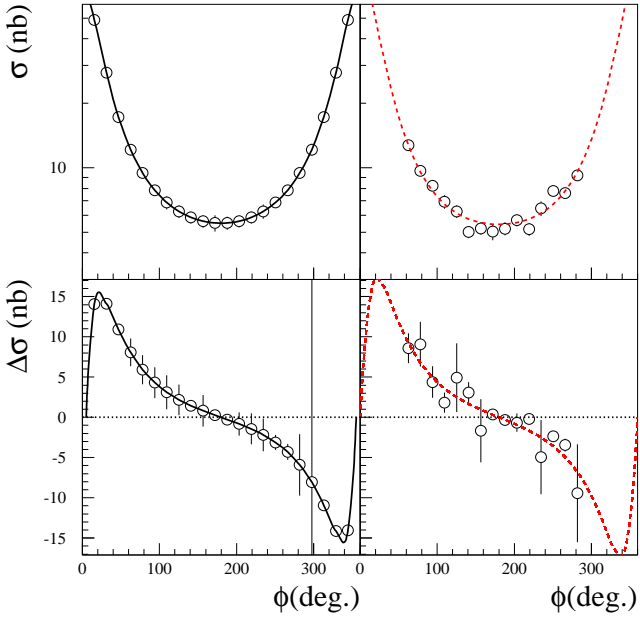}
\caption{Pseudo-data, generated from 8 randomly generated CFFs (see Eqs.~\ref{eq:cffset} and \ref{eq:vggcffs}) for the kinematics ($x_B$, $Q^2$, $t$)=($0.126, 1.1114$ GeV$^2, -0.1078$ GeV$^2$) corresponding to one bin measured by the CLAS experiment. The unpolarized cross section (top) and the difference of beam-polarized cross section (bottom) are shown unsmeared (left) and smeared (right). The solid lines show the originally generated distribution. The dashed line show the results of the fits (see text for details).}
\label{fig:exam_gen}
\end{figure}

The goal of this study is to find out if, by fitting the generated $\phi$ pseudo-data distribution, we are able to retrieve, or constrain, the 8 original randomly generated CFFs multipliers, or at least some of them, under realistic experimental conditions. For the latter, we smear the theoretically calculated cross sections according to the experimental uncertainties of the Hall A and CLAS experiments. Figure~\ref{fig:exam_gen} shows the $\phi$ dependence of the $e p\to e p \gamma$ unpolarized cross section and difference of beam-polarized cross sections (top and bottom panels respectively), unsmeared and smeared (left and right panels respectively), generated with the 8 random CFFs multipliers of Eq.~\ref{eq:cffset}, multiplied by the 8 VGG CFFs of Eq.~\ref{eq:vggcffs}.

In Fig.~\ref{fig:exam_gen}, the 24 $\phi$ points superimposed on the theoretical curves are equidistant. This corresponds approximately to the $\phi$ binning of the experimental data. We added on those points the error bars corresponding to the published experimental uncertainties of the CLAS data. For this particular bin, they range from $\approx$ 5\% to $\approx$ 9\% for the unpolarized cross section and from $\approx$ 20\% to more than 100\% for the difference of beam-polarized cross section. On the left panels of Fig.~\ref{fig:exam_gen}, the three lowest $\phi$ and the three largest $\phi$ points have no error bar. 
This means that these $\phi$ regions were actually not measured experimentally, likely for detector acceptance issues. Thus, these 6 $\phi$ don't appear on the right panels of Fig.~\ref{fig:exam_gen}, which are meant to mimic real data with the use of smearing (we however recall that the cross sections in Fig.~\ref{fig:exam_gen} are not the measured ones since they have been generated with random CFFs here).
The error bar values and the accessible $\phi$ regions vary for each ($x_B$, $Q^2$, $t$) bin, and differ for the Hall-A and CLAS experiments. 
%Fig.~\ref{fig:exam_gen} shows only the $\phi$ coverage and cross sections uncertaintites
%of the particular ($0.126, 1.1114, -0.1078$) CLAS bin for illustration (while we recall that
%the cross sections in Fig.~\ref{fig:exam_gen} are not the measured ones, here we have
%randomly generated them).

The smearing of the points of the right part of Fig.~\ref{fig:exam_gen} has been done via a Gaussian distribution, centered at the theoretically computed value, with a standard deviation corresponding to the experimental uncertainties (i.e. the error bars of the points of the left part of the figure). 
%In other words, the cross section of each $\phi$ point was shifted to another value, picked randomly about the 
%theoretically calculated value of the left part of the figure following a Gaussian distribution. 
Each $\phi$ point was smeared independently of the other $\phi$ points.
The right part of Fig.~\ref{fig:exam_gen} shows one particular instance of such a series of smearings. 
In the following, we will carry out our studies for several random smearings so
that we are not biased by one particular smearing. 
%In the plots of the right part of Fig.~\ref{fig:exam_gen}, 
%we display only the points which have a non-zero error
%bar, i.e. which can be accessed and measured in the CLAS experimental conditions
%and thus enter in our fit. 
Under these conditions, we deem that in the following we will perform our fits in rather realistic conditions, taking into account the $\phi$-coverage of the data, their dispersion and their uncertaintities.

\subsection{Pseudo-data Fitting}

The second stage of the study consists in fitting the generated $\phi$ distributions leaving the 8 CFFs as free parameters. This should be done, ideally, in ``blind" conditions, i.e. not making use of the knowledge of the originally generated CFF values. However, as was mentioned earlier, the condition for the fitting procedure to converge is to limit the hyperspace in which the 8 CFFs are allowed to vary. The choice of the values of these boundaries is the only model-dependent input in our approach. 
We take the same hyperspace in which the 8 CFFs were originally generated, i.e. $\pm 5$ times the VGG CFFs. 
%In such framework, where the parameter hyperspace is centered on the VGG CFFs, 
Like for the generation of the CFFs, we take as the free parameters of the fit, rather than the absolute CFFs themselves, the relative deviations from the reference VGG CFFs. We will therefore fit in the following the multipliers of the VGG CFFs, with the goal to recover the originally generated ones.

For the minimization we use the least squares method. We minimize $\chi^2$, defined as follows: 

\begin{equation}
\chi^2=\sum_{i=1}^{n}
\frac{(\sigma^{theo}_i-\sigma^{data}_i)^2}{(\delta\sigma^{exp}_i)^2}
+\frac{(\Delta\sigma^{theo}_i-\Delta\sigma^{data}_i)^2}{(\delta(\Delta\sigma^{data}_i))^2}
\label{eq:chi2}
\end{equation}

In Eq.~\ref{eq:chi2}, $\sigma^{theo}$ ($\Delta\sigma^{theo}$) is the theoretical 
DVCS+BH cross section (difference of beam-polarized cross section), which depend on the CFFs multipliers, which are the free parameters of the fit. The quantities $\sigma^{data}$, $\delta\sigma^{data}$, and $\Delta\sigma^{data}$, $\delta\Delta\sigma^{data}$, are, respectively, the values and the uncertainties of the pseudo- or experimental data. 
The index $i$ runs over all the available $\phi$-points for a given ($x_B$,$Q^2$,$t$) bin. 
We use the well-known MINUIT code from CERN~\cite{james} with the MINOS option. With this option, 
MINUIT calculates $\chi^2$ at multiple points of the multi-dimensional 
hyperspace of the free parameters. Thus,
step by step, the full phase space of the free parameters is explored. 
This method is costly in terms of computing power and time but it allows, numerical precision and step-size issues aside, to find the global minimum (or minima) of the problem, reducing the risk of falling into local minima. 
In parallel, it allows to determine the errors on the fitting parameters. The 1-$\sigma$ uncertainty on a given parameter corresponds to the value of
this parameter for $\Delta\chi^2=+1$ above $\chi^2_{min}$, the minimum $\chi^2$ value. 
When the problem is not linear and when the $\chi^2$ shape is not a simple parabola or a simple function, as in our case, this is the only way to determine this error.

\begin{figure}[htbp] 
\includegraphics[width=0.50\textwidth]{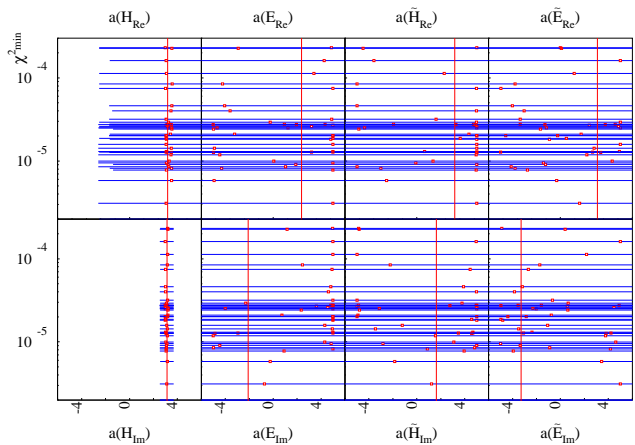}
\vskip -2cm
\caption{Results of a series of fits, differing by their randomly generated starting values, of the $\sigma$ and $\Delta\sigma_{LU}$ pseudo-data of the left part of Fig.~\ref{fig:exam_gen}, i.e. without smearing. The red dots show for each fit, on the $x$-axis, the values of the CFFs multipliers which minimize the problem and, on the $y$-axis, the corresponding $\chi^2_{min}$ value. The blue bars indicate the 1-$\sigma$ uncertainty corresponding to $\chi^2_{min}+1$.
The non-finite error bars observed for the CFFs other than $H_{Im}$ mean that the $\chi^2_{min}+1$ value lies out of the $\pm 5$ times VGG CFF range.
The red vertical lines indicate the CFF-multiplier values used for the generation of the pseudo-data (see Eq.~\ref{eq:cffset}).}
\label{fig:himvschi2_sim_nosmear}
\end{figure}

\subsubsection{Non-smeared pseudo-data}
We start from the simplest case: fitting the pseudo-data of the left part of Fig.~\ref{fig:exam_gen}, $\sigma$ and $\Delta\sigma_{LU}$, without smearing. It is important to make sure that the result of the fit is not dependent on the particular starting values of the 8 CFFs. Indeed, by selecting or favoring specific starting points in the 8-dimensional CFF hypervolume, one can end up in a particular local minimum. We therefore carried out the fits several hundreds of times with arbitrary starting points, randomly selected in the $\pm 5$ times VGG CFF hypervolume. Figure~\ref{fig:himvschi2_sim_nosmear} shows with the red dots the results of the fits for the 8 CFFs (or rather their multipliers) as a function of $\chi^2_{min}$, for a random sample of hundreds of starting points. The blue bars indicate the 1-$\sigma$ uncertainty corresponding to $\chi^2_{min}+1$. The $\chi^2_{min}$ values of the fits are very low, of the order of $10^{-5}$. We recall that in this first exercise no smearing was applied to the pseudo-data. Thus, all the fits go exactly through the data points. Therefore, the precise $\chi^2_{min}$ values and their dispersion are not very meaningful in this case (incidentally, note that the plotted $\chi^2_{min}$ values here are not normalized, i.e. they are not divided by the number of degrees of freedom). 

What is apparent in Fig.~\ref{fig:himvschi2_sim_nosmear} is that, out of the eight CFFs, only $H_{Im}$ emerges from the fit with a quite well-nailed minimum and finite error bars (of the order of $\approx$20\%). This happens systematically and invariably, which\-ever the starting point in the 8-dimen\-sional CFF-multiplier hypervolume. 
All $H_{Im}$ minima lie very closely to the originally generated $a(H_{Im})=3.124754$ (see Eq.~\ref{eq:cffset}), which is indicated by the vertical red line in Fig.~\ref{fig:himvschi2_sim_nosmear}.
One can also note that, in most cases, the error bars of $a(H_{Im})$ appear asymmetric. We will encounter such asymmetric errors often in the following. This is the signature of a non-parabolic $\chi^2$ profile and of a non-linear problem. This is expected as CFFs contribute in a bilinear way to the unpolarized cross section (although in a linear way to the beam-polarized cross section)~\cite{Belitsky:2001ns}.
The non-finite error bars observed for the other seven CFFs mean that the $\chi^2_{min}+1$ value lies out of the $\pm 5$ times VGG CFF range. 
Some partial information can nevertheless be extracted for $H_{Re}$ as, while the positive error bar is infinite, the negative one appears to be finite. Also, the minimum $\chi^2_{min}$ values for $H_{Re}$ lie, with some dispersion, around the originally generated value. This is not the case for the remaining six CFFs which have both negative and positive error bars non-finite, and for which the values of $a()$ which minimize the problem are essentially randomly distributed between $-5$ and $+5$. There is in some cases a tendency for some of these non-converging CFFs to have their multipliers clustering near the edges
of the allowed domain, i.e. $-5$ and $+5$. We will come back to this point further down.

In summary, these first results show that $\sigma$ and $\Delta\sigma_{LU}$ are dominantly sensitive to the $H_{Im}$ and $H_{Re}$ CFFs and that these two CFFs seem, in the present ideal (i.e. unsmeared) conditions, to be recoverable, albeit only partially for $H_{Re}$, from the simultaneous fit of $\sigma$ and $\Delta\sigma_{LU}$. 
 
\subsubsection{Smeared pseudo-data}\label{sect_smeared_ps}
\begin{figure}[htbp] 
\includegraphics[width=0.5\textwidth]{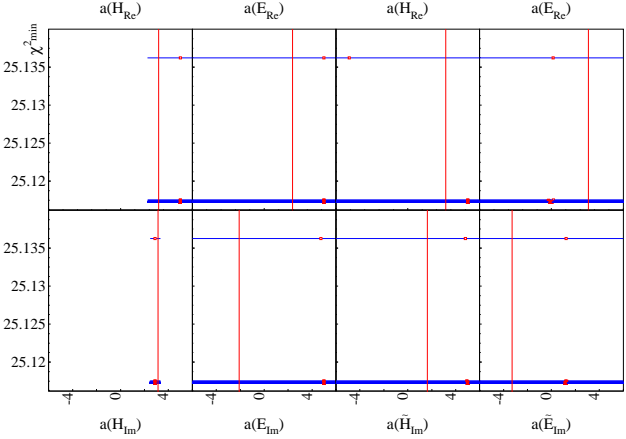}
\vskip -2cm
\caption{Results of a series of fits, differing by their randomly generated starting values, of the $\sigma$ and $\Delta\sigma_{LU}$ pseudo-data of the right part of Fig.~\ref{fig:exam_gen}, i.e. with smearing. The figure shows, for each fit, on the $x$-axis the values of the CFFs multipliers which minimize the problem, and on the $y$-axis the corresponding $\chi^2_{min}$ value. 
The red vertical lines indicate the CFF-multiplier values used for the generation of the pseudo-data (see Eq.~\ref{eq:cffset}).}
\label{fig:himvschi2_sim_smear}
\end{figure}

Figure~\ref{fig:himvschi2_sim_smear} shows the result of the same kind of study on smeared pseudo-data, such as those in the right part of Fig.~\ref{fig:exam_gen}. For this particular smearing of the data, we also performed the fits with many starting points in the 8-dimensional CFF hypervolume. Fig.~\ref{fig:himvschi2_sim_smear} shows that all fits led to the same set of 8 CFFs solution. 
Indeed, compared to Fig.~\ref{fig:himvschi2_sim_nosmear}, there is here no dispersion of the solutions for the non-dominant CFFs.
We tend to attribute the dispersion of the solutions that was observed in Fig.~\ref{fig:himvschi2_sim_nosmear} to the very
low $\chi^2_{min}$ values, which were, we recall, of the order of $10^{-5}$. Such low values reflect the ill-nature of the problem of fitting
data points which are not smeared. Then $\chi^2_{min}$ values, at the limit of the numerical precision of the minimizing algorithms, have little significance.
In Fig.~\ref{fig:himvschi2_sim_smear}, which correspond to fits of smeared data, the unnormalized-$\chi^2$ values are indeed now of the order of 25. 
This is consistent with the observation that there are 30 data points which are fitted in the right part of Fig.~\ref{fig:exam_gen}.
In this latter figure, the dashed curves on the smeared data (right part of the figure) actually show the results of the fits with the values 
of the 8 CFFs multipliers extracted from Fig.~\ref{fig:himvschi2_sim_smear}.  

Regarding the results for the $H_{Im}$ and $H_{Re}$ CFFs, from Fig.~\ref{fig:himvschi2_sim_smear} we reach conclusions which are almost similar to the previous case, with the unsmeared pseudo-data. Namely, all the fits, independently of their starting values, allow to recover the originally generated value of $H_{Im}$ 
(at the $\approx$ 20\% level) and partially that of $H_{Re}$, with its finite negative error bar.
For most of the other (non-dominant) CFFs, the fits find values on the edge of the allowed CFF range, i.e. $\pm$ 5. 
We will come back to this point further down.
 
A closer look at Fig.~\ref{fig:himvschi2_sim_smear} reveals that the values of $a(H_{Im})$ and $a(H_{Re})$ corresponding to $\chi^2_{min}$
(red points in Fig.~\ref{fig:himvschi2_sim_smear}) are not exactly
centered on the originally generated values (red lines). In particular, $a(H_{Re})$, is clearly shifted to the right compared to the generated value
(which nevertheless lies well within the negative blue error bar). The origin of such shift is the particular smearing of the data that we introduced 
and can accidentally bias the $\phi$ distributions in a given direction (overall decrease or increase of the $\phi$ distributions). 

\begin{figure}[htbp] 
\includegraphics[width=0.5\textwidth]{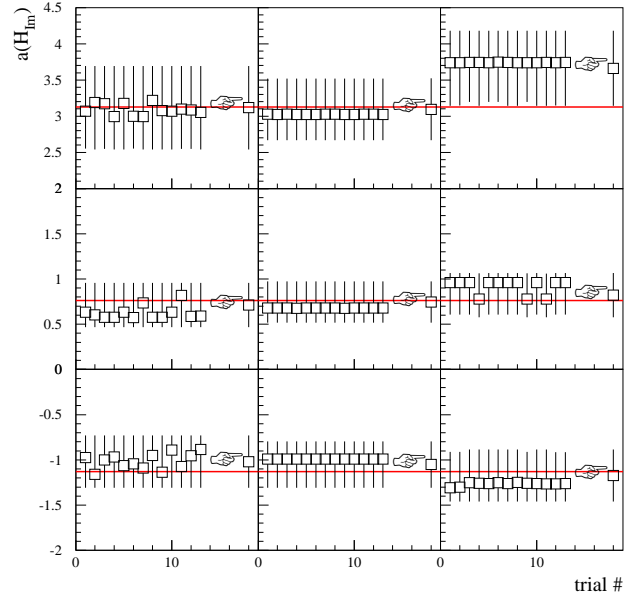}
\caption{The result of several fits for $a(H_{Im})$, differing by their randomly
generated starting values (``trial''). The results are shown for a selection of
three different random sets of generated CFFs (rows) and, for each of
these, three different random smearings of the pseudo-data (columns). The 
original value of the CFF multiplier is marked by the red line. The point indicated by the ``hand" shows the final unique solution according to the prescription that we advocate and describe in Section~\ref{subsub_prescription}: taking the largest error bars of all solutions and their middle as most probable value.}
\label{fig:sim_him}
\end{figure}

Indeed, the smearing of the data that we adopted in Fig.~\ref{fig:exam_gen} was a particular random one. It has to be checked for other smearings that our
fit procedure is also able to recover well the $H_{Im}$ CFF (in particular),
from the simultaneous fit of $\sigma$ and $\Delta\sigma_{LU}$, in order to confirm
the robustness of the method. Figure~\ref{fig:sim_him} shows, still for the CLAS kinematics 
($x_B$, $Q^2$, $t$)=($0.126, 1.1114, -0.1078$), a sample of fit results
for various smearings of the $\phi$ distributions and different generated CFF values. 
Each column corresponds to a different smearing, the first column having no smearing, and each row to a different set of generated CFF multipliers for $a(H_{Im})$. 

The abscissa represents different``trials", i.e. different randomly generated starting points. We plot in the figure only a small sample for sake of visibility.

Among the hundreds of different smearings and CFFs choices, we chose the nine particular cases of Fig.~\ref{fig:sim_him} as they illustrate different typical situations. 
Fig.~\ref{fig:sim_him} shows the ideal no-smearing case on the left column, one recognizes the small dispersion of the fitted $a(H_{Im})$'s, which lie close to the originally generated ones. This generalizes what we observed
in Fig.~\ref{fig:himvschi2_sim_nosmear}. Every single fit, differing only
by its starting values, leads to a slightly different solution, always
close to the originally generated value (with a $\chi^2_{min}$ value of the order 
of $10^{-5}$, not shown in Fig.~\ref{fig:sim_him}). 
It is remarkable that even though the solutions slightly vary between trials, the range defined by the positive and negative error bars always remains the same. In other words, even if the $\chi^2_{min}$ value happen to fluctuate, the $\chi^2_{min}+1$ values seem to be well delineated. As illustrated by the three rows of the first column of the figure, this is in general the case independently of the originally generated $a(H_{Im})$, be it positive or negative, close to 0 or not. 

The next two columns of Fig.~\ref{fig:sim_him} illustrate the solutions that
one typically finds for non-zero smearings. Like we noticed and discussed with Fig.~\ref{fig:himvschi2_sim_smear}, when smearing is involved, there is quite less dispersion of the solutions.
All trials, only differing by their starting points, converge in general to one or a couple of stable values, which have very similar $\chi^2_{min}$ values (of the order of 25, like in Fig.~\ref{fig:himvschi2_sim_smear}). In particular, in the right-column/central-row plot, one clearly distinguishes two values of $a(H_{Im})$ which minimize the problem and which are attained depending on the starting values of the fit parameters. There is almost no difference in the $\chi_{min}$ between the two solutions: the $a(H_{Im})\approx0.96$ solution has $\chi^2_{min}\approx 27.14$ while the $a(H_{Im})\approx0.77$ solution has $\chi^2_{min}\approx 27.46$. 
As a matter of fact, the solution that has the slightly larger $\chi^2_{min}$ value is the one which has the fitted $H_{Im}$ value the closest to the originally generated one ($a(H_{Im})=0.761933$ in this particular case). The range of the error bars of the two solutions is very similar, although the negative error bar appears slightly larger for one solution than for the other. In general, be it for single or multiple solutions cases, the error bars
are very similar from one trial to the other. Again, even though the $\chi^2_{min}$ value might not be unique and well defined, the $\chi^2_{min}+1$ values appear to be rather well specified. 

In all plots of Fig.~\ref{fig:sim_him}, the red horizontal line indicates the originally generated $a(H_{Im})$ value. It is remarkable that it is always contained in the largest error bars of the fitted values. It is admittedly at the very edge for the top right plot; among our hundreds of smearings, we selected this particular one, which is not at all a general case, as an illustration of an ``extreme" case.

\begin{figure}[htbp] 
\includegraphics[width=0.5\textwidth]{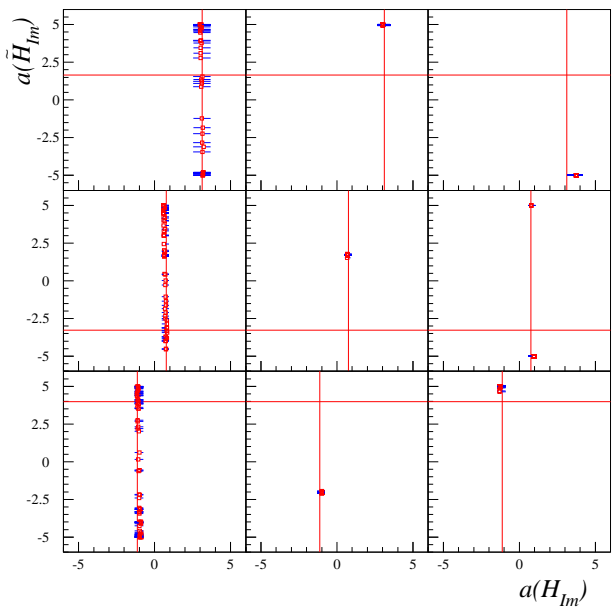}
\caption{Results of the fitted $a(H_{Im})$'s and $a(\tilde H_{Im})$'s for the same fits as in Fig.~\ref{fig:sim_him}. Each square corresponds to a fit with different starting values. The red lines indicate the originally
generated values. The error bars of $\tilde{H}$ are not shown, as they are all
infinite (i.e. they extend beyond the $\pm 5$ range) and would clutter the plot 
too much.}
\label{fig:sim_himhtim}
\end{figure}

One can better understand some of these behaviors by examining Fig.~\ref{fig:sim_himhtim}. For the same nine conditions of Fig.~\ref{fig:sim_him}, the figure shows to which $a(\tilde H_{Im})$ value the $a(H_{Im})$ solution corresponds to. We consider this correlation since $\tilde H_{Im}$ is expected to be the next dominant contributor to $\Delta\sigma_{LU}$ after $H_{Im}$~\cite{Belitsky:2001ns}.
The upper left plot of Fig.~\ref{fig:sim_himhtim} shows that the apparently randomly distributed $a(H_{Im})$ solutions around the originally generated value of the upper left plot of Fig.~\ref{fig:sim_him} actually correspond each to a different value of $a(\tilde H_{Im})$, all distributed along the whole allowed $\pm$ 5 range
(error bars on $\tilde{H}$ extend beyond the $\pm 5$ range and only the central
values are plotted in Fig.~\ref{fig:sim_himhtim}). 
It reveals (confirms) the strong correlation between these two CFFs. Depending on the starting point, the fitter code ends up in ($a(H_{Im})$, $a(\tilde H_{Im})$) correlated solutions. 
One notices that while $\tilde H_{Im}$ is not constrained at all within the $\pm 5$ range, $H_{Im}$ is always contained in a very limited range. This latter range is defined by the $\chi^2_{min}+1$ error bar, whose projection is displayed in Fig.~\ref{fig:sim_him}. 
We actually see that what determines the error bar on $H_{Im}$ is the range of variation allowed for $\tilde H_{Im}$ (this effect was studied in detail in Ref.~\cite{Boer:2014kya}). Were $\tilde H_{Im}$ allowed to vary in a domain larger than $\pm 5$ times the VGG CFF hyperspace, the error bar on $H_{Im}$ would be bigger (and conversely). This is why the error bars on $H_{Im}$ that we obtained so far are in general of the order of 20 to 30\% (see Fig.~\ref{fig:sim_him}), i.e. somewhat larger than the experimental precision of the data. Once again, this is because they reflect the influence of the other CFFs (mostly $\tilde H_{Im}$ in the present case) and their correlation with $H_{Im}$. Therefore, the value of $H_{Im}$ will be better determined by having some extra constraint on $\tilde H_{Im}$ such as additional 
observables.

When smearing is introduced (second and third columns of Figs.~\ref{fig:sim_him} and \ref{fig:sim_himhtim}) the well-defined single or double $a(H_{Im})$ values correspond to, also, well-defined single or double values for $a(\tilde H_{Im})$. In several cases, these $a(\tilde H_{Im})$ values are actually on the edge of the allowed phase space, i.e. $\pm 5$. In particular, the double solution for $a(H_{Im})$ that is found for the right-column/central-row plot of Figs.~\ref{fig:himvschi2_sim_smear} and~\ref{fig:sim_him} corresponds to two extreme values for $a(\tilde H_{Im})$, i.e. $\pm 5$. They are anyway far from the originally generated values (indicated by the horizontal red lines in Fig.~\ref{fig:sim_himhtim}), and have infinite error bars. Still, this does not prevent the fitting code from finding the right solution for $H_{Im}$.

\subsubsection{Prescription for central value and error bars}\label{subsub_prescription}
\begin{figure}[htbp] 
\includegraphics[width=0.5\textwidth]{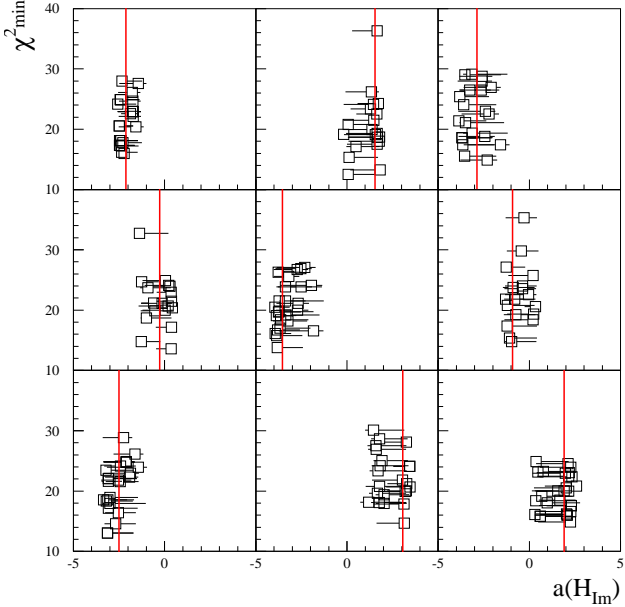}
\caption{Results of the fitter code, with the 8 CFFs taken
as free parameters, for $a(H_{Im})$ as a function of $\chi^2_{min}$.
The nine plots correspond to nine randomly generated values of 8 CFFs in the
$\pm 5$ times VGG CFFs hyperspace.
The red lines indicate the originally randomly generated $a(H_{Im})$ values.
The black points with their error bars indicate the results of the fitter code.
The ($x_B$, $Q^2$, $t$) kinematics at which this study was done 
is (0.375,1.964 GeV$^2$, -0.278 GeV$^2$), one of those measured by the Hall A. In each case, we show, in order not to overcrowd the figure, $\approx$ 20 fit results, out of hundreds. For each fit, both the initial parameters and the smearing on the data are different.}
\label{fig:checksim_8cff}
\end{figure}

Figure \ref{fig:checksim_8cff} shows another test of our fitting procedure. 
The study is done this time for a kinematics measured
in Hall A: ($x_B$, $Q^2$, $t$)=(0.375,1.964 GeV$^2$,-0.278 GeV$^2$). As before,
we generate $\phi$ distributions from several random sets of 8 CFFs,
smear the distributions according to Gaussians with standard deviations
corresponding to the experimental Hall A data uncertainties, and fit them,
taking randomly chosen starting points in the $\pm 5$ times the VGG-CFFs hypervolume.
Figure~\ref{fig:checksim_8cff} illustrates with nine plots, taken out of hundreds, the results for the reconstructed $a(H_{Im})$ CFF as a function of the unnormalized $\chi^2$. The vertical red lines indicate the originally generated $a(H_{Im})$ values. For a given set of 8 CFFs, each fit yields a different solution and different $\chi^2_{min}$ values. 
This is due to the random starting point and to the random smearing of the cross sections, which are both different for each fit.
These two individual effects can be seen separately in Fig.~\ref{fig:sim_him}.
Figure~\ref{fig:checksim_8cff} mixes the two effects and shows them for more cases. What is remarkable in Fig.~\ref{fig:checksim_8cff} is that for all fits, whatever the set of 8 CFFs,
the smearing of the $\phi$ points and the starting values of the CFFs, the originally generated
$a(H_{Im})$ value always lies within the error bars of the fitted $a(H_{Im})$'s.

When we fit real data, and extract $H_{Im}$ in particular, the only feature that we can change in our fit procedure is the starting point of the fit, the smearing of the data being imposed by the experiment. We saw in Fig.~\ref{fig:sim_him} that, in some cases, the solution $a(H_{Im})$ corresponding to $\chi^2_{min}$, was not unique: there could be ``double" (or a few more) solutions or ``single" solutions but with fluctuations. 
In many cases, the multiple solutions obtained are apart by insignificant 
$\chi^2$ differences, as we saw, and the $\chi^2_{min}$ solution cannot be clearly determined.
The starting point can also have an influence on the error bar of the solution: although error bars ranges are
almost always the same, one can distinguish in Fig.~\ref{fig:sim_him} in some cases small differences 
between error bars. It is not satisfactory to have several solutions for a fit and we have to devise 
a way to define a final, unique and reliable result, which should not depend on the particular
starting values and which should always contain the ``true" (generated) solution.

It seems that a good and conservative ad-hoc prescription is to take, among our series of solutions, 
the range between the maximum value of all error bars and the minimum value of all error bars in order to define
an effective error bar and take as the most probable value the middle of this interval. 
This recipe is indicated, in Fig.~\ref{fig:sim_him}, by the hand symbol, where the most probable value according to our prescription is the empty square. 
This ``middle value" that we advocate does not in general correspond to any of the $\chi^2_{min}$ values of the fit. However, we saw for example in Fig.~\ref{fig:sim_him} that the $\chi^2_{min}$ values are actually not corresponding to the originally generated value. The latter lies within the error bars of the $\chi^2_{min}$ solution. 
The $\chi^2_{min}$ values are thus not a better guess of the ``true" value than the ``middle'' value we propose. Since the smearing of the data (on which we have obviously no control when dealing with true experimental data) can shift the fitted $H_{Im}$ above or under the ``true" $H_{Im}$, taking the middle point of the biggest error bars as the most probable value provides an improved evaluation of the true value. 
Also, we saw that in most instances we obtained asymmetric error bars. These asymmetric error bars are typically defined by extreme (edge) values of the subdominant CFFs. For instance, one can see $\tilde H_{Im}$ in the top right plot of Fig.~\ref{fig:sim_himhtim}. 
These subdominant CFFs are in general not constrained, i.e. they are only restrained by the domain 
over which they are allowed to vary in the fit (i.e. $\pm$ 5 times the VGG CFFs). 
Thus, the $\chi^2_{min}$ solution corresponding to such an extreme value for these unconstrained CFF is actually not significantly more probable than any other. Choosing the central value of the error bars for $H_{Im}$ corresponds to setting $\tilde H_{Im}$, and, more generally, the unconstrained CFFs, around 0. This seems a reasonable choice, especially when these latter tend to lie at the edges of our fitting range. 

\begin{figure}[htbp] 
\includegraphics[width=0.5\textwidth]{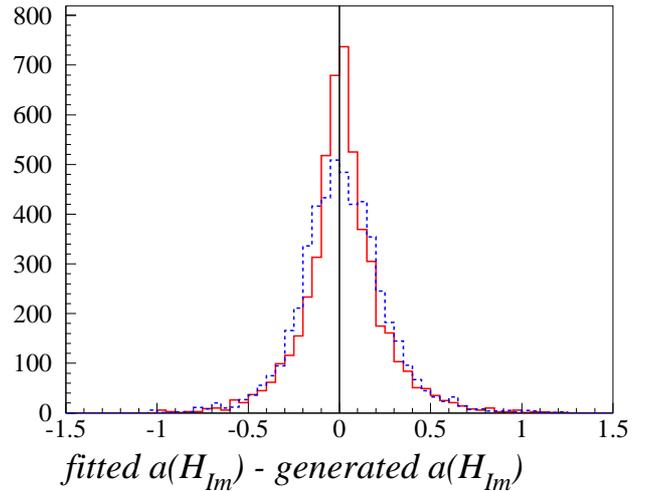}
\caption{Solid line: difference between the ``middle value" calculated from the largest error bars of all solutions and the generated value. Dashed line: difference between the $\chi^2_{min}$ solution and the generated value.}
\label{fig:cent}
\end{figure}

Figure~\ref{fig:cent} justifies this prescription. The solid-line distribution shows, for thousands of events like in Fig.~\ref{fig:checksim_8cff}, i.e. mixing randomly smearings and starting values, the difference between the ``middle value" calculated from the largest error bars of all solutions and the generated value. As a comparison, the dashed-line distribution shows the difference between the $\chi^2_{min}$ solution and the generated value.
Both distributions are well-centered around 0, which shows that both solutions are meaningful. However, it is clear that the ``middle value" distribution is significantly narrower than the $\chi^2_{min}$ one. 

To summarize this sub-section, we carried out our simulation studies for hundreds of cases, mixing sets of 8 CFFs, different starting points and cross section smearings and different JLab-type kinematics. The cross-examination of all these cases made us reach the general conclusion that in a 8-CFFs fit of the $\sigma$
and $\Delta\sigma_{LU}$ observables, using realistic experimental precisions, albeit largely underconstrained our fitter code appears to always manage to recover the originally generated $H_{Im}$, as the ``true" generated solution always lies in the $\chi^2_{min}+1$ error bar of the fitted solution.
Obviously we could not explore every combination of starting points, generated sets of 8 CFFs and cross sections smearings, and we cannot exclude the possibility that there are exceptions to this conclusion which escaped our scrutiny. We feel nevertheless rather confident that our procedure is reliable and robust. We finally advocate that, since there are cases where it is difficult to define exclusively
the $\chi^2_{min}$ solution and therefore the $\chi^2_{min}+1$ value, 
it is the most appropriate to take as final and unique solution the largest error bar
solution and the associated ``middle" point, as illustrated in Fig.~\ref{fig:sim_him}.

\subsection{Fitting with four CFFs}
\label{lab:4cffs}

We conclude this section on Monte-Carlo studies by a last exercise.
Since the GPDs $H$ and, to a lesser extent, $\tilde H$, 
are the dominant contributors to $\sigma$ and $\Delta\sigma_{LU}$, 
an idea is to investigate the outcome of a fit with only these two GPDs,
i.e. only 4 CFFs as free parameters.
This effectively means setting the 4 CFFs $E_{Im}$, $E_{Re}$, $\tilde E_{Im}$
and $\tilde E_{Re}$ to 0 in the fit, while they are not null in the generation
of the distributions to be fitted. This technique had been adopted previously to extract information on the kinematic dependence of $H_{Im}$ and $H_{Re}$ in Ref.~\cite{Jo:2015ema}. 
We used the same series of simulated $\phi$ distributions as before, generated by 8 CFFs taken randomly in the $\pm 5$-times-VGG CFFs hyperspace, and smeared according to the experimental uncertainties. For the present simulation, we use the same kinematics as in Fig.~\ref{fig:checksim_8cff}, i.e. the Hall A kinematics ($x_B$, $Q^2$, $t$)=(0.375, 1.964 GeV$^2$, -0.278 GeV$^2$), with its associated experimental uncertainties on the cross sections.
This time, we fit the smeared $\phi$ distributions by only the 4 CFFs $H_{Im}$, $H_{Re}$, $\tilde H_{Im}$ and $\tilde H_{Re}$, instead of the 8 CFFs as before. 

\begin{figure}[htbp] 
\includegraphics[width=0.5\textwidth]{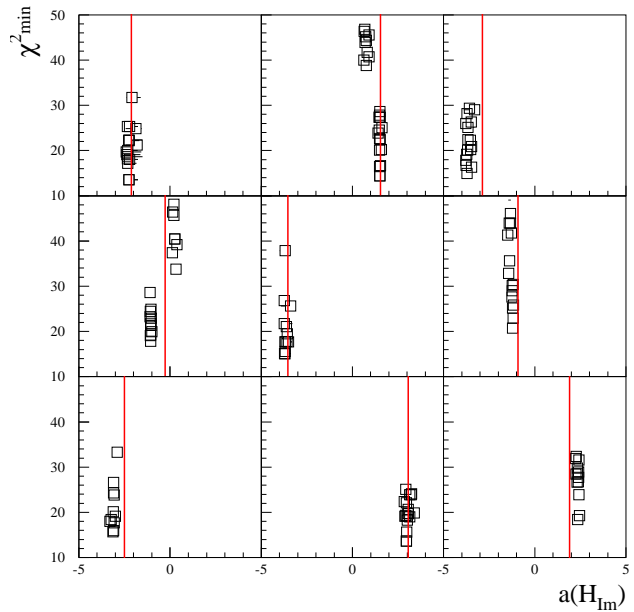}
\caption{Results of the fitter code for $a(H_{Im})$ as a function of $\chi^2$, 
with only the 4 CFFs $H_{Im}$, $H_{Re}$, $\tilde H_{Im}$ and $\tilde H_{Re}$ as free parameters.
The nine plots correspond to nine randomly generated values of 8 CFFs in the
$\pm 5$ times VGG CFFs hyperspace.
The red lines indicate the originally randomly generated $a(H_{Im})$ values.
The squares with their error bars indicate the results of the fitter code
for a sample of $\approx$ 20 fits, each fit differing by its starting values and smearings.}
\label{fig:checksim_4cffeet0}
\end{figure}

The results for $a(H_{Im})$ are displayed in Fig.~\ref{fig:checksim_4cffeet0}, which is the analog for 4 CFFs of Fig.~\ref{fig:checksim_8cff}.
We first observe that the error bars on the fitted $a(H_{Im})$'s are in general smaller than for the 8-parameters case. This decrease of the error bars can be simply understood as there are less free parameters (4 instead of 8) entering the problem and therefore less correlations.
However, we now observe several types of results. 
For the left top-row plot, the central mid-row plot and the bottom mid-row plot, the results of the fits can be considered satisfactory as the squares lie relatively well along the red lines, which indicate the originally generated values.
However, we also observe cases where the solutions are clearly systematically shifted, by 30 to 50\% w.r.t. the red lines.
Although the fitted solutions are always relatively ``close" to the true solutions, the latter are quite often outside the error bar of the former, defined as usual by $\chi^2_{min}+1$. 
We shall therefore conclude that the 4-CFFs free-parameters fit based on the $H$ and $\tilde H$ GPDs is not fully reliable. At best, it can provide a flavor for the solution at the 30 to 50\% level, i.e. the relative shifts between the fitted solutions and the generated one. This 30 to 50\% relative uncertainty will however not be reflected in the error bars coming out of the fitter, which are much smaller.

\begin{figure}[htbp] 
\includegraphics[width=0.5\textwidth]{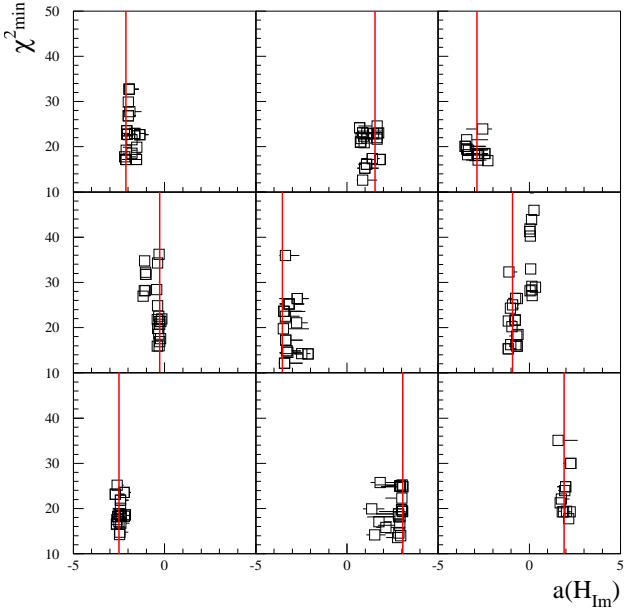}
\caption{Results of the fitter code for $a(H_{Im})$ as a function of $\chi^2$, 
with the 4 CFFs $H_{Im}$, $H_{Re}$, $\tilde H_{Im}$
and $\tilde H_{Re}$ taken as free parameters and $E_{Im}$, $E_{Re}$, $\tilde E_{Im}$
and $\tilde E_{Re}$ set to their originally generated values.
The nine plots correspond to nine randomly generated values of 8 CFFs in the
$\pm 5$ times VGG CFFs hyperspace.
The red lines indicate the originally randomly generated $a(H_{Im})$ values.
The black points with their error bars indicate the results of the fitter code
for a sample of $\approx$ 20 fits, each fit differing by its starting values.}
\label{fig:checksim_4cffeettrue}
\end{figure}

We also studied the case of fitting $\sigma$ and $\Delta\sigma_{LU}$ with the 4 CFFs $H_{Im}$, $H_{Re}$, $\tilde H_{Im}$ and $\tilde H_{Re}$ as free parameters and
$E_{Im}$, $E_{Re}$, $\tilde E_{Im}$ and $\tilde E_{Re}$ set to their original values
(as before, randomly generated), instead of 0 as in the previous case.
For the same randomly generated sets of CFFs as before, Fig.~\ref{fig:checksim_4cffeettrue}
shows the results for $a(H_{Im})$ in this configuration. We observe 
that in general we are able, within error bars, to recover the originally generated values for $H_{Im}$
(while the three other CFFs don't come out in general with finite error bars, both the positive and the negative one). 
This means that, if the unfitted CFFs are set to their true values, a fit
with only the 4 CFFs based on the $H$ and $\tilde H$ GPDs might be meaningful (at least for $H_{Im}$).
With the (strong) assumption that VGG (or, more generally, any other model) gives a reasonable description of the $E$ and $\tilde E$ GPDs, this gives a motivation to fit real data with only $H_{Im}$, $H_{Re}$, $\tilde H_{Im}$ and $\tilde H_{Re}$ as free parameters and setting $E_{Im}$, $E_{Re}$, $\tilde E_{Im}$ and $\tilde E_{Re}$ to their VGG values. 
The merit of this 4 CFF fit approach is that this provides smaller error bars. 
This is however clearly at the price of introducing some model dependence since,
in the most general case, a 4-CFFs fit is not fully reliable as seen earlier.

\section{Real Data Fitting}

Being convinced of the soundness and reliability of our fitting approach after our Monte-Carlo pseudo-data tests, we now apply our method to real data. The JLab Hall A and CLAS collaborations have recently released new sets of unpolarized and beam-polarized cross sections ($\sigma$ and $\Delta\sigma_{LU}$)~\cite{Defurne:2015kxq,Jo:2015ema}. 
At the light of the simulations of the previous section, we therefore expect to 
extract constraints on the $H_{Im}$ CFF and, partially, on $H_{Re}$.
In addition, the CLAS collaboration has measured, using a longitudinally
polarized target, the single and double target-spin asymmetries $A_{UL}$ and $A_{LL}$~\cite{Pisano:2015iqa,Seder:2014cdc}. 
The $\tilde H_{Im}$ CFF being a strong contributor to $A_{UL}$, we expect to extract constraints on this CFF as well. The analysis of $A_{UL}$ will also allow to improve the precision on $H_{Im}$ due to its strong correlation with $\tilde H_{Im}$, as we saw in the previous section.

We start our study with the Hall A data and then proceed with the CLAS data.

\subsection{Hall A data}

The JLab Hall-A collaboration has measured the two observables $\sigma$ and $\Delta\sigma_{LU}$
at four average kinematical settings ($x_B$, $Q^2$): (0.36, 1.90 GeV$^2$), (0.36, 2.3 GeV$^2$), (0.39, 2.06 GeV$^2$) and (0.34, 2.17 GeV$^2$). In Ref.~\cite{Defurne:2015kxq} they are called KIN2, KIN3, KINX2 and KINX3, respectively. The latter two kinematics are actually a subset, 
obtained with tighter cuts, of the first two. For each of these four ($x_B$, $Q^2$) kinematics, 
the $\phi$ distribution has been measured for five $t$ bins.
 
We fit simultaneously the $\sigma$ and $\Delta\sigma_{LU}$ 
$\phi$-distributions, for each of these 20 ($x_B$, $Q^2$, $t$) bins.
We use either the eight CFFs as free parameters or only the
four $H_{Im}$, $H_{Re}$, $\tilde H_{Im}$ 
and $\tilde H_{Re}$, the other CFFs being set to their VGG values, as invoked
in the previous section. We carry out
our fits with hundreds of different starting values
randomly generated in the $\pm 5$-times-VGG-CFF hyperspace,
in order to make sure that the results are stable, as discussed previously.

\begin{figure}[htbp] 
\includegraphics[width=0.48\textwidth]{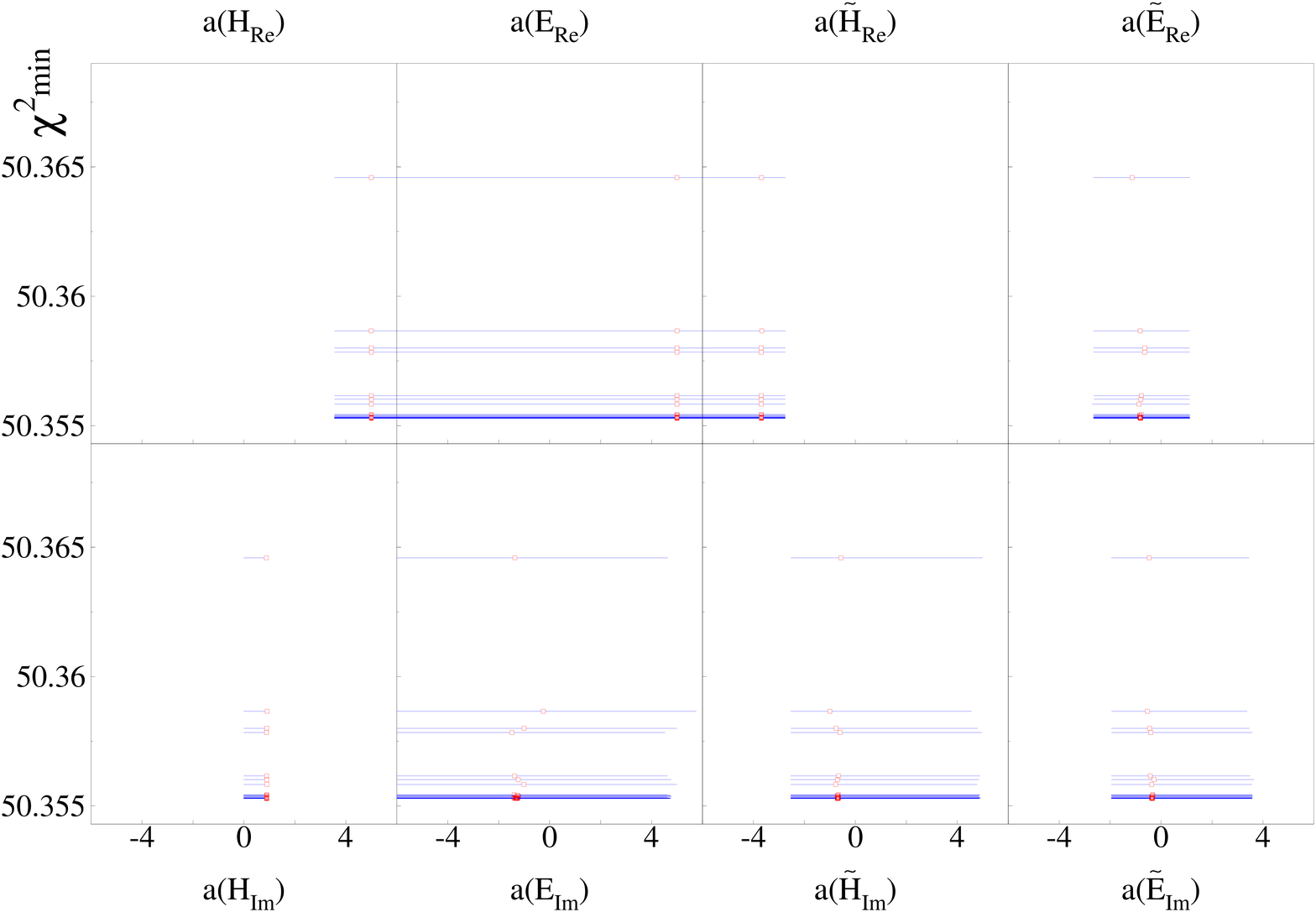}
%\vskip -2cm
\caption{Result of the fits for the 8 CFF multipliers $a(H_{Im})$, $a(E_{Im})$,
$a(\tilde H_{Im})$, $a(\tilde E_{Im})$,
$a(H_{Re})$, $a(E_{Re})$, $a(\tilde H_{Re})$ and $a(\tilde E_{Re})$, as a function
of $\chi^2_{min}$, for 50 trials differing only by the starting values of the fit. 
The value of the CFF multiplier corresponding to the $\chi^2_{min}$ value 
for a given trial is in red and its associated error bar 
corresponding to $\chi^2_{min}+1$ is in blue. 
This example is for the third $t$-bin of the KIN2 JLab Hall A kinematics.}
\label{fig:himvschi2_108}
\end{figure}

Analogously to Fig.~\ref{fig:himvschi2_sim_smear},
Fig.~\ref{fig:himvschi2_108} shows an example of the 8-CFFs fit results for one of the 20 ($x_B$, $Q^2$, $t$) bins, namely the third $t$-bin of the KIN2 kinematics: ($x_B$, $Q^2$, $t$)=(0.375, 1.964 GeV$^2$, -0.278 GeV$^2$). The figure shows the result of the fit, for 50 different starting points,  for the 8 CFF multipliers with the associated $\chi^2_{min}$ values. The red points indicate the minimum $\chi^2_{min}$ solutions and the blue bars the errors corresponding to $\chi^2_{min}+1$.

We observe that all trials
end up with essentially the same set of solutions, all with very similar
$\chi^2_{min}$ values. The $\chi^2_{min}$ values in Fig.~\ref{fig:himvschi2_108}
range from 50.3553 to 50.3587. These $\chi^2$ values are unnormalized. For normalized values, 
one has to divide by 48 (corresponding to the number of data points: 24 for $\sigma$ and 
24 for $\Delta\sigma_{LU}$) minus 8 (corresponding to the number of free
parameters), i.e. 40.

Taking the solution which yields the minimum 
of all $\chi^2_{min}$'s, i.e. 50.3553, the results of the 8 fitted CFF multipliers are:

\begin{equation}
\begin{aligned}
a(H_{Im})&=0.89322_{-0.90729}^{0.065256}, & a(E_{Im})&=-1.3109_\infty^\infty, \\
a(\tilde H_{Im})&=-0.68653_{-1.8512}^\infty, & a(\tilde E_{Im})&=-0.35243_{-1.5984}^{3.9312},\\
a(H_{Re})&=5.0000_{-1.4469}^\infty, & a(E_{Re})&=5.0_\infty^\infty, \\
a(\tilde H_{Re})&=-3.6919_\infty^{0.94013}, & a(\tilde E_{Re})&=-0.81330_{-1.8439}^{1.9356}.
\end{aligned}
\label{eq:cffresult_108}
\end{equation}

We recall that the $a()'s$ measure the deviation from the VGG CFFs. Thus, the
interpretation of $a(H_{Im})=0.89322$ is that the value of $H_{Im}$ that best
fits the Hall A data is $\approx$89\% of that given by the VGG model. 
In Eq.~\ref{eq:cffresult_108}, the $\infty$ error values mean that the $\chi^2_{min}+1$ value could not be reached and that it therefore lies outside the $\pm 5$-times-VGG-CFF hypervolume. In some cases, $a(E_{Re})$ for instance, both positive and negative error bars are infinite. Then, no constraint at all can be drawn on such CFF. In some other cases, $\tilde H_{Im}$
for instance, one of the two errors is finite and then a lower (or upper)
limit on the CFF can be drawn. The most favorable case is when the two error
bars are finite and lie in the $\pm 5$-times-VGG-CFF range. This is, for the present kinematics, the case of the $H_{Im}$, $\tilde E_{Im}$ and $\tilde E_{Re}$ CFFs. $H_{Im}$ is the most constrained by far. Its negative error bar is of the order of 100\% while the positive one is only of a few percent. We could also observe in the simulations in the previous section at several instances such asymmetric error bars  for $H_{Im}$, which reflect the non-linearity of the problem. 

\begin{figure}[htbp] 
\includegraphics[width=0.385\textwidth]{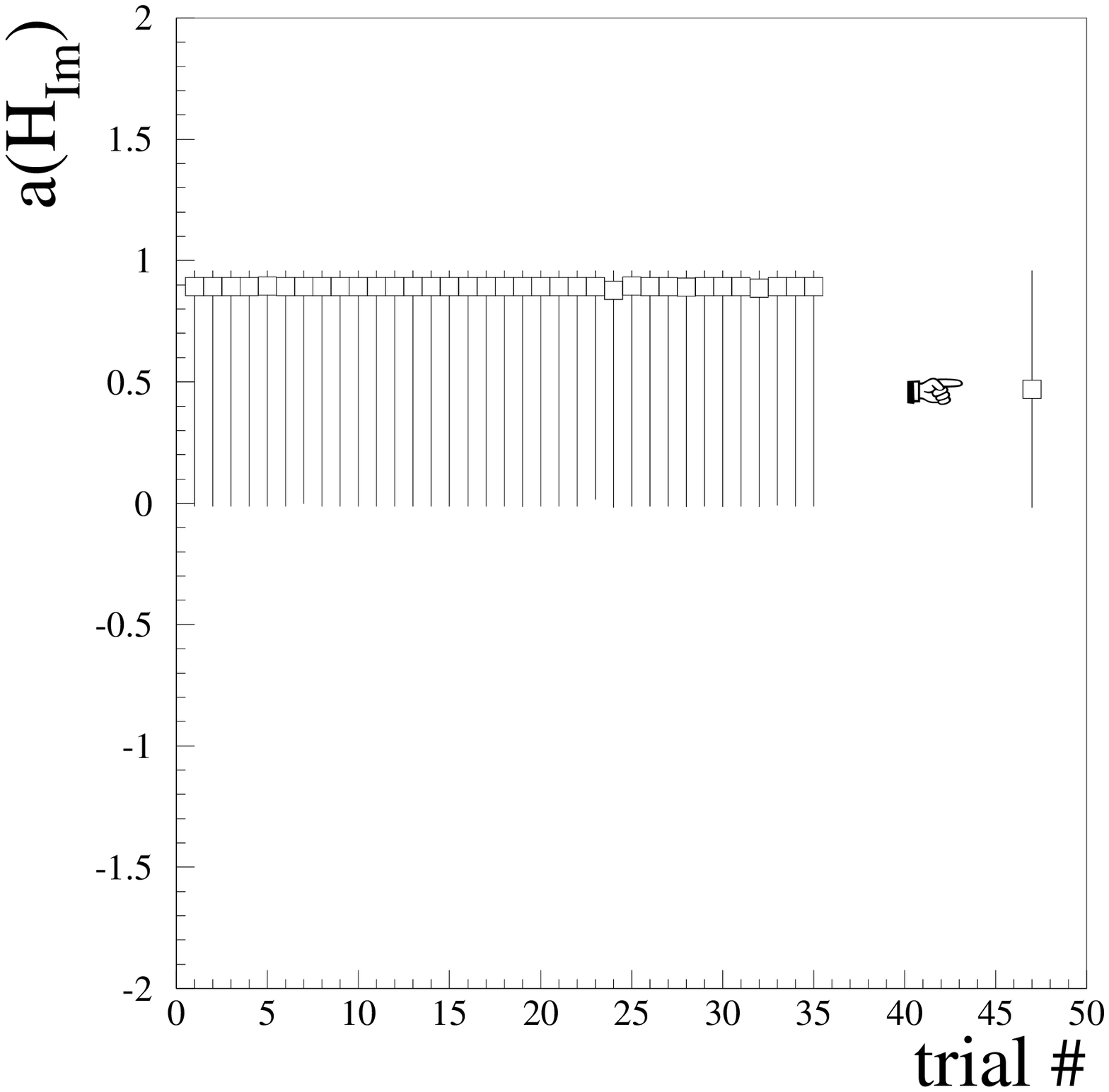}
\includegraphics[width=0.385\textwidth]{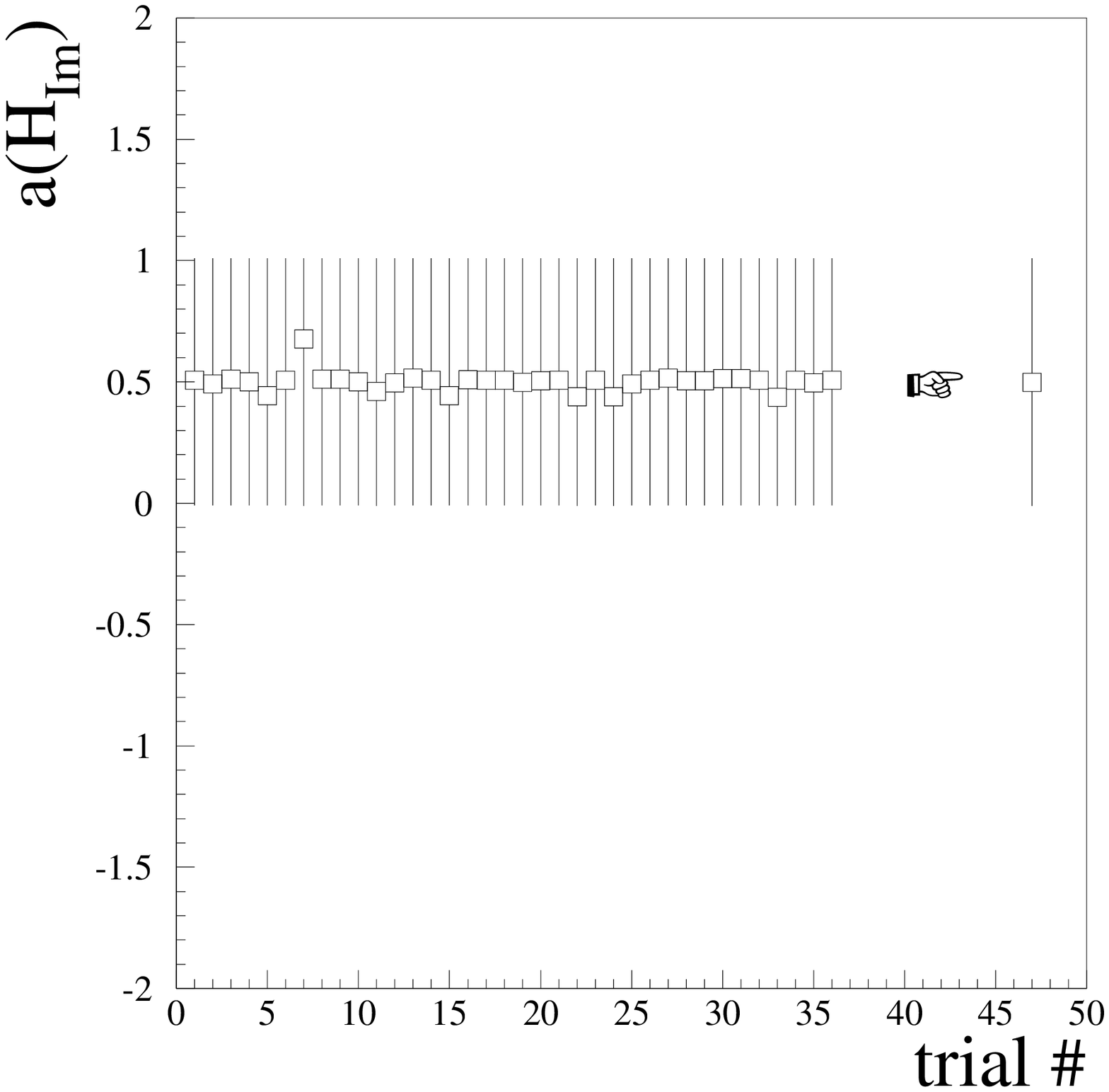}
\includegraphics[width=0.385\textwidth]{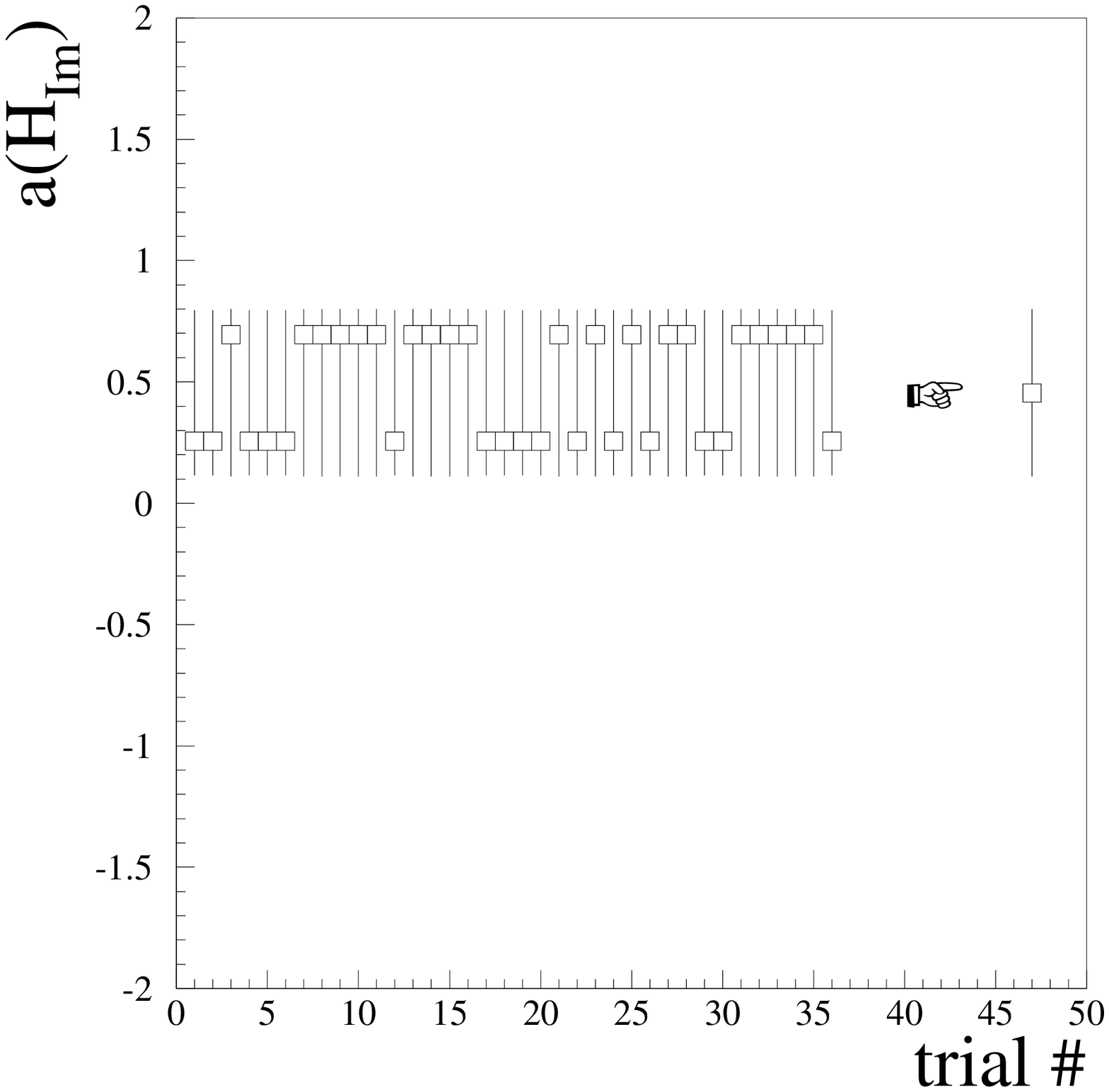}
\caption{Result of the fitted multiplier $a(H_{Im})$ for 
several fits differing on their starting values. Top: 8-CFFs fit
for the third $t$-bin of KIN2 (same bin as in Fig.~\ref{fig:himvschi2_108}). 
Center: 8-CFFs fit for the third $t$-bin of KINX3.
Bottom: 4-CFFs fit ($H_{Im}$, $\tilde H_{Im}$, $H_{Re}$ and $\tilde H_{Re}$) with the four other CFFs set to their VGG value, for the fourth $t$-bin of KIN2. The points indicated by the hand show the solutions that we advocate and that we will finally retain.}
\label{fig:trials}
\end{figure}

The top plot of Fig.~\ref{fig:trials} displays in a more visible way the results 
of Fig.~\ref{fig:himvschi2_108} for only $a(H_{Im})$. The results are shown
for different trials differing only by their starting values.
In the central plot of Fig.~\ref{fig:trials}, we display the results for another Hall A bin
(third $t$-bin of KINX3), to illustrate the variety of types of results, depending on 
the kinematics which are studied. While for the top plot the error bars, which are constant, are very asymmetric w.r.t. the $a(H_{Im})$ values which minimize the problem, in the central plot the values of $a(H_{Im})$ corresponding to $\chi^2_{min}$ lie, with a few fluctuations, around the center of the error bars, which are also constant.
Then, as an illustration of a 4 CFF fit, we show in the bottom plot of Fig.~\ref{fig:trials}
the result of a fit with only $H_{Im}$, $\tilde H_{Im}$, 
$H_{Re}$ and $\tilde H_{Re}$ as free parameters, the four other
CFFs being set to their VGG value. We observe double solutions. Depending on the starting point, the fitter code ends up in one or in the other of two solutions.
The unnormalized $\chi^2_{min}$ values of the $a(H_{Im})\approx 0.25$ and $a(H_{Im})\approx 0.69$ solutions are, respectively, $\approx 61.95$ and $\approx 62.01$.
It is clearly not meaningful to favor one solution rather than the other. We also notice that the error bar ranges of the two solutions are identical.
We already encountered such a situation in the previous section dedicated to simulations. We saw that the ``true" value was actually likely to lie
between these two solutions. 

For unique final results, we learned from our Monte-Carlo
studies that a good and safe policy was to take as most probable point
the middle of the maximal error bars of all trials.
We illustrate the prescription in the right part of each plot of Fig.~\ref{fig:trials} where we plot the final central value and error bars that we will retain.

\begin{figure}[htbp] 
\includegraphics[width=0.45\textwidth]{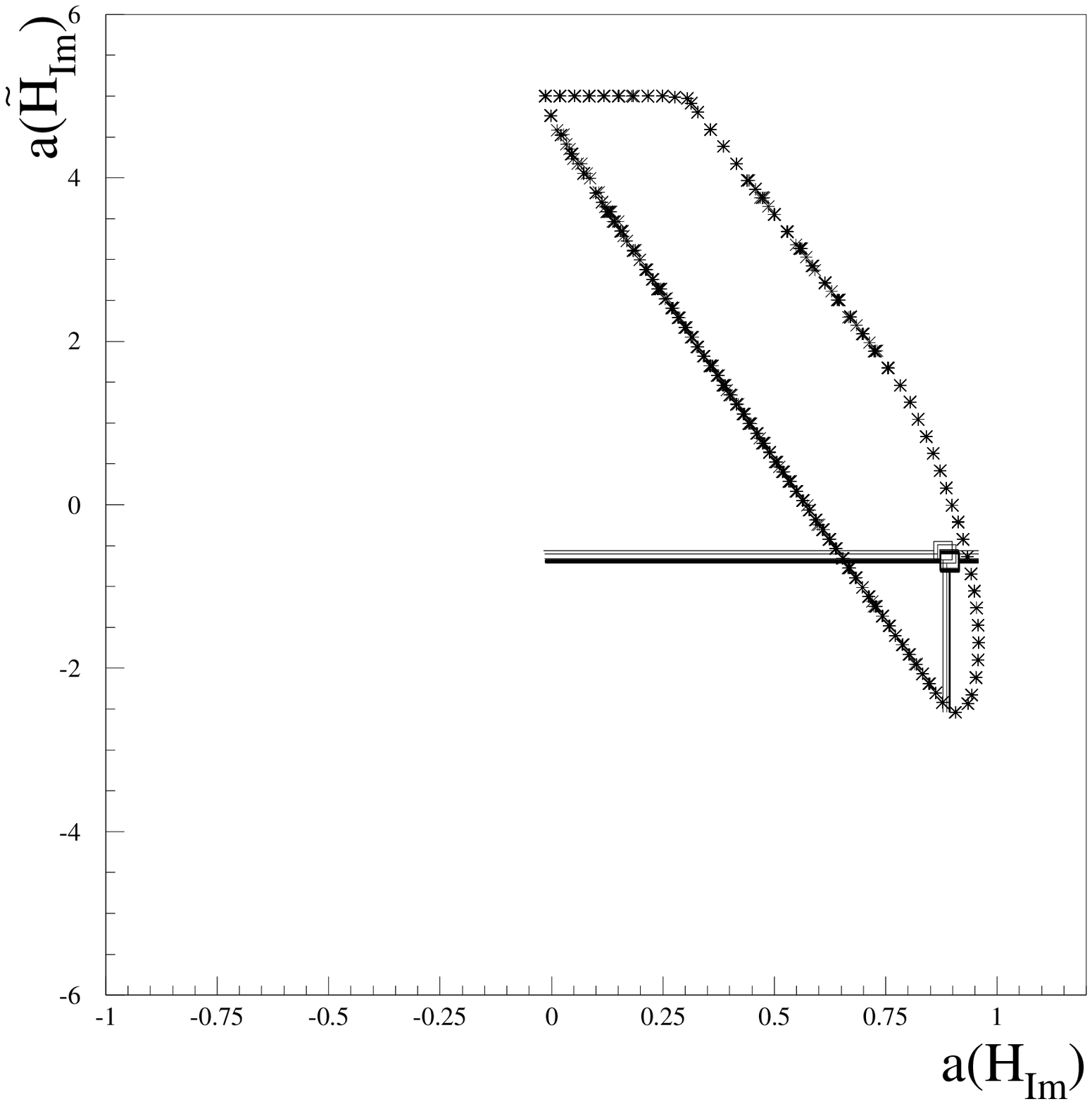}
\includegraphics[width=0.45\textwidth]{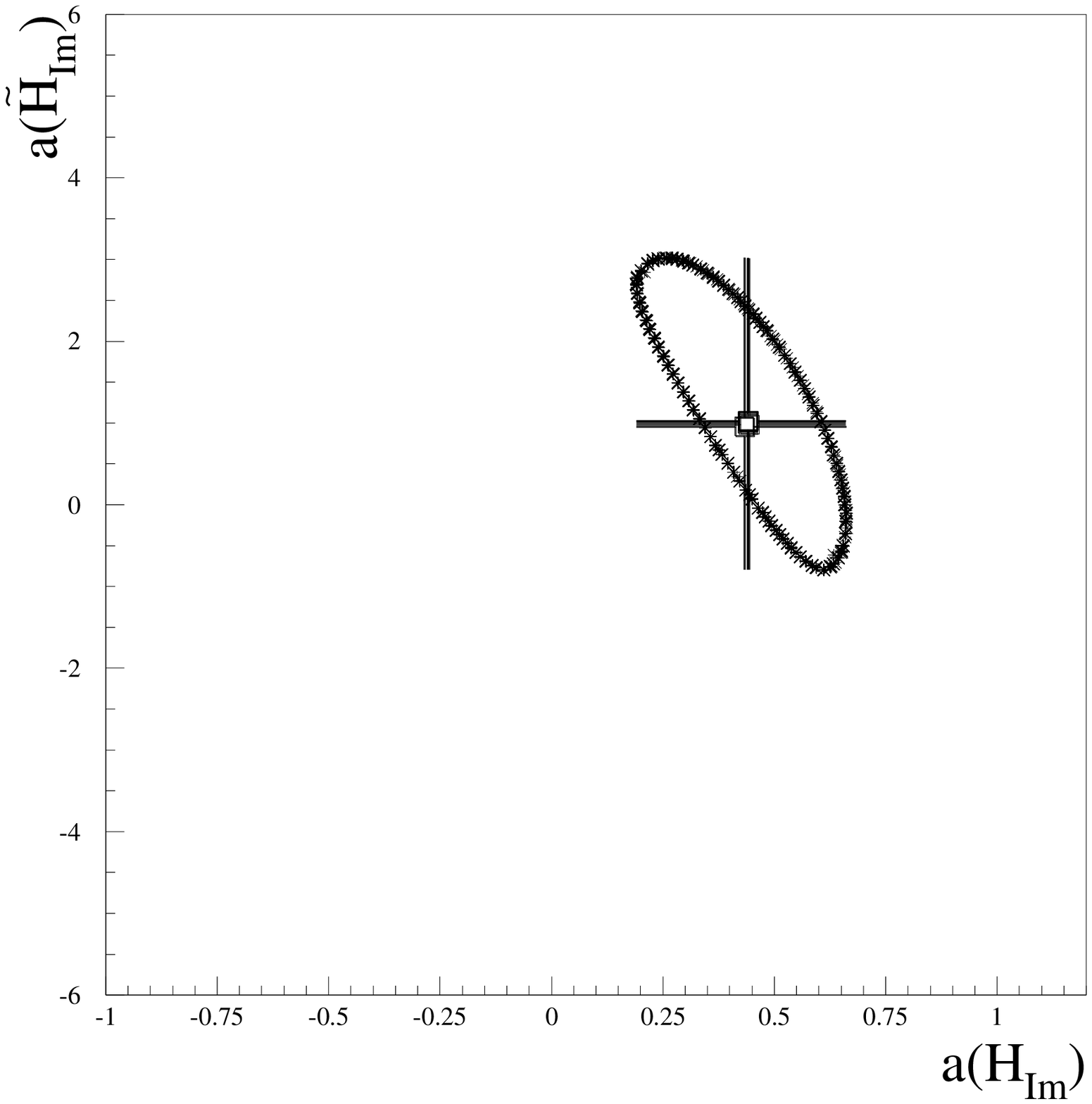}
\caption{Contour plot of the fit results in the ($a(\tilde H_{Im})$, $a(H_{Im})$ plane).
Top: third bin in $t$ of KIN2 (same as top plot of Fig.~\ref{fig:trials}). 
Bottom: third $t$-bin of KINX3 (same as central plot of Fig.~\ref{fig:trials}).
The open squares show the values of $a(H_{Im})$ and $a(\tilde H_{Im})$
corresponding to the minimum $\chi^2$ values of the fit. The ``asterisk curves"
show the contour corresponding to $\chi^2_{min}+1$. The plots have been produced by superposing
the results of 50 fits differing only by their starting values.}
\label{fig:contourhalla}
\end{figure}

As we already discussed, the rather large error bars in Fig.~\ref{fig:trials}
do not reflect the statistical error of the data. They reflect the influence
of the sub-dominant CFFs on the dominant $H_{Im}$ CFF and more
generally the underconstrained nature of the problem. This is illustrated 
in Fig.~\ref{fig:contourhalla} where we display the correlation (contour plot) between 
the $a(H_{Im})$ and $a(\tilde H_{Im})$ multipliers for two Hall A bins. 
The open squares show the values of $a(H_{Im})$ corresponding to the minimum $\chi^2$ values 
of the fit. All these solutions correspond to different starting values
in the $\pm 5$-times-VGG-CFFs hypervolume. We plot in Fig.~\ref{fig:contourhalla} a sample of 50 fits results. The ``asterisk curves" are the associated
contours corresponding to $\chi^2_{min}+1$.
The top plot corresponds to the third bin in $t$ of KIN2, i.e. the same kinematics as in Fig.~\ref{fig:himvschi2_108} and as the top plot of Fig.~\ref{fig:trials}. The bottom plot of Fig.~\ref{fig:contourhalla} corresponds to the third $t$-bin of KINX3, i.e. the same kinematics as in the central plot of Fig.~\ref{fig:trials}. 
One sees that the one-dimensional error bars that are displayed in Fig.~\ref{fig:trials} correspond to the projections on the $a(H_{Im})$-axis of the ellipse-like contours of Fig.~\ref{fig:contourhalla}.
For the top plot of Fig.~\ref{fig:contourhalla}, one should note that the ellipse 
is truncated on the upper side of 
the $a(\tilde H_{Im})$ axis. Thus, no positive error bar on $\tilde H_{Im}$ can be defined.
This explains the $\infty$ positive error bar of $a(\tilde H_{Im})$ in
Eq.~\ref{eq:cffresult_108}. In this case, this truncation on $a(\tilde H_{Im})$ defines and
influences the negative error bar of $a(H_{Im})$. Were the range of $\pm 5$-times-VGG CFFs
larger, the negative error
bar on $H_{Im}$ would be larger as well. This is the only model dependency of this approach in the 8-CFFs case, as we already underlined. 

Such a truncation is not always happening. For the kinematics
of the bottom plot of Fig.~\ref{fig:contourhalla}, all fits,
differing only by their starting values, converge to a quasi-unique ($a(H_{Im})$, $a(\tilde H_{Im})$) solution. The full contour ellipse
holds in the ($-5<a(\tilde H_{Im})<5$, $-5<a(H_{Im})<5$) surface. This means that 
constraints on $\tilde H_{Im}$ can also be drawn for this particular bin.

\begin{figure}[tbp] 
\includegraphics[width=0.5\textwidth]{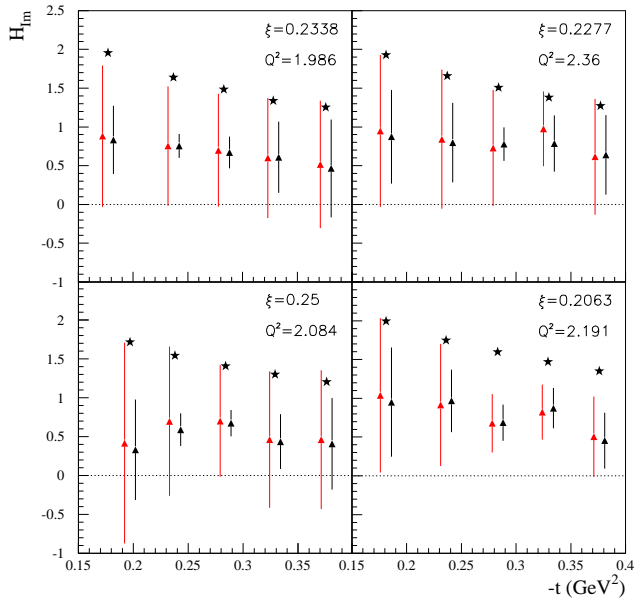}
\caption{Result of the fitted $H_{Im}$ as a function of $t$, for Hall-A kinematics, with the 8 CFFs as free parameters in red triangles and with 4 CFFs as free parameters ($H_{Im}$, $\tilde H_{Im}$, $H_{Re}$ and $\tilde H_{Re}$, the other four CFFs being fixed at their VGG values) in black triangles. The black triangles have been slightly shifted to the right of the red triangles for visibility.
Upper left plot: KIN2; upper right plot: KIN3; lower left plot: KINX2; lower right plot: KINX3. Here we have converted the $x_B$ values into $\xi$ values using Eq.~\ref{xi_def}.}
\label{fig:fithalla_all}
\end{figure}

We now display in Fig.~\ref{fig:fithalla_all} the outcome of the fits 
for the dominant $H_{Im}$ CFF for the 20 ($x_B$, $Q^2$, $t$) Hall A bins. For each of the 20 bins, hundreds of starting points have been randomly chosen, leading to results for the 8 CFFs of the form of Figs.~\ref{fig:himvschi2_108} and~\ref{fig:trials}. 
The $H_{Im}$ CFF is the one always coming out with finite negative and positive error bars. 
Figure~\ref{fig:fithalla_all} shows our fit results in
the two approaches: 8 CFFs free parameters with red triangles and 4 CFFs free 
parameters ($H_{Im}$, $\tilde H_{Im}$, $H_{Re}$ and $\tilde H_{Re}$ with the four other CFFs set to their VGG values) with black triangles. 
The two sets of results are very compatible, with of course significantly smaller error bars in the case of the 4-CFFs fit. 
Except maybe for the bin of the lower left plot of Fig.~\ref{fig:fithalla_all},
one can in general discern a decreasing trend for $H_{Im}$ as $-t$ increases.
For comparison, we also plot in Fig.~\ref{fig:fithalla_all} the values of $H_{Im}$ 
from the VGG model, with black stars. The model exhibits, indeed, such a decrease with $-t$.
However, the VGG model, with the valence (sea) quark profile parameter choice $b_v = 1$ ($b_s = 1$) respectively~\cite{Vanderhaeghen:1998uc}, seems to overestimate by a factor $\approx$ 2 the outcome of the fits.

The error bars that we obtain on $H_{Im}$ are rather large.
They are of the order of 100\% for the 8 CFFs fits
and of 50\% for the 4 CFF fits. This prevents to draw strong conclusions
at this stage. With additional constrains, like the measurement of new
observables, which is expected to come in the near future, the situation shall improve.
We are paving the way for those days.

\subsection{CLAS data}

The CLAS collaboration has measured the $\phi$ distribution of the two observables $\sigma$ and $\Delta\sigma_{LU}$ for 21 ($x_B$, $Q^2$) bins in the range $0.12\lesssim x_B\lesssim 0.50$, $1.11\lesssim Q^2\lesssim 3.90$, with 6 $t$-bins (in most cases), ranging up to $-t=0.5$ GeV$^2$. The CLAS collaboration has also measured the $\phi$ distribution of the $A_{UL}$ and $A_{LL}$ asymmetries for 5 ($x_B$, $Q^2$) bins, in a roughly equivalent phase space to the $\sigma$ and $\Delta\sigma_{LU}$ case, with 4 $t$-bins (in most cases), ranging up to $-t\approx 1.3$ GeV$^2$. 
Among these $\approx 20$ ($x_B$, $Q^2$, $t$) bins, 15 have common kinematics with the $\sigma$ and $\Delta\sigma_{LU}$ measurements. It should be noted that $A_{UL}$ and $A_{LL}$ have been measured up to larger $-t$ values than $\sigma$ and $\Delta\sigma_{LU}$. 

\subsubsection{Fits of $\sigma$ and $\Delta\sigma_{LU}$.}

\begin{figure}[htbp] 
\center
\includegraphics[width=0.385\textwidth]{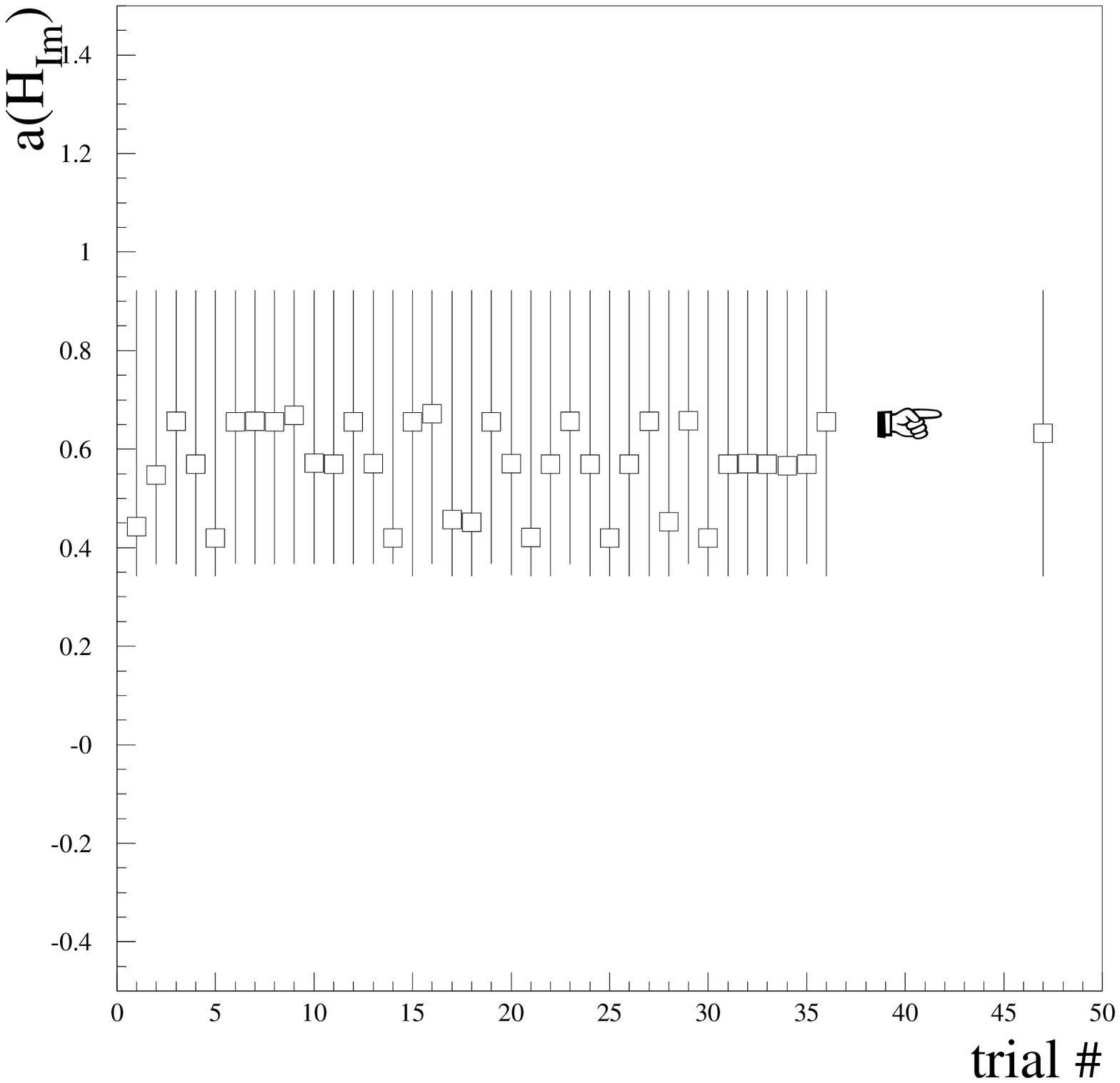}
\includegraphics[width=0.385\textwidth]{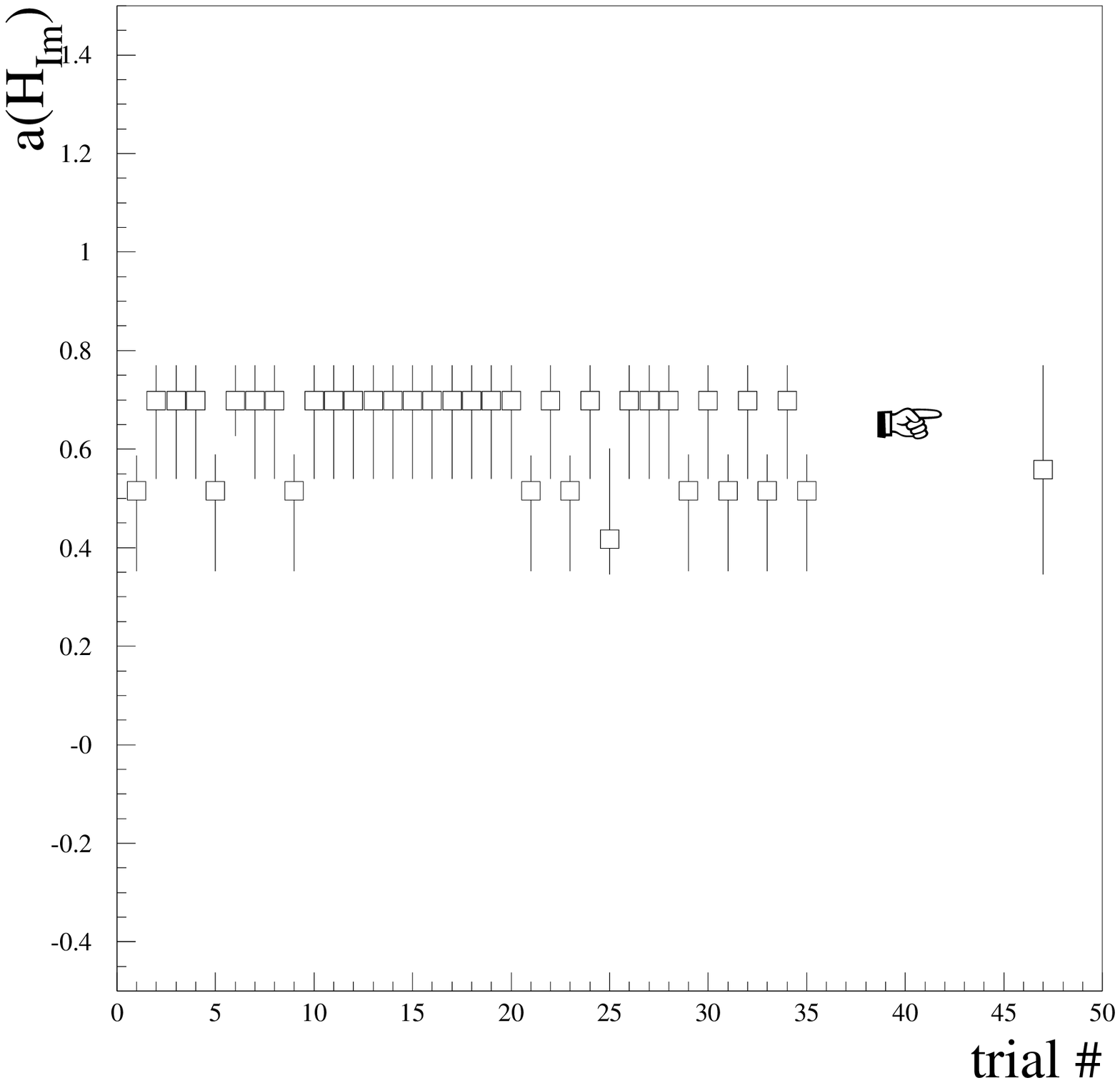}
\includegraphics[width=0.385\textwidth]{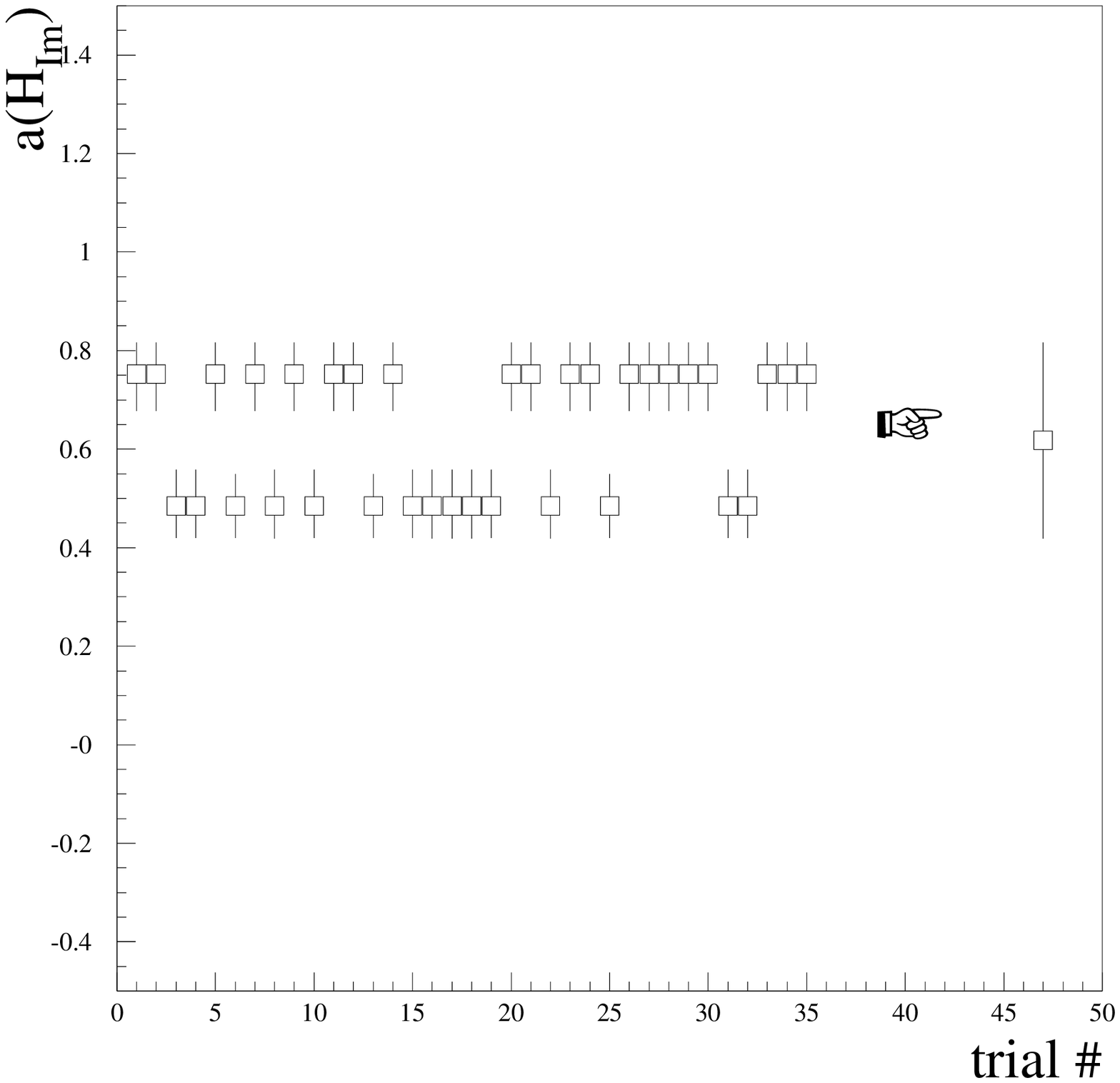}
\caption{Examples of results for the fitted multiplier $a(H_{Im})$ for 
several fits, differing only by their starting values. Top plot: 8-CFFs fit for the CLAS kinematics ($x_B$, $Q^2$, $t$)=(0.1541, 1.2656 GeV$^2$, -0.1526 GeV$^2$). Center plot: 8-CFFs fit for the CLAS kinematics (0.126, 1.1114 GeV$^2$, -0.1078 GeV$^2$). Bottom plot: 4-CFFs fit ($H_{Im}$, $\tilde H_{Im}$, $H_{Re}$ and $\tilde H_{Re}$, the other four CFFs being fixed at their VGG values) for the CLAS kinematics (0.1541, 1.2652 GeV$^2$, -0.1082 GeV$^2$).}
\label{fig:trialshallb}
\end{figure}

In a first stage, we extract $H_{Im}$ out of $\sigma$ and $\Delta\sigma_{LU}$, as we did for the Hall A data, for all the CLAS ($x_B$, $Q^2$, $t$) bins.
Most of the results of our fits look like those we obtained for Hall A (Fig.~\ref{fig:trials}). In particular, $H_{Im}$ always comes out of the fit with finite error bars. 
However, in some cases, we encounter new features such as those shown in Fig.~\ref{fig:trialshallb}. The figure shows a few examples of the $a(H_{Im})$
multipliers that were extracted for different randomly generated
starting points for three particular ($x_B$, $Q^2$, $t$) CLAS bins. 
The first example (top plot of Fig.~\ref{fig:trialshallb}) 
shows a case where the results for $a(H_{Im})$ have constant error bars 
but large fluctuations for the values corresponding to $\chi^2_{min}$. The next two examples
(central and bottom plots of Fig.~\ref{fig:trialshallb})
show cases where double solutions occur. 
In the bottom plot, resulting from a 4 CFF fit (with $H_{Im}$, $\tilde H_{Im}$, 
$H_{Re}$ and $\tilde H_{Re}$ as free parameters and the other four CFFs being fixed at their
VGG values), the error bars do not even overlap.
Such feature was also found in Ref.~\cite{Kumericki:2015lhb} which 
also explored and considered in part the present local fitting method and these
new JLab data.
As done previously, based on our simulations studies, for all those cases, 
we will take as most probable point the middle of the maximal error bars of all trials. This is illustrated by the point indicated by the hand in Fig.~\ref{fig:trialshallb}.

\begin{figure}[tbp] 
\includegraphics[width=0.5\textwidth]{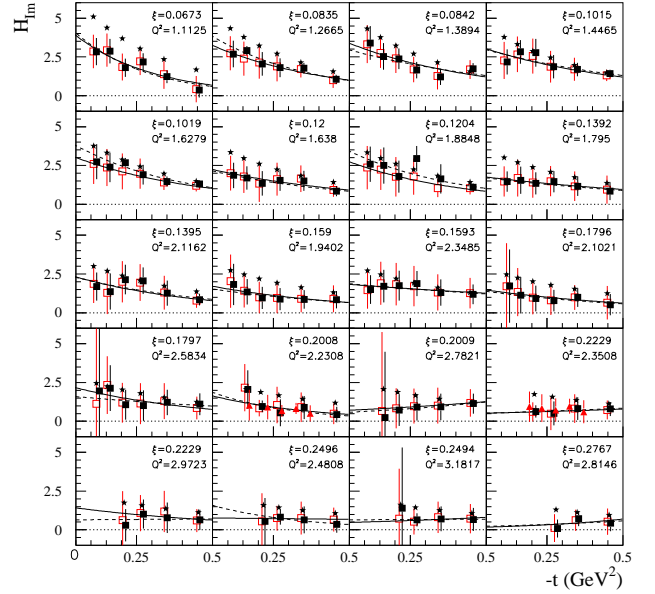}
\caption{The $H_{Im}$ CFF as a function of $t$ for the 20 CLAS ($x_B$, $Q^2$) bins,
fitting only $\sigma$ and $\Delta\sigma_{LU}$. Red open squares: results of the CLAS data fit with the 8 CFFs as free parameters.
Black solid squares: results of the CLAS data fit with the 4 CFFs $H_{Re}$, $\tilde H_{Re}$,
$H_{Im}$ and $\tilde H_{Im}$ as free parameters, the other 4 CFFs being 
set to their VGG value. Red triangles (($x_B/\xi$, $Q^2$)=(0.3345/0.2008, 2.2308 GeV$^2$)
and (0.3646/0.2229, 2.3508 GeV$^2$) bins): results of the Hall-A data fit with the 8 CFFs as free parameters (taken from Fig.~\ref{fig:fithalla_all}).
Stars: VGG predictions. The black solid square points have been slightly
shifted to the right of the red open square points for visibility. 
The solid line shows an exponential fit of the red open 
squares and the dashed line an exponential fit of the black solid squares.}
\label{fig:fithallb_all}
\end{figure}

With such prescription, Fig.~\ref{fig:fithallb_all} shows our results for $H_{Im}$ with the two approaches that we considered: 8 CFFs as free parameters (red open squares) and the 4 CFFs $H_{Re}$, $\tilde H_{Re}$, $H_{Im}$ and $\tilde H_{Im}$ as free parameters, with the others set to their VGG value (black solid squares). 
We notice the good agreement between the 8-CFFs and 4-CFFs fit results. The latter have in general smaller error bars, as expected. We also insert in the figure the Hall-A results for $H_{Im}$ with the 8 CFFs as free parameters that we obtained for the KIN3 and KINX3 bins (red solid triangles). These two bins correspond almost exactly to the CLAS ($x_B/\xi$, $Q^2$)=(0.3345/0.2008, 2.2308 GeV$^2$) and (0.3646/0.2229, 2.3508 GeV$^2$) bins. There is a good general agreement between the $H_{Im}$ values between the two experiments.
For reference, we also show the VGG predictions in Fig.~\ref{fig:fithallb_all} 
with stars. We published a similar figure in Ref.~\cite{Dupre:2016mai}, where the 4-CFFs fit results were not present and to which we had added the fit results
obtained when the $A_{UL}$ and $A_{LL}$ observables entered in the fit. We will discuss these latter results in the next subsection. 

We observe the general trend that $H_{Im}$ decreases with increasing $-t$. To quantify this, we fit these $t$-dependences with an exponential function $Ae^{Bt}$, with $A$ and $B$ as free parameters. The solid lines in Fig.~\ref{fig:fithallb_all} show the fit of the red empty squares and the dashed lines the fit of the black solid squares.
We will discuss the results for the amplitude $A$ and for the slope $B$ in the next section.

\begin{figure}[htbp] 
\includegraphics[width=0.5\textwidth]{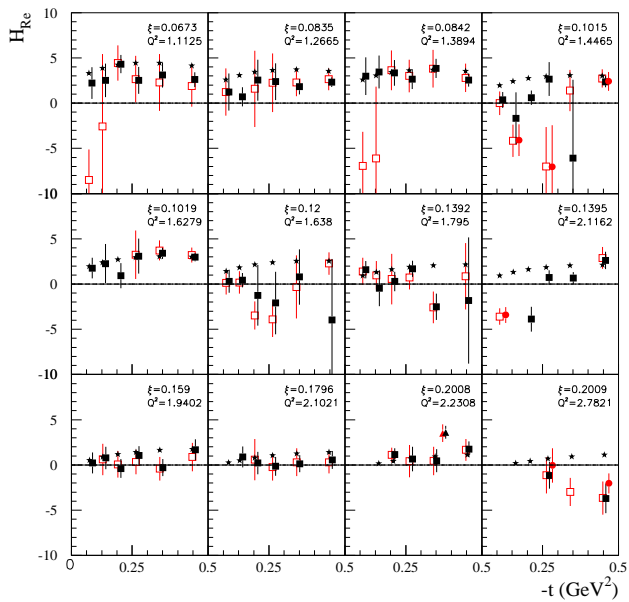}
\caption{$H_{Re}$ as a function of $t$ for 12 CLAS ($x_B$, $Q^2$) bins,
fitting only $\sigma$ and $\Delta\sigma_{LU}$.
Red open squares: results of the CLAS data fit with the 8 CFFs as free parameters.
Black solid squares: results of the CLAS data fit with the 4 CFFs $H_{Re}$, $\tilde H_{Re}$,
$H_{Im}$ and $\tilde H_{Im}$ as free parameters, the other 4 CFFs being set to their
VGG value. Red solid circles: results of the fit with the 8 CFFs as free parameters,
fitting in addition $A_{UL}$ and $A_{LL}$. The black solid squares,
as well as the red solid circles, have been slightly
shifted to the right of the red open squares for visibility.
Red solid triangle (lowest row, third column): result of the Hall A data fit 
with the 8 CFFs as free parameters.
Black solid triangle (lowest row, third column): result of the Hall A data fit 
with the 4 CFFs $H_{Re}$, $\tilde H_{Re}$,
$H_{Im}$ and $\tilde H_{Im}$ as free parameters, the other 4 CFFs being set to their
VGG value.
Stars: VGG predictions. In the figure, we have converted the $x_B$ values to $\xi$ values.}
\label{fig:hre}
\end{figure}

As we saw with our simulation studies in the previous section, fitting $\sigma$ and $\Delta\sigma_{LU}$ can also lead to some constraints on the $H_{Re}$ CFF
(in Figs.~\ref{fig:himvschi2_sim_nosmear} and~\ref{fig:himvschi2_sim_smear}, 
lower limits could be obtained).
We obtained for this CFF results with both error bars finite, for 12 CLAS ($x_B$, $Q^2$) bins, 
out of 20. Figure~\ref{fig:hre} shows these results. While for the vast majority of points there is good agreement between the results of the 8-CFFs (red open squares) and of the 4-CFFs (black solid squares) fits, for a few points there are disagreements between the results of the two approaches.
This is the case for instance for the first $t$ point of the upper left plot in Fig.~\ref{fig:hre}. 
Such differences had not been observed previously
for $H_{Im}$. We notice that this disagreement actually occurs 
when the 8 CFFs fit yields a result far from the VGG prediction.
For the first $t$ point of the upper left plot in Fig.~\ref{fig:hre},
the 8 CFFs fit result has actually an opposite sign to the VGG prediction.
We saw in Section~\ref{lab:4cffs} that the 4-CFFs fit was reliable when
the 4 non-fitted CFFs were set to their true value. For real data, we
assumed that VGG could make up a good guess for such ``true" value. 
However, the important disagreement between the 8-CFFs fit and the VGG prediction for a few particular ($x_B$, $Q^2$, $t$) bins hints that VGG actually does not estimate correctly the ``true" values for these unfitted CFFs, for these
specific kinematics. We shall therefore conclude that the 4-CFFs fits,
which, we recall, are model-dependent, are not reliable for these few bins
where there is an important disagreement between the results of the 8-CFFs and the 4-CFFs fits.

We also show in Fig.~\ref{fig:hre} the only $H_{Re}$ value, i.e.
with finite negative and positive error bars, that
we could get out of the Hall A $\sigma$ and $\Delta\sigma_{LU}$ data.
It lies in the third column plot of the lowest row in Fig.~\ref{fig:hre}, which is 
the CLAS ($x_B$, $Q^2$) bin which approximately matches the Hall A KINX3 bin.
It is represented by the red (black) solid triangle for the 8 (4) CFFs
free parameters fit. Both the 8-CFFs and the 4-CFFs fits give similar values.
There seems to be an incompatibility between these Hall A $H_{Re}$ values and the neighboring CLAS $H_{Re}$ values. It was pointed out in Ref.~\cite{Jo:2015ema}
that there was probably some tension between the Hall A and the CLAS unpolarized
cross sections. This discord in the data might explain the
difference in the $H_{Re}$ fitted values between the two experiments,
as $H_{Re}$ is one important contributor to the unpolarized 
cross section~\cite{Belitsky:2001ns}. 
We notice that there is not such conflict in the
beam-polarized cross sections. This may explain why the $H_{Im}$ values
were found compatible between the Hall A and CLAS experiments
(see Fig.~\ref{fig:fithallb_all}). 

We finally display in Fig.~\ref{fig:hre}, with red circles, the results that we obtain for $H_{Re}$ when we fit, with 8 CFFs, $A_{UL}$ and $A_{LL}$ from CLAS in addition to $\sigma$ and $\Delta\sigma_{LU}$. We discuss these $A_{UL}$ and $A_{LL}$ fits in more details in the next subsection.
For the moment being, we observe that these points are in very good agreement 
with the $H_{Re}$ values obtained from the fit of the CLAS $\sigma$ and 
$\Delta\sigma_{LU}$ data only. 

The $t$-dependence of $H_{Re}$ doesn't appear simple. There seems to be
several structures, in particular changes of signs. 
We notice that such zero-crossings for $H_{Re}$ are predicted 
by models (at least for HERMES kinematics,
see Refs.~\cite{Guidal:2009aa,Guidal:2010de,Kumericki:2012yz}). The $H_{Re}$ CFF is in general not easy to interpret and model, as it results from a weighted integral of $x$ over its whole range ($-1$ to $+1$). We expect that our fit results
will permit to constrain significantly the models.

\subsubsection{Fits of $\sigma$, $\Delta\sigma_{LU}$, $A_{UL}$
and $A_{LL}$.}

We now take into account the longitudinally polarized target asymmetries measured by CLAS, fitting simultaneously the four observables $\sigma$, $\Delta\sigma_{LU}$, $A_{UL}$ and $A_{LL}$. There are 15 ($x_B$, $Q^2$, $t$) bins for which the kinematics is approximately common between the $\sigma$, $\Delta\sigma_{LU}$ and the $A_{UL}$ and $A_{LL}$ measurements.

\begin{figure}[htbp] 
\includegraphics[width=0.4\textwidth]{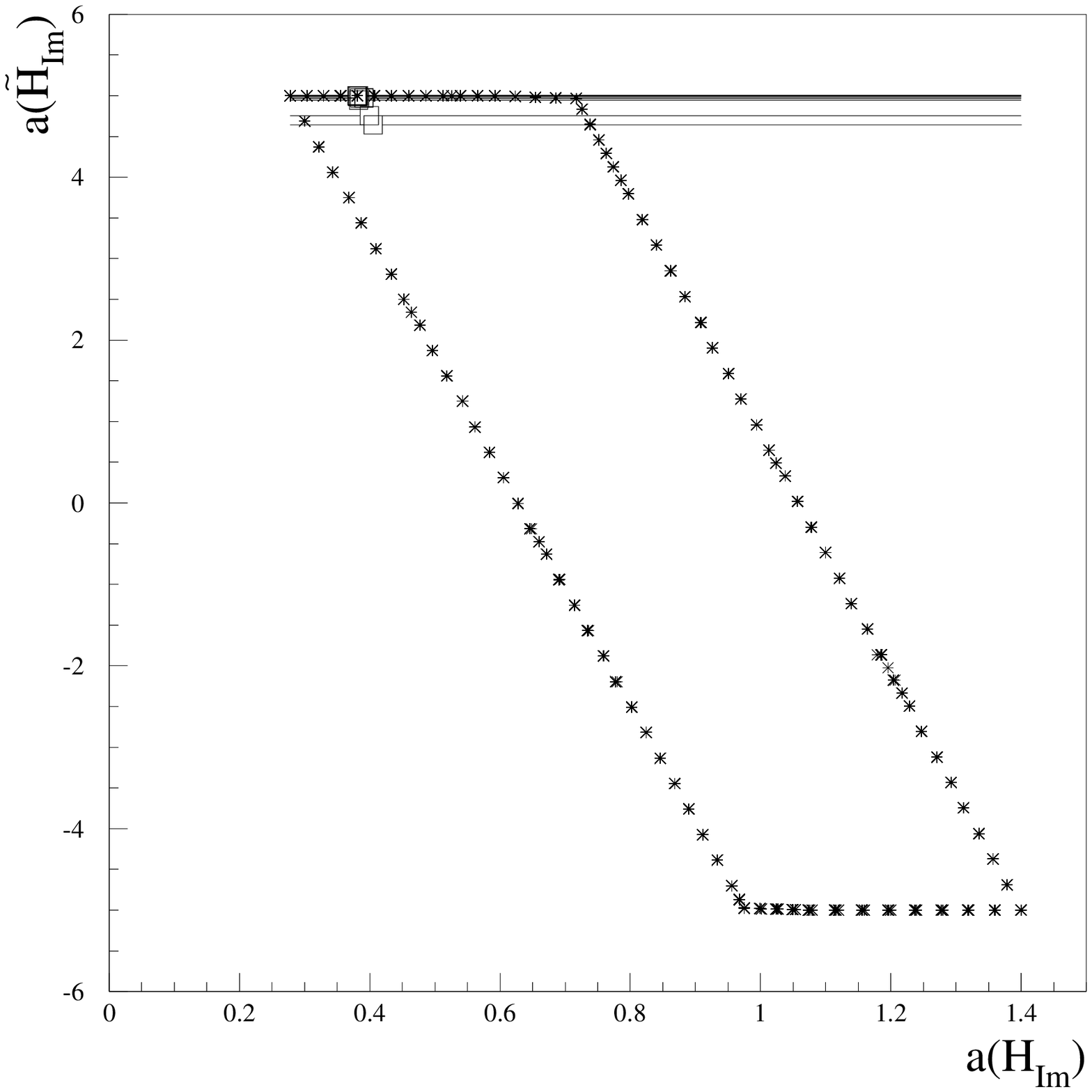}
\includegraphics[width=0.4\textwidth]{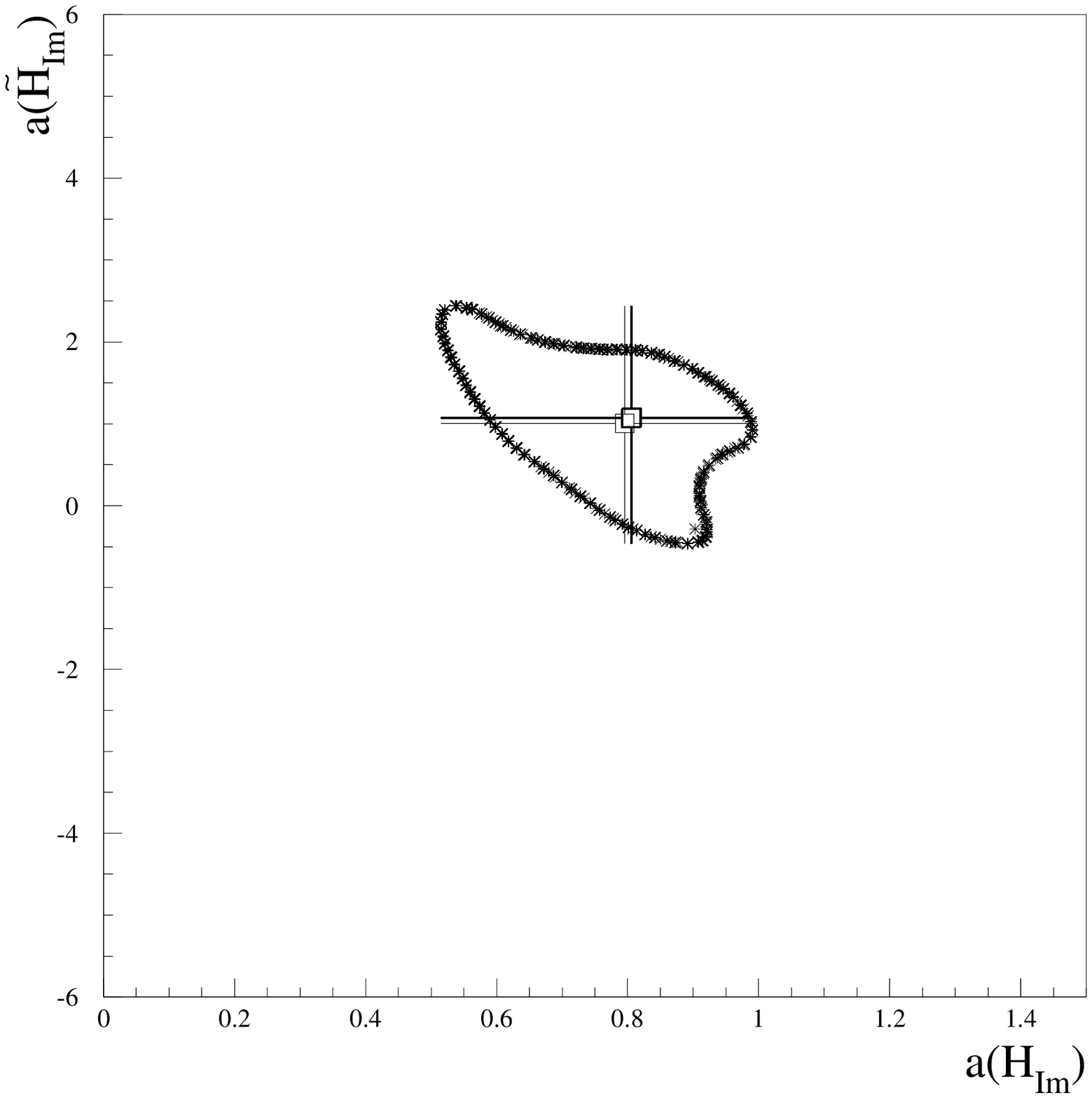}
\caption{Top: contour plot of $\tilde H_{Im}$ vs $H_{Im}$ when only
$\sigma$ and $\Delta\sigma_{LU}$ are fitted (the error bars on $\tilde H_{Im}$ are not shown here
as they are infinite). Bottom: contour plot of $\tilde H_{Im}$ 
vs $H_{Im}$ when $\sigma$, $\Delta\sigma_{LU}$, $A_{UL}$ and $A_{LL}$ are fitted.
The open squares show the minimum $\chi^2$ values and the ``asterisk curves"
the contour corresponding to $\chi^2_{min}+1$. The plots have been produced by superimposing the results ($\chi^2_{min}$ points and contours) of 50 fits differing by their starting points. The ($x_B$, $Q^2$, $t$) kinematics of the left plot are (0.2448, 2.1168 GeV$^2$, 0.2032 GeV$^2$) and the one of the right plot are (0.2556, 1.9700 GeV$^2$, 0.2343 GeV$^2$).}
\label{fig:cont_claseg1}
\end{figure}

We present in Fig.~\ref{fig:cont_claseg1} the comparison, for one given
($x_B$, $Q^2$, $t$) bin, of the $a(\tilde H_{Im})$ vs $a(H_{Im})$
contour plots when one fits only $\sigma$ and $\Delta\sigma_{LU}$ (top plot)
and one fits $\sigma$, $\Delta\sigma_{LU}$, $A_{UL}$ and $A_{LL}$ (bottom plot).
This comparison is done for ($x_B$, $Q^2$, $t$) bins at approximately the same 
kinematics: (0.2448, 2.1168 GeV$^2$, -0.2032 GeV$^2$) for $\sigma$ and $\Delta\sigma_{LU}$ 
and (0.2556, 1.9700 GeV$^2$, -0.2343 GeV$^2$) for $A_{UL}$ and $A_{LL}$.
Both plots are obtained with 8-CFFs fits.
When only $\sigma$ and $\Delta\sigma_{LU}$ enter the fit (top plot),
one sees that $a(\tilde H_{Im})$ is not constrained and can take
any value between -5 and +5. These limits on $a(\tilde H_{Im})$ determine the error on $a(H_{Im})$, as was mentioned in Section~\ref{sect_smeared_ps}. If $a(\tilde H_{Im})$ were allowed to vary beyond $\pm$ 5, the error on $a(H_{Im})$ would be larger. The correlation between the two CFFs $H_{Im}$ and $\tilde H_{Im}$ is clear from this plot. The bottom plot
of Fig.~\ref{fig:cont_claseg1} shows that the introduction
of $A_{UL}$ in the fit constrains $a(\tilde H_{Im})$ and,
as a consequence, strongly reduces the error bars on $a(H_{Im})$. 
$\tilde H_{Im}$ is indeed known to be an important contributor to $A_{UL}$~\cite{Belitsky:2001ns}.

\begin{figure}[tbp] 
\includegraphics[width=0.5\textwidth]{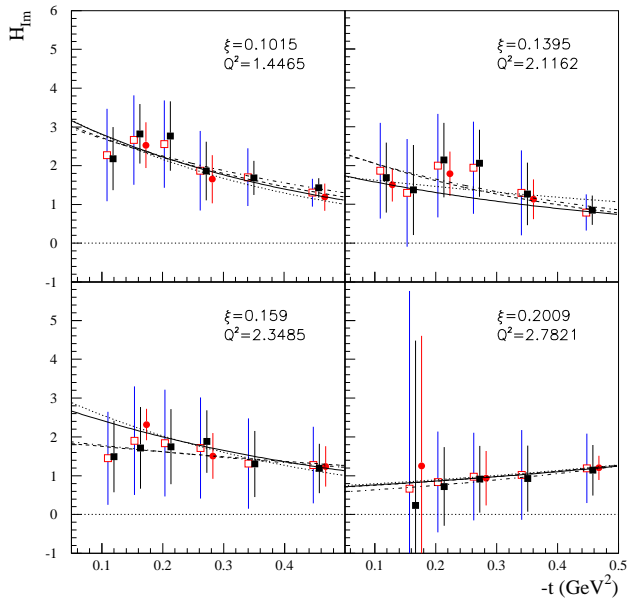}
\caption{The $H_{Im}$ CFF as a function of $t$ for 4 CLAS ($x_B$, $Q^2$) bins
where the four observables $\sigma$, $\Delta\sigma_{LU}$, $A_{UL}$ and $A_{LL}$
can be fitted simultaneously.
Red open squares: results of the fit of $\sigma$ and $\Delta\sigma_{LU}$
with the 8 CFFs as free parameters. Black solid squares: results of the fit of $\sigma$ 
and $\Delta\sigma_{LU}$with the 4 CFFs $H_{Re}$, $\tilde H_{Re}$,
$H_{Im}$ and $\tilde H_{Im}$ as free parameters, the other 4 CFFs being set to their
VGG values. Red circles: results of the fit of $\sigma$, 
$\Delta\sigma_{LU}$, $A_{UL}$ and $A_{LL}$ with the 8 CFFs as free parameters.
The black solid squares and, in some cases the red circles, are shifted to the right of the red open square points for visibility.
The dashed line shows the fit of the 6 red open squares
(i.e. the 8-CFFs fit of $\sigma$ and $\Delta\sigma_{LU}$). 
The dash-dotted line shows the fit of the 6 black solid squares
(i.e. the 4-CFFs fit of $\sigma$ and $\Delta\sigma_{LU}$). 
The dotted line shows the fit of the 3 red circles
(i.e. the 8-CFFs fit of $\sigma$, $\Delta\sigma_{LU}$, $A_{UL}$ and $A_{LL}$). 
The solid line shows the fit of the 3 red circles and the 3 red open squares whose $t$-values are different from the red circles (i.e. the 8-CFFs fit of $\sigma$, $\Delta\sigma_{LU}$, $A_{UL}$ and $A_{LL}$ and of $\sigma$, $\Delta\sigma_{LU}$ when only these two observables are available).}
\label{fig:fithallb_eg1}
\end{figure}

Figure~\ref{fig:fithallb_eg1} shows with the red circles the results for $H_{Im}$ at the 4 ($x_B$, $Q^2$) bins (corresponding to 12 ($x_B$, $Q^2$, $t$) bins) for which the 4 observables $\sigma$, $\Delta\sigma_{LU}$, $A_{UL}$ and $A_{LL}$ can be simultaneously fitted. 
There is in principle a fifth ($x_B$, $Q^2$) bin where such a simultaneous fit can be done but the fitted $H_{Im}$ has infinite error bars due to the large uncertaintities in the experimental data. 

We also display in Fig.~\ref{fig:fithallb_eg1} with red open and black solid
squares the results from the fit of only $\sigma$ 
and $\Delta\sigma_{LU}$, which are taken from Fig.~\ref{fig:fithallb_all}.
We observe in general an excellent compatibility between all the points:
the 8-CFFs fit of $\sigma$ and $\Delta\sigma_{LU}$ (red open squares),
the 4-CFFs fit of $\sigma$ and $\Delta\sigma_{LU}$ (black solid squares)
and the 8 CFFs fit of $\sigma$, $\Delta\sigma_{LU}$, $A_{UL}$ and $A_{LL}$
(red solid circles). 

In Fig.~\ref{fig:fithallb_all}, the red triangles have in general smaller error bars than the squares. This can easily be understood from Fig.~\ref{fig:cont_claseg1}: adding the extra constraints from $A_{UL}$ and $A_{LL}$ reduces the correlation between $H_{Im}$ and $\tilde H_{Im}$ and therefore the error on both CFFs. A particularly illustrative example is the
red solid circle at the smallest $| t |$-value in the lower left plot of Fig.~\ref{fig:fithallb_eg1} (($x_B/\xi$, $Q^2$)=(0.2744/0.1590, 2.3485 GeV$^2$)), where one goes from a precision of $\approx$85\% (red open square) to $\approx$70\% (black solid square) to $\approx$20\% (red solid circle) in the extraction of the $H_{Im}$ CFF.

We fit in Fig.~\ref{fig:fithallb_eg1}, for each ($x_B$, $Q^2$) bin,
the $t$ dependence of the $H_{Im}$ values that we extracted.
We use an exponential function of the form $Ae^{Bt}$ with $A$ and $B$ as free parameters.
The dashed line shows the fit of the 6 red open squares (i.e. the 8-CFFs fit of $\sigma$ and $\Delta\sigma_{LU}$). 
The dash-dotted line shows the fit of the 6 black solid squares (i.e. the 4-CFFs fit of $\sigma$ and $\Delta\sigma_{LU}$). 
The dotted line shows the fit of the 3 red circles (i.e. the 8-CFFs fit of $\sigma$, $\Delta\sigma_{LU}$, $A_{UL}$ and $A_{LL}$). 
The solid line shows the fit of the 3 red circles and of the 3 red open squares whose $t$-values are different from the red circles (i.e. the 8 CFFs fit of $\sigma$, $\Delta\sigma_{LU}$, $A_{UL}$ and $A_{LL}$ and of $\sigma$, $\Delta\sigma_{LU}$ when only these two observables are available). 
We will discuss the results of the $A$ and $B$ values and their
interpretation in the next section.

\begin{figure}[htbp] 
\includegraphics[width=0.5\textwidth]{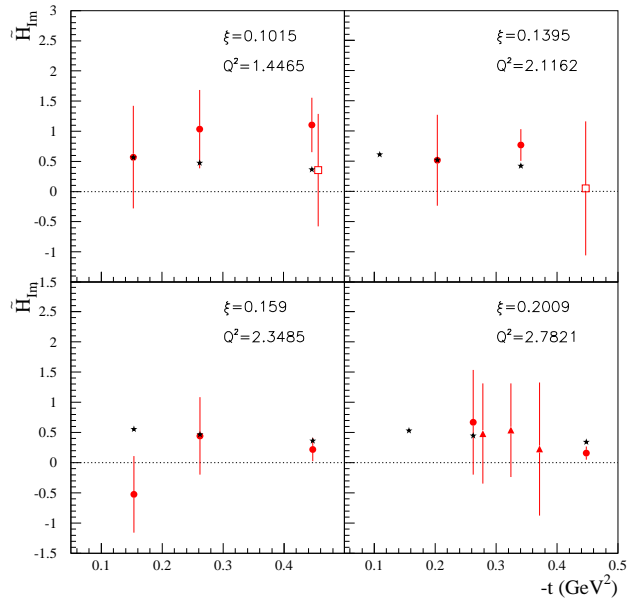}
\caption{The $\tilde H_{Im}$ CFF as a function of $t$ for 4 CLAS ($x_B$, $Q^2$) bins.
Red circles: results of the fit of $\sigma$, 
$\Delta\sigma_{LU}$, $A_{UL}$ and $A_{LL}$ with the 8 CFFs as free parameters.
Red empty squares: results of the fit of $\sigma$ and $\Delta\sigma_{LU}$ only, from CLAS.
Red triangles: results of the fit of $\sigma$ and $\Delta\sigma_{LU}$ only, from Hall A.
For visibility, the red empty square of the upper left plot has been 
slightly shifted to the right of the red circle. Stars: VGG predictions.}
\label{fig:htim}
\end{figure}

From the simultaneous fit of $\sigma$, $\Delta\sigma_{LU}$,
$A_{UL}$ and $A_{LL}$, we can also extract the $\tilde H_{Im}$ CFF.
The red circles in Fig.~\ref{fig:htim} show the results that we obtained.
We didn't obtain results for $\tilde H_{Im}$ with both error bars finite for each of the 12 ($x_B$, $Q^2$, $t$) bins of Fig.~\ref{fig:fithallb_eg1}. 
As seen in the simulation section, in some cases and particular kinematics it is also possible to get a constraint on $\tilde H_{Im}$ only from the fit of $\sigma$ and $\Delta\sigma_{LU}$. We show the $\tilde H_{Im}$ values resulting
from the fit of the CLAS $\sigma$ and $\Delta\sigma_{LU}$ with red empty squares in Fig.~\ref{fig:htim}.
Similarly, three $\tilde H_{Im}$ values (in the lower right
plot of Fig.~\ref{fig:htim}) can be obtained from the fit
of the Hall A $\sigma$ and $\Delta\sigma_{LU}$'s. These results obtained
from the fit of only two observables are well compatible with those obtained
from the fit of four observables. Still, the gain of using $A_{UL}$ and $A_{LL}$ in the fit is obvious: more precise results on $\tilde H_{Im}$ and more kinematics for which $\tilde H_{Im}$ can be extracted. For reference, we also show in Fig.~\ref{fig:htim} the VGG prediction for $\tilde H_{Im}$ 
with stars. When there are VGG predictions and no fit result for $\tilde H_{Im}$,
it means that there were $A_{UL}$ and $A_{LL}$ data but that the fit didn't 
converge and/or ended up with non-finite error bars.
Given the scarce data and their unertainties, we do not carry out a 
fit of the $t$-dependence. However, it is clear by eye that the $t$-dependency is quite flat, 
much more than for $H_{Im}$.

In addition to the $H_{Im}$ and $\tilde H_{Im}$ CFFs, the $H_{Re}$ CFF was also obtained in the simultaneous fit of $\sigma$, $\Delta\sigma_{LU}$, $A_{UL}$ and $A_{LL}$. In principle, the $A_{LL}$ observable has sensitivity to the real part of the DVCS amplitude and to $H_{Re}$ in particular~\cite{Belitsky:2001ns}.
In Fig.~\ref{fig:hre} the results that we obtained with these additional observables in the fit are shown by red solid circles, for the few ($x_B$, $Q^2$, $t$) bins for which both error bars of $H_{Re}$ are finite. 
In general, the results confirm those obtained with
the fit of only $\sigma$ and $\Delta\sigma_{LU}$ (red open squares). The 
experimental precision on $A_{LL}$ doesn't seem to be sufficient to
dramatically change the $H_{Re}$ results obtained by the fit of only
$\sigma$ and $\Delta\sigma_{LU}$. Only for the largest $x_B$ bin
(lower right plot of Fig.~\ref{fig:hre}), one can see an effect
as the red solid circles show a somewhat smaller $H_{Re}$ magnitude and smaller error bars than the red open squares, although all values are compatible within error bars.

In conclusion of this section, we have obtained constraints on the $H_{Im}$ CFF from the simultaneous fit of $\sigma$ and $\Delta\sigma_{LU}$. The relative error bars range from $\approx$40\% to $\approx$100\%, depending on the kinematics and on the experiment (CLAS or Hall A), in the case of the quasi-model-independent 8-CFFs fit. The 4-CFFs approach can decrease these uncertainties to $\approx$10\% in some cases, but this is at the price of a model-dependent input (i.e. fixing the four non-varying CFFs to a model value).
An important improvement is achieved by introducing the additional 
$A_{UL}$ and $A_{LL}$ observables in the 8-CFFs fit. The drawback is the limited amount of data available as it is more challenging to measure
polarized-target observables. In addition to the $H_{Im}$ CFF, some
constraints on the $H_{Re}$ CFF can be extracted from the simultaneous
fit of $\sigma$ and $\Delta\sigma_{LU}$ (with very little improvement from
the $A_{UL}$ and $A_{LL}$ observables input) as well as on the $\tilde H_{Im}$ CFF
with the input of $A_{UL}$.

\section{Physics interpretation}

In this section, we will discuss how to obtain a tomographic image of the proton, i.e. the $x$-dependence of the charge radius of the proton, from
the $\xi$ and $t$-dependencies of the $H_{Im}$ CFF that we just extracted with our fitting procedure.

\begin{figure}[htbp] 
\begin{center}
\includegraphics[width=0.45\textwidth]{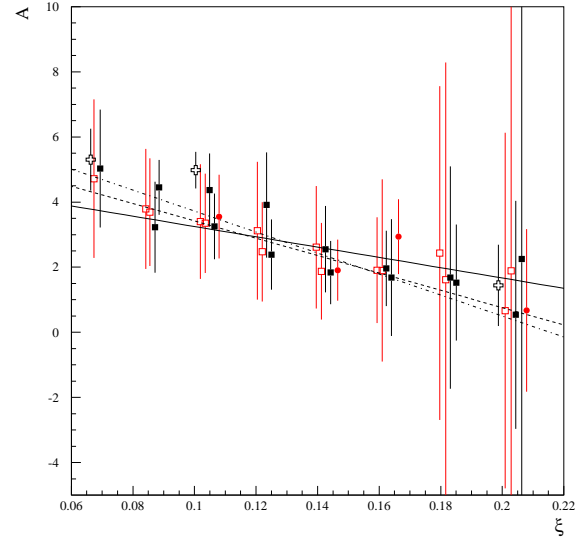}
\includegraphics[width=0.45\textwidth]{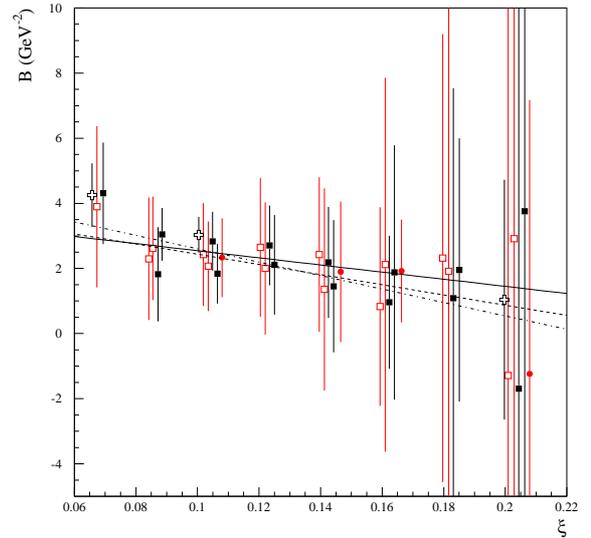}
\caption{Top: Amplitude $A$ of the exponential fit of $H_{Im}$ as a function of $\xi$, corresponding to the extrapolated value of $H_{Im}$ at $t = 0$, as a function of $\xi$. Bottom: $t$-slope $B$ of the exponential fit of $H_{Im}$, as a function of $\xi$. 
Red open squares: 8 CFFs fit of CLAS $\sigma$ and $\Delta\sigma_{LU}$'s. Black solid squares: 4 CFFs fit of CLAS $\sigma$ and $\Delta\sigma_{LU}$'s with the other CFFs set to their VGG values. Red circles: 8 CFFs fit of CLAS $\sigma$, $\Delta\sigma_{LU}$, $A_{LL}$ and $A_{UL}$'s. Black open crosses: results quoted in Ref.~\cite{Jo:2015ema},
i.e. obtained with a 4 CFFs fit with the other CFFs set to 0.
When there are two points for the same $\xi$ value, one of the red solid squares, black solid squares and red circles have been slightly shifted to the right for sake of better visibility.
The black open crosses have been slightly shifted to the left of the red open squares also for better visibility.
The dashed lines show a linear fit of only the red open squares. 
The dash-dotted lines show a linear fit of only the black solid squares. 
The solid lines show a linear fit of the 4 red solid circles and of the 
4 red open squares when their $t$-values are different from the red solid circles. The latter corresponds with the 8 CFFs fit of 
the CLAS $\sigma$, $\Delta\sigma_{LU}$, $A_{UL}$ and $A_{LL}$'s
and of the CLAS $\sigma$, $\Delta\sigma_{LU}$'s when the $t$ values are different.
When there are two points for the same $\xi$ value, both are included in the linear fits.
}
\label{fig:slope}
\end{center}
\end{figure}

In the following, we will parametrize the data for $H_{Im}$ of Eq.~(\ref{eq:eighte}) in the following way:
\begin{eqnarray}
H_{Im}(\xi, t) = A(\xi) e^{B(\xi) t}.   
\label{eq:him}
\end{eqnarray}
Fig.~\ref{fig:slope} shows the $\xi$-dependences of the slope $B$ and amplitude $A$
determined from the exponential fits of the $t$-dependence of $H_{Im}$ 
displayed in Figs.~\ref{fig:fithallb_all} and~\ref{fig:fithallb_eg1}. 
In this figure, we have decided to limit the upper range in $\xi$ to 0.22 as, at large
$\xi$ values, the uncertainties in $B$ and $A$ become too large to be useful
and to make an impact. 
The red open squares correspond to the 8 CFFs fit of the CLAS $\sigma$ 
and $\Delta\sigma_{LU}$'s as obtained from the solid curves of Fig.~\ref{fig:fithallb_all}.
For most of the CLAS bins, there are two $Q^2$ values for one $\xi$ value, which explains why the red open squares generally come in pairs in Fig.~\ref{fig:slope}. 
We notice, in passing, the good compatibility, within admittedly rather large error bars, of the paired points. This is a hint that $H_{Im}$ is quite independent of $Q^2$, 
and supports our starting hypothesis of working in the QCD leading-order and leading-twist framework.
In Fig.~\ref{fig:slope}, the black solid squares correspond to the 4-CFFs fit of the CLAS $\sigma$ and 
$\Delta\sigma_{LU}$'s as obtained from the dashed curves of Fig.~\ref{fig:fithallb_all}.
The red solid circles correspond to the 8 CFFs fit of the CLAS $\sigma$, $\Delta\sigma_{LU}$, $A_{LL}$
and $A_{UL}$'s,  obtained from the solid curves of Fig.~\ref{fig:fithallb_eg1}.

In spite of the large size of the errors, one can discern that, for all fit configurations, both the 
$t$-slope $B$ and the amplitude $A$ of the exponentials tend to increase as $\xi$ decreases.
To quantitatively support this qualitative impression, we fit the different sets of points by 
straight lines. The dashed curves in Fig.~\ref{fig:slope} show the fit of only the red open squares, whereas the dash-dotted curves show the fit of only the black solid squares. 
The solid curves show the fit of the 4 red solid circles and of the 4 red open squares whose $t$-values are different from the red solid circles. It is clear in Fig.~\ref{fig:slope} that all the slopes of the curves are negative, i.e. that the both $A$ and $B$ increase as $\xi$ decreases.
The numerical results of the linear fits in $\xi$ are displayed in Tables~\ref{tab:fitslopeA} and~\ref{tab:fitslopeB}.

It is important to underline the systematic nature of the error bars to properly assess the significance of these results. The errors encode the level of unknown in the subleading CFFs, therefore a solution with flat distributions would have to be compensated with significantly stronger opposite slopes for other CFFs.
At the price of more model dependence, global fits should be able to clarify how much flexibility the GPDs can have in this regard.

\begin{table}[t]
\center
\begin{tabular}{|c||c|c||c|c|}
\hline
& $i_A$ & $\Delta i_A$ & $s_A$ & $\Delta s_A$\\
\hline
8 CFFs fit of $\sigma$, $\Delta\sigma_{LU}$ & 6.09 & 2.21 & -26.6 & 17.8 \\\hline
4 CFFs fit  of $\sigma$, $\Delta\sigma_{LU}$ &  &  &  &  \\ 
(others set to VGG)  & 6.95 & 1.38 & -32.3 & 11.1 \\ 
\hline
8 CFFs fit of  &  &  & &  \\
$\sigma$, $\Delta\sigma_{LU}$, $A_{UL}$, $A_{LL}$ & 4.89 & 2.21 & -15.8 & 16.8 \\\hline
\end{tabular}
\caption{Fit results of $A$ as a function of $\xi$ (Fig.~\ref{fig:slope} top)
by the function $A=i_A+s_A \xi$ with the associated errors $\Delta i_A$
and $\Delta s_A$.}
\label{tab:fitslopeA}
\end{table}

\begin{table}[t]
\center
\begin{tabular}{|c||c|c||c|c|}
\hline
& $i_B$ & $\Delta i_B$ & $s_B$ & $\Delta s_B$\\
\hline
8 CFFs fit of $\sigma$, $\Delta\sigma_{LU}$ &  4.00 & 2.77 & -15.6 & 25.2 \\\hline
4 CFFs fit of $\sigma$, $\Delta\sigma_{LU}$ &  &  &  & \\ 
(others set to VGG)  & 4.67  & 1.74 & -20.6 & 16.1 \\ \hline
8 CFFs fit of  & &  &  &  \\
$\sigma$, $\Delta\sigma_{LU}$, $A_{UL}$, $A_{LL}$ & 3.64  & 2.44 & -11.0 & 20.0 \\ \hline
\end{tabular}
\caption{Fit results of $B$ as a function of $\xi$ (Fig.~\ref{fig:slope} bottom)
by the function $B=i_B+s_B \xi$ with the associated errors $\Delta i_B$
and $\Delta s_B$.}
\label{tab:fitslopeB}
\end{table}

For comparison purposes, we display in Fig.~\ref{fig:slope}, 
with black open crosses, the slopes and amplitudes quoted in Ref.~\cite{Jo:2015ema},
i.e. obtained with a 4 CFFs fit and the others set to 0
at the three $\xi$ values where they were extracted. Although this
method should certainly not be pursued in light of what our
simulations taught us, notably the underestimation of error bars,
we see that it allows to give some first general trends. In particular,
it allowed to first suggest the conclusions that we now corroborate in a more meticulous way, namely the rise of the amplitude $H_{Im}$ at $t=0$ with decreasing $\xi$, as well as the rise of the $t$-slope of 
$H_{Im}$ with decreasing $\xi$.

Physically, the behaviors of $A$ and $B$ can be understood as follows.
The parameter $A$ can be associated to the density of quarks in the nucleon.
So the rise of $A$ as $\xi$ decreases reflects an increase of the quark (and anti-quark)
density as smaller longitudinal momentum fractions are probed.
Furthermore, we already mentioned in the introduction that $t$ is the conjugate variable
of the transverse localization of the quarks in the nucleon (in the light-front frame).
Thus, the rise of $B$ as $\xi$ decreases reflects an increase of the
transverse size of the proton as smaller longitudinal momentum fractions
are probed. 

With these considerations, one can find a more 
physically motivated ansatz for the $\xi$-dependences
of $A$ and $B$ as compared to the linear fits shown in 
Fig.~\ref{fig:slope}.  
At small $\xi$, one expects $A$ to rise steeply as $1 / \xi$ due to the sea-quark contribution. Furthermore, $A$  
is expected to vanish in the limit $\xi \to 1$, when one valence quark takes all longitudinal momentum. 
Therefore, one can parametrize the $\xi$-dependence of $A$ by the simple one-parameter form which embodies both features: 
\begin{eqnarray}
A(\xi) &=& a_A (1 - \xi)/ \xi.
\label{eq:fita}
\end{eqnarray}
 
For the slope $B$, we expect it to sharply decrease from a Regge-type behavior when $\xi \to 0$ to 
a flat $t$-dependence in the limit $\xi \to 1$, reflecting the 
pointlike coupling to a valence quark carrying all longitudinal momentum. 
To encompass both limits, one can parametrize the $\xi$-dependence of $B$ by the following one-parameter 
ansatz in $\xi$:
 
\begin{eqnarray}
B(\xi) = a_B \, \mathrm{ln}(1/\xi). 
\label{eq:fitb}
\end{eqnarray} 

\begin{figure}[htbp]
\begin{center} 
\includegraphics[width=0.45\textwidth]{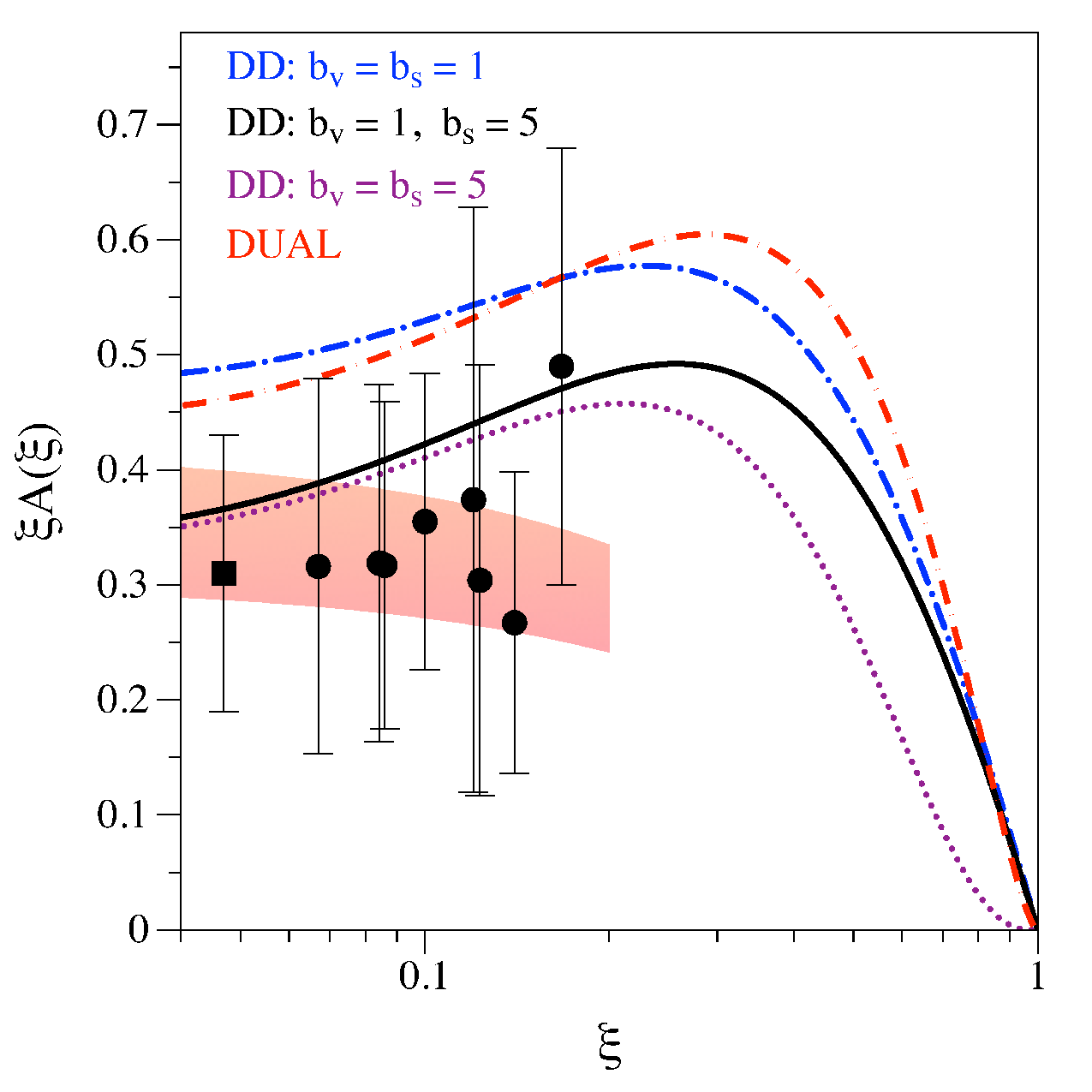}
\includegraphics[width=0.45\textwidth]{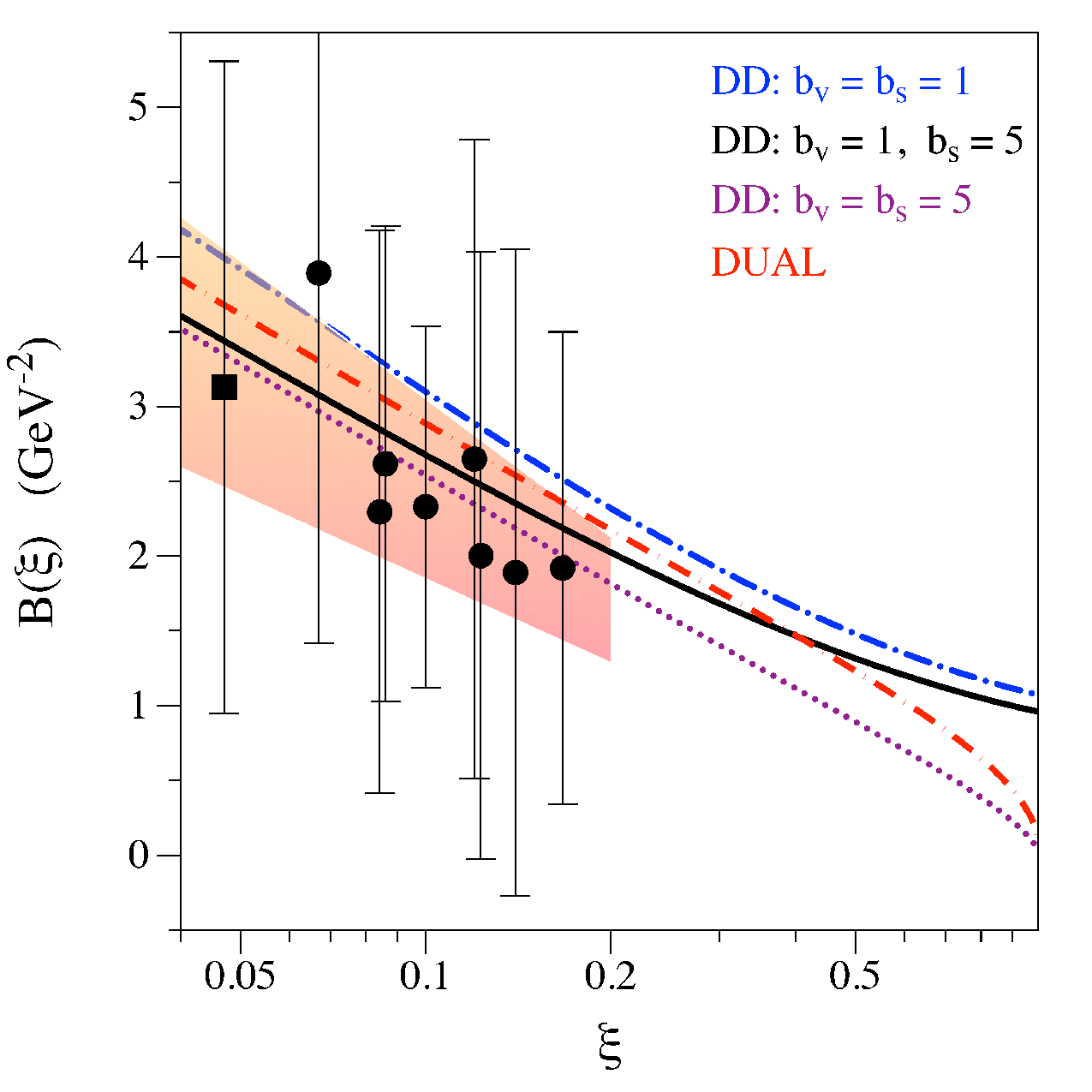}
\caption{Top: Amplitude $A$ of $H_{Im}$, multiplied by $\xi$, 
as a function of $\xi$. Bottom: $t$-slope $B$ of $H_{Im}$ as a function of $\xi$. Data points: 8 CFFs fit from CLAS (circles) as extracted in the present work and from HERMES (square) as extracted in Refs.~\cite{Guidal:2009aa,Guidal:2013rya}. 
The one-parameter fits to these data points according to Eqs.~(\ref{eq:fita}, \ref{eq:fitb}) are shown by the bands, corresponding to a $1 \sigma$ variation of $a_A$ and $a_B$, whose fit values are given by Eq.~(\ref{eq:fitAB}). 
When there are two points for the same $\xi$-value, both are included in the fits.      
The theory curves  correspond to the dual model and to the double distribution (DD) model for three choices of the valence (sea) profile parameters $b_v$ ($b_s$).}
\label{fig:fitab}
\end{center}
\end{figure}

The parameters $a_A$ and $a_B$ can be determined from a fit to the $A$ and $B$ data of Fig.~\ref{fig:slope}.
In the following, we will keep only the set of data corresponding to the 8 CFFs fit of $\sigma$, $\Delta\sigma_{LU}$, 
$A_{UL}$, $A_{LL}$, i.e. the 4 red solid circles and the 4 red open squares 
whose $t$-values are different from the red solid circles in Fig.~\ref{fig:slope}. This corresponds to the
most precise model-independent set of data in our approach. To further constrain our parametrization,
one can also add the $H_{Im}$ value that was extracted for HERMES kinematics in Refs.~\cite{Guidal:2009aa,Guidal:2013rya}
with the same technique as in the present work. This corresponds to fitting the points that we show in 
Fig.~\ref{fig:fitab}. In this figure, the black solid circles correspond to the 6 lowest $\xi$ bins of the CLAS data set of Fig.~\ref{fig:slope} and the black solid square corresponds to the HERMES point. 
Given their uncertainties larger than $100$\%, the largest $\xi$ bins 
of the CLAS data set don't bring significant information, 
and were omitted in the following discussion. Notice also that we decided to adopt a logarithmic scale for the horizontal axis (i.e. $\xi$)
and to plot $\xi A$ for the amplitude in the top plot. 
A fit to these data with the functional forms of Eqs.~(\ref{eq:fita}, \ref{eq:fitb}) yields the values:
\begin{eqnarray}
a_A =  0.36  \pm 0.06 , \quad 
a_B &=& 1.06 \pm 0.26 \; \mathrm {GeV}^{-2}.
\label{eq:fitAB}
\end{eqnarray}
The resulting fits are shown by the bands in Fig.~\ref{fig:fitab}.

We also compare in Fig.~\ref{fig:fitab} the experimentally extracted values of the amplitude $A$ and the $t$-slope $B$ with the expectations from GPD models. We use two GPD models: the dual model~\cite{Polyakov:2008xm} and the VGG double distribution model~\cite{Goeke:2001tz,Vanderhaeghen:1998uc,Vanderhaeghen:1999xj,Guidal:2004nd}. 
In the following, we will tag the latter DD to underline that it belongs to the 
generic double distribution family. We will use three choices of the valence (sea) profile 
parameters $b_v$ ($b_s$) respectively. 
For large values of these profile parameters ($b \to \infty$), the corresponding GPD $H(x,\xi,t)$ tends to the GPD $H(x,0,t)$, where the effect of the skewness (i.e. its $\xi$-dependence) disappears. The three parameter combinations are chosen to correspond with the cases where both valence and sea distributions show strong skewness ($b_v = b_s = 1$), 
where only the valence distributions shows a strong skewness ($b_v = 1, b_s = 5$), and  
where neither the valence nor the sea distributions show any strong skewness ($b_v = b_s = 5$).
For the dual model, we have used the lowest forward-like function~\cite{Polyakov:2008xm}. 
For both models, we use the same empirical forward parton distributions as input and use in both cases a 
Regge parameterization for the $t$-dependence with Regge slope parameter $1.05$~GeV$^{-2}$. The latter value is 
obtained from the requirement that the first moment of the valence GPD is fixed by the slope at $t=0$ of the proton 
Dirac form factor. We refer the reader to the review of Ref.~\cite{Guidal:2013rya} for details of these parameterizations.

Comparing the extracted data for the amplitude $A$ with theory, we notice from Fig.~\ref{fig:fitab} that in the region 
$0.05 \lesssim \xi \lesssim 0.2$ the data tend to lie systematically below the result of the dual model 
(with lowest forward-like function) and the DD models where sea quarks display a strong skewness ($b_s = 1$). 
The DD models with small skewness effects of sea-quarks ($b_s = 5$) are in good agreement with the data. 
To distinguish for the valence quarks between the cases of strong skewness ($b_v = 1$) and weak skewness ($b_v = 5$) 
will require data in the region $\xi \gtrsim 0.3$.  Such data are expected from the forthcoming dedicated DVCS program of JLab at 12 GeV. We also notice from Fig.~\ref{fig:fitab}
that the GPD models predict a maximum for $\xi A(\xi)$ around $\xi \approx 0.3$, which is due to the $x$-dependence of 
the underlying valence quark distributions. 
At present, the available data only allow to fit one parameter. Therefore, the one-parameter fit of Eq.~(\ref{eq:fita}), 
shown by the band in Fig.~\ref{fig:fitab}
shows a monotonic decrease from its constrained value at small $\xi$ to its (imposed) vanishing behavior at $\xi \to 1$. 
Once data will become available around $\xi \approx 0.3$, one can try more elaborate fit functions encompassing 
the intermediate structures in the valence region as predicted by the GPD models.

For the exponential $t$-slope $B(\xi$), both the data as well as the models follow a $\ln (1/\xi)$ behavior, 
thus leading to an increase of the slope as $\xi$ decreases. Only for $\xi \gtrsim 0.5$, which is beyond the reach of the current data, some differences between the models appear. 

We now seek to relate the increasing $t$-slope $B(x)$ when $x$ decreases with the 
variation of the spatial size of the proton when probing partons with 
different longitudinal momentum fraction $x$. 
For this purpose, we relate it to the helicity-averaged transverse 
charge distribution in the proton, denoted by $\rho$, 
which is obtained through a 2-dimensional 
Fourier transform of the FF $F_1$ as~\cite{Burkardt:2000za}:
\begin{equation}
\rho({\bf b_\perp})=\int
\frac{d^2 {\bf \Delta_\perp}}{(2\pi)^2}e^{- i {\bf b_\perp \cdot \Delta_\perp}} 
F_1(-{\bf \Delta}^2_\perp).
\label{eq:fourier1}
\end{equation}
Here ${\bf b_\perp}$ denotes the quark position in the plane transverse to the 
longitudinal momentum of a fast moving proton, and the conjugate momentum variable 
${\bf \Delta_\perp}$ denotes the momentum transfer towards the proton.
The squared radius of this unpolarized 2-dimensional transverse charge distribution 
in the proton is then defined as:
\begin{equation}
\langle b^2_\perp \rangle  = \int d^2 {\bf b_\perp} {\bf b}^2_\perp \rho({\bf b_\perp}). 
\label{eq:cr1}
\end{equation}
The squared radius of the proton FF $F_1$, denoted by $\langle r_1^2 \rangle$, 
is usually defined through its Taylor expansion:
\begin{equation}
F_1(- {\bf \Delta}^2_\perp) = 1 - \langle r_1^2 \rangle  
{\bf \Delta}^2_\perp / 6 + {\cal O} ({\bf \Delta}^4_\perp),
\end{equation}
which allows to readily identify $\langle b^2_\perp \rangle  = 2/3 \langle r_1^2 \rangle$.  
The experimental extraction of $\langle r_1^2 \rangle $ based on elastic 
electron-proton scattering data 
yields~\cite{Bernauer:2013tpr}: $\langle r_1^2 \rangle = 0.65 \pm 0.01~\mathrm{fm}^2$, 
resulting in the empirical value for the squared radius of the proton's transverse 
charge distribution:
\begin{equation}
\langle b^2_\perp \rangle = 0.43 \pm 0.01~\mathrm{fm}^2 = 11.05 \pm 0.26~\mathrm{GeV}^{-2}.   
\label{eq:cr4}
\end{equation}

Similarly to the FFs, the $t$ variable in the GPDs is the conjugate variable
of the impact parameter. For $\xi=0$ (where one identifies $t=-\Delta_\perp^2$), 
one therefore has an impact parameter version of GPDs through a Fourier integral in 
tranverse momentum $\Delta_\perp$, which for a parton of flavor $q$ reads as~:
\begin{equation}
\rho^q(x, {\bf b_\perp})=\int
\frac{d^2 {\bf \Delta_\perp}}{(2\pi)^2}e^{- i {\bf b_\perp \cdot \Delta_\perp}}
     H^q_-(x,0,-{\bf \Delta_\perp}^2).
\label{eq:fourier}
\end{equation}
Here $H^q_-(x, 0, t)$ is the so-called non-singlet or valence GPD combination, 
defined as:
\begin{eqnarray}
H^q_-(x,0, t) \equiv H^q(x,0,t) + H^q(-x,0,t),   
\end{eqnarray}
with $0 \leq x \leq 1$.  
At $\xi$=0, the function $\rho^q(x, {\bf b_\perp})$ can then be interpreted 
as the number density of quarks of flavor $q$ with {\it longitudinal} momentum 
fraction $x$ at a given {\it transverse} distance ${\bf b_\perp}$ (relative 
to the transverse c.m.) in the proton~\cite{Burkardt:2000za}. 
Note that the transverse position of the quarks and their longitudinal momenta
are independent variables which can be determined simultaneously.  
 
Generalizing Eq.~(\ref{eq:cr1}), one can define the $x$-dependent squared radius of the quark density in the transverse plane as:
\begin{equation}
\langle b^2_\perp \rangle^q (x) = \frac{ \int d^2 {\bf b_\perp} {\bf b}^2_\perp 
\rho^q(x, {\bf b_\perp})}{\int d^2 {\bf b_\perp}  \rho^q(x, {\bf b_\perp})}. 
\label{eq:cr5}
\end{equation}
Inserting Eq.~(\ref{eq:fourier}) in Eq.~(\ref{eq:cr5}) 
allows one to express the $x$-dependent squared radius as:
\begin{equation}
\langle b^2_\perp \rangle^q (x)= - 4 \frac{\partial}{\partial {\bf \Delta}^2_\perp} 
\ln H^q_-(x,0,-{\bf \Delta_\perp}^2) \biggr| _{{\bf \Delta_\perp} = 0}.
\label{eq:crgpd}
\end{equation}
Assuming the $t$-dependence of the valence GPD $H^q_-(x,0,t)$ to be exponential of the form:
\begin{eqnarray}
H^q_-(x,0, t) = q_v(x) e^{B^0_{-}(x) t},   
\label{eq:hzeroxi}
\end{eqnarray}
with $q_v(x)$ the corresponding valence quark distribution, 
Eq.~(\ref{eq:crgpd}) then yields for each flavor $q$:
\begin{eqnarray}
\langle b^2_\perp \rangle^q (x)= 4 B^0_{-}(x). 
\label{eq:bperp1}
\end{eqnarray}
The $x$-independent squared radius is obtained from 
$\langle b^2_\perp \rangle^q (x)$ through the following 
average over $x$:
\begin{eqnarray}
\langle b^2_\perp \rangle^q 
=  \frac{1}{N_q} \int_0^1 dx \,q_v(x) \, \langle b^2_\perp \rangle^q (x), 
\label{eq:bperp2}
\end{eqnarray}
with the integrated number of valence quarks $N_u = 2$ and $N_d = 1$, for the proton. The Dirac squared radius  $\langle b^2_\perp \rangle $ 
is then obtained as the charge weighted sum over the valence quarks:
\begin{eqnarray}
\langle b^2_\perp \rangle =  2 e_u \langle b^2_\perp \rangle^u + e_d \langle b^2_\perp \rangle, 
\label{eq:bperpproton}
\end{eqnarray}
with quark electric charges $e_u = +2/3$ and $e_d = -1/3$.
A Regge ansatz for the $t$-dependence of $H_-^q(x,0,t)$ yields: 
\begin{eqnarray}
B^0_{-}(x) = a_{B^0_{-}} \ln(1/x), 
\label{eq:B0}
\end{eqnarray}
with $a_{B^0_-}$ the Regge slope. When evaluating the corresponding 
integral of Eq.~(\ref{eq:bperp2}), 
using the  empirical constraint of Eq.~(\ref{eq:cr4}) for 
$\langle b^2_\perp \rangle$, we obtain the estimate:
\begin{eqnarray}
a_{B^0_{-}} = \left(1.05 \pm 0.02 \right) \mathrm{GeV}^{-2} . 
\label{eq:aB0}
\end{eqnarray}

\begin{figure}
\begin{center} 
\includegraphics[width=0.45\textwidth]{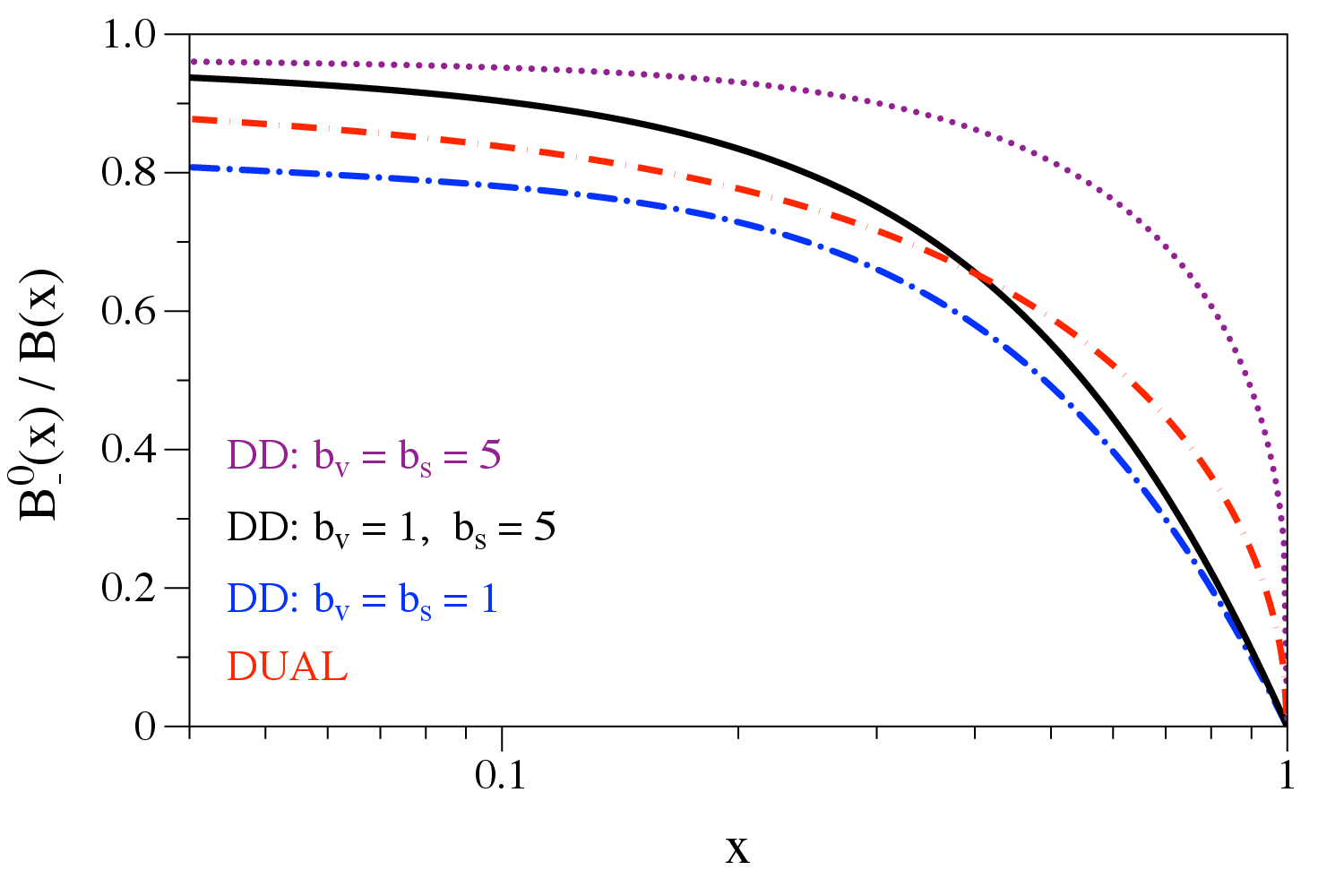}
\caption{$x$-dependence of the ratio $B^0_-(x) / B(x)$, with $B^0_-$ the exponential 
$t$-slope of $H^p_-(x,0,t)$ according to Eq.~(\ref{eq:hzeroxi}), 
and $B$ the exponential $t$-slope of $H^p_+(x,x,t)$ according to Eq.~(\ref{eq:him}).  
The theory curves correspond to the dual model (red dashed curve) and the double 
distribution (DD) model for three choices of the  valence (sea) profile parameters 
$b_v$ ($b_s$), as indicated.
}
\label{fig:B0overB}
\end{center}
\end{figure}

\begin{figure}
\begin{center} 
\includegraphics[width=0.45\textwidth]{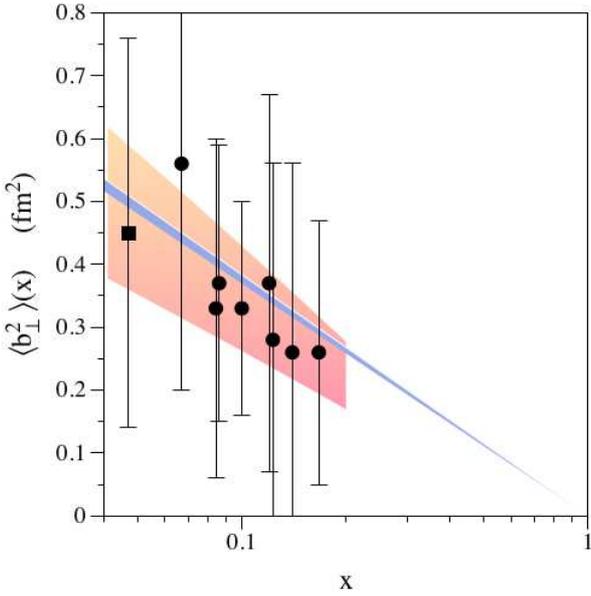}
\caption{$x$-dependence of  $\langle b^2_\perp \rangle$ for quarks in the proton. The data points correspond to the results obtained in this work for $B(x)$, as displayed in Fig.~\ref{fig:slope}. 
They have been multiplied by the correction factor $B^0_- / B $ in the $x$-range of the data, as obtained from the black curve in Fig.~\ref{fig:B0overB}. The total model uncertainty originating  from the red band for $B(x)$ in Fig.~\ref{fig:fitab}, 
and from the conversion of $B^0_-$  to $B$ (using the black solid curves in Fig.~\ref{fig:B0overB}) is shown by the red band.  
The narrow purple band shows the empirical result using the logarithmic ansatz for $B^0_-(x)$ 
of Eqs.~(\ref{eq:B0}, \ref{eq:aB0}) with the parameter $a_{B^0_-}$ 
determined from the proton Dirac radius. 
%The inset is a graphical presentation of the proton's transverse extent based on the central value of the data shown. 
}
\label{fig:bperp}
\end{center}
\end{figure}

To quantitatively compare this with the $t$-slope of $H_{Im}$ defined through 
Eq.~(\ref{eq:him}), we need to be aware of a difference. 
The experimentally measured $t$-slope $B(x)$ is for the singlet GPD combination $H_+(x,x,t)$. 
On the other hand, the $t$-slope $B^0_-(x)$ of Eq.~(\ref{eq:B0}, \ref{eq:aB0}) 
is for the valence GPD in the limit $\xi = 0$, i.e. for the function 
$H^q_-(x,0, t)$ for a quark of flavor $q$. In our analysis we will assume that the 
function $B^0_-(x)$ is the same for $u$ and $d$ quarks, in agreement with the 
observed universality of the Regge slopes for meson trajectories. 
To get some quantitative idea how large the difference between the flavor-independent slopes 
$B^0_-$ and $B$ is, we perform a study within GPD models. In Fig.~\ref{fig:B0overB}, 
we show the $x$-dependence of the ratio $B^0_-(x) / B(x)$ within the same dual and 
DD GPD models which we previously had compared to data (Fig.~\ref{fig:fitab}). 
One sees from Fig.~\ref{fig:B0overB} that $B^0_-$ is smaller than $B$, approaching the latter for small $x$. We 
also notice that $B^0_-(x)$ decreases much faster than $B(x)$ in the limit $x \to 1$. 
For the $x$ range of the available data,   $0.05 \lesssim x \lesssim 0.2$, we notice that 
the GPD models with $b_s = 5$, which were found to be compatible with both the data for 
$A$ and $B$, yield:
$B^0_-/B \simeq 0.90 - 0.95$. 
Opportunely, in the $x$-range of the data studied in this work, this correction factor is 
close to 1, and therefore the model error in passing from $B(x)$ to $B^0_-(x)$ is much smaller than the experimental error. In our 
extractions we will use the DD model for $b_v = 1$ and $b_s = 5$ (black curves in Figs.~\ref{fig:fitab}, ~\ref{fig:B0overB}) which was found to yield a good description of the available data. 
As a result, we can use the data on $B(x)$ 
to obtain a value for 
$\langle b^2_\perp \rangle(x)$  using Eq.~(\ref{eq:bperp1}), as shown in Fig.~\ref{fig:bperp} (black data points and red bands). 
These data are also compared with the result assuming the logarithmic ansatz for $B^0_-(x)$ 
of Eq.~(\ref{eq:B0}), with parameter $a_{B^0_-}$ determined from the proton Dirac radius, 
according to Eq.~(\ref{eq:aB0}). One sees that within errors both determinations are perfectly compatible. 

\begin{figure}[h]
\begin{center} 
\includegraphics[width=0.45\textwidth]{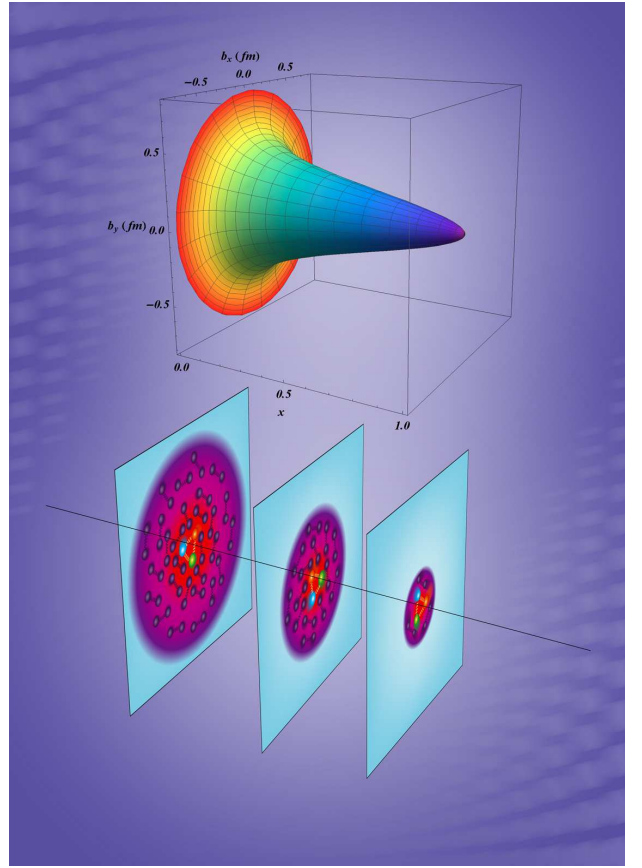}
\caption{Top panel: three-dimensional representation of the
function of Eq.~(\ref{eq:bperp1}) fitted to the data of Fig.~\ref{fig:bperp},
showing the $x$-dependence of the proton's transverse charge radius. 
Bottom panel: artistic illustration of the corresponding
rising quark density and transverse extent as a function of $x$.  
}
\label{fig:illus}
\end{center}
\end{figure}

The upper plot Fig.~\ref{fig:illus} shows a 3-dimensional representation of the fit of Fig.~\ref{fig:bperp}.
The bottom plot is an artistic view of the tomographic quark content of the proton, with
the charge radius and the density of the quarks increasing as smaller and smaller quark momentum
fractions are probed.
 
We have here extracted the $x$-dependence of the squared radius of the quark distributions in the transverse plane, 
demonstrating an increase of this radius with decreasing value of the longitudinal quark momentum fraction $x$. 
The hypotheses which have entered our work are the general framework of QCD leading-twist and leading-order, a maximum 
deviation of the values of the ``true" GPDs by a factor 5 w.r.t. to 
the VGG GPDs, and a model-dependent $\xi$-dependent correction factor to convert
the $t$-slope of the singlet to the non-singlet distributions. 
We deem that the uncertainties associated to these assumptions are included in our systematic error bars.

At this stage, we don't carry out such study for the axial charge radius because of the
quite large error bars that we obtained for $\tilde H_{Im}$ (Fig.~\ref{fig:htim}), which make it difficult to extract a precise $t$-slope.
Qualitatively, we can nevertheless say that the $t$-slope is apparently quite flat
for $\tilde H_{Im}$. This leads us to say that the axial charge
of the nucleon seems to be very concentrated, at least more than the 
electric charge, in the core of the nucleon at the currently probed $\xi$ values.

Finally, we also provide a sketch of the information which can be extracted from 
the CFF $H_{Re}$ of Eq.~(\ref{eq:eighta}). For this purpose we analyze this 
CFF using a fixed-t once-subtracted dispersion relation, which can 
be written as: 
\begin{eqnarray}
H_{Re}(\xi, t) = - \Delta(t) + {\cal P} \int_0^1 dx \, H_+(x, x, t) \, C^+(x,\xi),
\label{eq:disp}
\end{eqnarray}
where $\Delta(t)$ is the subtraction constant, which is directly related to the $D$-term form factor, see Ref.~\cite{Guidal:2013rya} for details. 
One notices that the dispersive term, corresponding to the second term on the rhs of Eq.~(\ref{eq:disp}), is in principle calculable provided one has empirical information on the CFF $H_{Im}$ over the whole $x$-range. 

\begin{figure*}[htbp] 
\begin{center} 
\includegraphics[height=7.15cm]{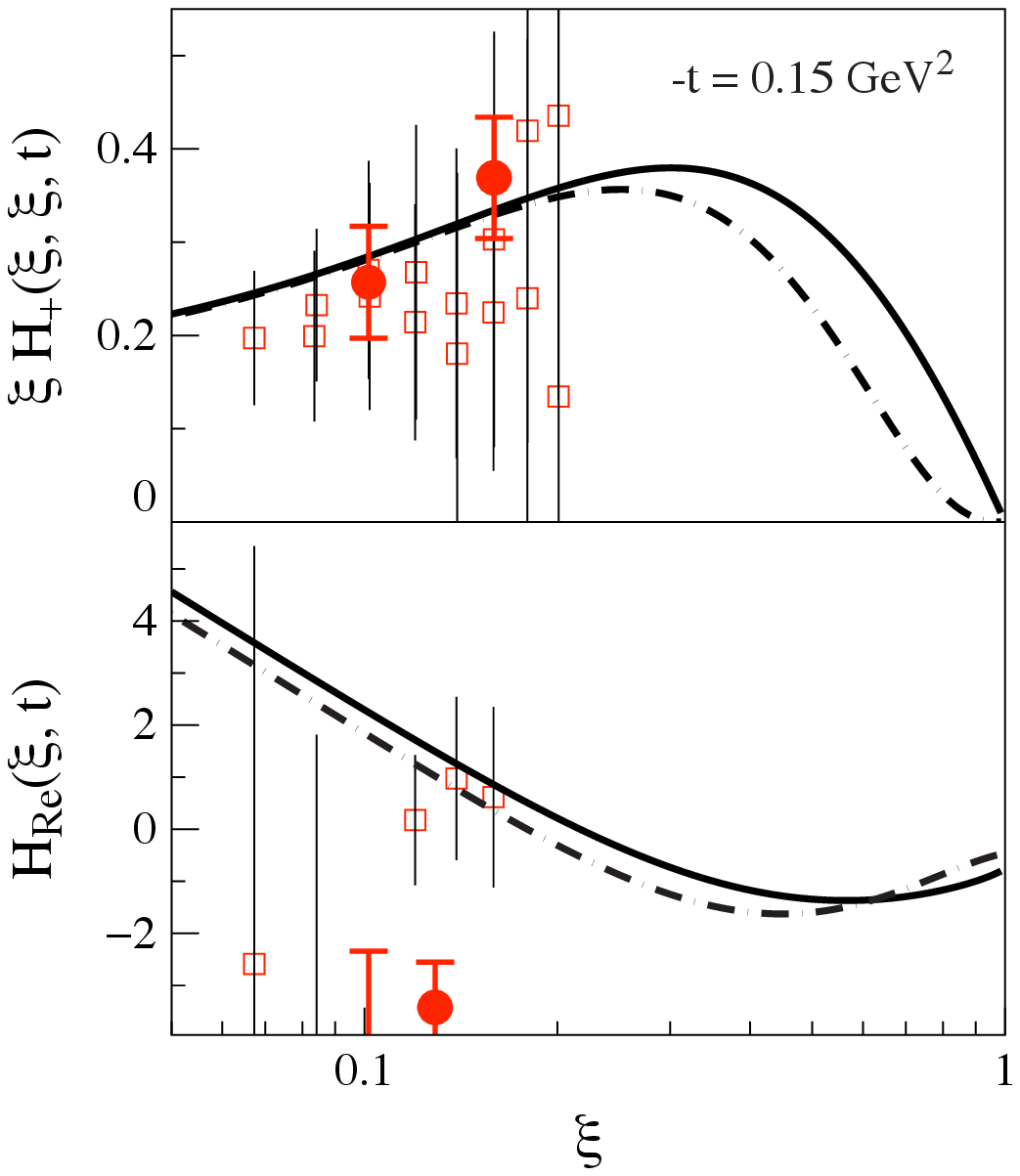}
\includegraphics[height=7.15cm]{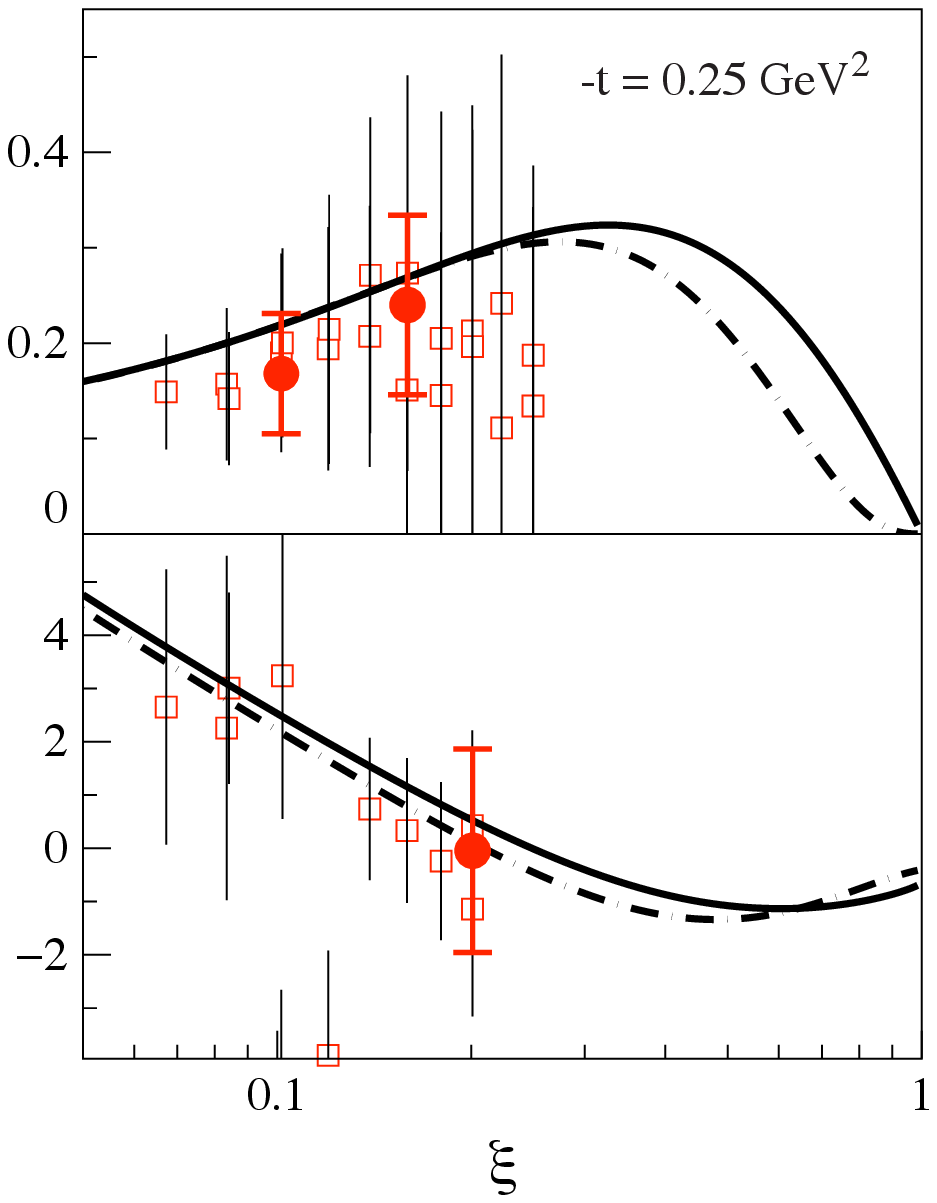}
\includegraphics[height=7.15cm]{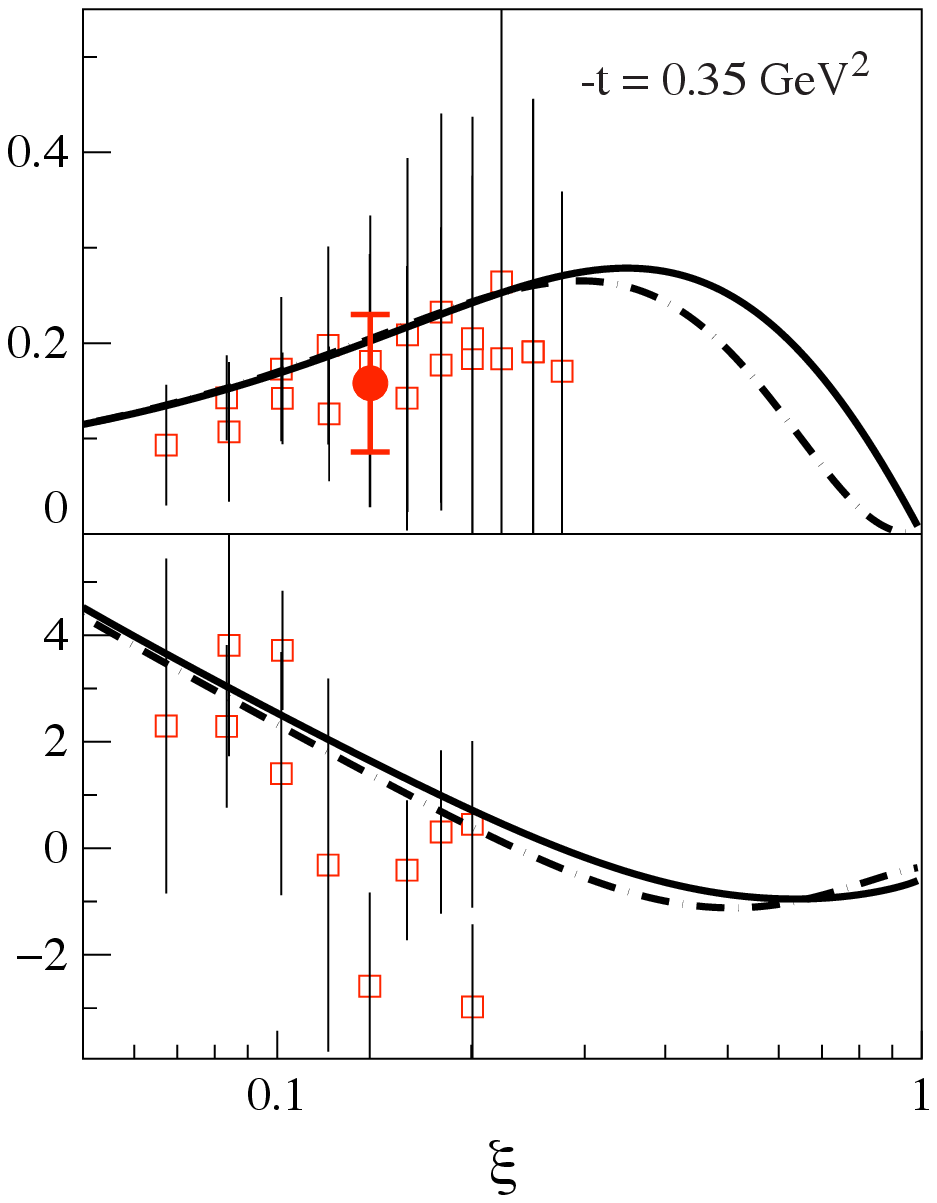}
\caption{Comparison of the $\xi$ dependence of the imaginary parts (upper plots) and real parts (lower plots) of the CFF related to the GPD $H$ for the proton for three values of $t$. 
The curves in the upper plots are based on two DD parameterizations. 
Solid curves: DD parameterization with $b_v = 1$ and $b_s = 5$; 
dashed curves: DD parameterization with $b_v = 5$ and $b_s = 5$. 
The curves in the lower plots are the dispersive calculations of the real parts according to Eq.~(\ref{eq:disp}), based on the input of the imaginary parts from the upper plots, 
and with subtraction function $\Delta(t)$ set equal to zero.  
Open squares: results of the CLAS $\sigma$ and $\Delta\sigma_{LU}$ fit. Solid circles: results of the fit to CLAS $\sigma$, $\Delta\sigma_{LU}$, $A_{UL}$, and 
$A_{LL}$ data.}
\label{fig:disphre}
\end{center}
\end{figure*}

To illustrate the power of the dispersion relation, 
we show an analysis in Fig.~\ref{fig:disphre} showing the CFFs $H_{Im}$ (top panels) and the CFFs $H_{Re}$ for three values of $-t$ for which CLAS data exist. 
We also show in the top panels two DD GPD parameterizations which give a good description of the CFF $H_{Re}$ data in the $\xi$-range of the CLAS data, but differ in the  $\xi > 0.3$ region, where no data exist at present. 
The GPD parameterization we use exactly satisfies a subtracted dispersion relation, and for the purpose of illustration we set the a-priori-unknown subtraction constant $\Delta(t)$ equal to zero. The corresponding dispersive results (second term of Eq.~(\ref{eq:disp})) are shown on the bottom panel of Fig.~\ref{fig:disphre}. We notice the importance of a large coverage in $x$ when performing the dispersion integral, because although the two GPD parameterizations are practically coinciding for $H_{Im}$ in the $\xi$-range of the data, they show a difference for $H_{Re}$ in the same $\xi$-range, which is due to their differences in the large $\xi$ region for $H_{Im}$. We compare our dispersive results for $H_{Re}$ with the direct extraction of the CFF $H_{Re}$ as performed in this work. Although the current error bars on the direct extraction of $H_{Re}$ are large due to systematics, we can observe that apart from the lowest bin in $-t$, the trend of the $\xi$ dependence which leads to a rise of $H_{Re}$ at smaller $\xi$ is well reproduced. Although our extraction method of $H_{Re}$  does not allow to extract a subtraction constant at this stage, we can see that this framework holds 
promise to extract $\Delta(t)$ once the systematic errors are reduced, through inclusion of  data which have a large sensitivity on $H_{Re}$. We also see that for the application of the dispersive framework it is important to measure the integrand $H_{Im}$ over a wide range in $\xi$, especially the $\xi > 0.3$ region, which will become possible with the forthcoming JLab 12 GeV data.  

\section{Conclusion}

In summary, we have analyzed in a GPD leading-twist and leading-order theoretical 
framework the latest $e p\to e p \gamma$ unpolarized cross sections, difference of 
beam-polarized cross sections, longitudinally polarized target single spin asymmetries
and beam-longitudinally polarized target double spin asymmetries measured
by the JLab Hall A and CLAS collaborations. We have extensively tested and validated on Monte-Carlo pseudo-data a quasi model independent
algorithm aimed at extracting CFFs from $e p\to e p \gamma$ observables. 
Applied to real data, this code has allowed us to extract constraints on the 
$H_{Im}$, $\tilde H_{Im}$ and $H_{Re}$ CFFs. From the $t$-dependence
of the $H_{Im}$ at various $x_B$ values, we have been able to derive the variation of the proton charge radius as a function of the quark's longitudinal momentum, on the domain covered by the JLab experiments.

\section*{Acknowledgments}

We are very thankful to D. Mueller and K. Kumericki for insightful 
discussions on this work.

%
% BibTeX users please use
% \bibliographystyle{}
% \bibliography{}

\end{document}